\documentclass[a4paper,11pt]{article}
\pdfoutput=1 
\usepackage{jheppub} 
\usepackage[T1]{fontenc} 
\usepackage{amsmath,amssymb,amsfonts,cases}
\usepackage{amsthm}
\usepackage{enumitem}
\usepackage{adjustbox}
\usepackage{graphicx,caption,subcaption}
\usepackage[T1]{fontenc} 
\usepackage{enumitem}  
\usepackage{empheq}
\usepackage{dsfont}
\usepackage{hyperref}
\usepackage[mathscr]{euscript}
\usepackage{dcolumn}
\usepackage{bm}
\usepackage{bbm}
\usepackage{slashed}
\usepackage{wrapfig}
\usepackage{mathtools}
\usepackage{tikz}

\newcommand{\tr}{\text{tr}}

\newcommand{\beq}{\begin{equation}}
\newcommand{\eeq}{\end{equation}}
\newcommand{\ba}{\begin{align}}
\newcommand{\ea}{\end{align}}
\newcommand{\bea}{\begin{eqnarray}}
\newcommand{\eea}{\end{eqnarray}}
\newcommand{\bi}{\begin{itemize}}
\newcommand{\ei}{\end{itemize}}
\newcommand{\ben}{\begin{enumerate}}
\newcommand{\een}{\end{enumerate}}

\def\Tr{{\rm Tr}}

\newcommand{\N}{{\mathcal N}}

\renewcommand{\a}{\alpha}
\renewcommand{\b}{\beta}

\newcommand{\e}{\epsilon}

\newcommand{\m}{\mu}

\newcommand{\Dslash}{D\!\!\!\!\slash\,}

\newcommand{\oD}{\ovl{D}}
\newcommand{\oB}{\ovl{\Box}}
\newcommand{\dww}{\delta\omega_2}
\newcommand{\dw}{\delta\omega_1}

\newcommand\eq[1]{eq.~(\ref{eq:#1})}
\newcommand\tbl[1]{tab.~\ref{tab:#1}}
\newcommand{\sn}[1]{sec.~\ref{sec:#1}}
\newcommand{\fig}[1]{fig.~\ref{fig:#1}}
\newcommand{\App}[1]{app.~\ref{app:#1}}

\newcommand\mf[1]{{\mathfrak{#1}}}
\newcommand\mc[1]{{\mathcal{#1}}}

\newcommand\CA{{\mc{A}}}
\newcommand\CB{{\mc{B}}}

\newcommand\CD{{\mc{D}}}
\newcommand\CE{{\mc{E}}}

\newcommand\CI{{\mc{I}}}
\newcommand\CJ{{\mc{J}}}
\newcommand\CK{{\mc{K}}}

\newcommand\CM{{\mc{M}}}
\newcommand\CN{{\mc{N}}}
\newcommand\CO{{\mc{O}}}

\newcommand\CR{{\mc{R}}}

\newcommand\CW{{\mc{W}}}
\newcommand\CX{{\mc{X}}}
\newcommand\CY{{\mc{Y}}}

\newcommand\veps{\varepsilon}
\newcommand\vphi{\varphi}

\newcommand\II{{\rm I\hspace{-0.02cm} I}}
\newcommand\oII{\mathring{\II}}
\newcommand\oK{\mathring{K}}
\newcommand\ovl[1]{{\overline{#1}}}

\newcommand\bc{a_\Sigma^{\text{\tiny(2d)}}}

\newcommand\dc[1]{{d_{#1}^{\text{\tiny(2d)}}}}
\newcommand\tdc[1]{{\tilde{d}_{#1}^{\text{\tiny(2d)}}}}
\newcommand\bcc{{a_{\Sigma}}}
\newcommand\dbcc[1]{{b_{#1}}}
\newcommand\dcc[1]{{d_{#1}}}
\newcommand\tdcc[1]{{\tilde{d}_{#1}}}
\newcommand\tdbcc[1]{{\tilde{b}_{#1}}}

\definecolor{cardinal}{rgb}{0.6,0,0}
\definecolor{darkgreen}{rgb}{0,0.5,0}
\definecolor{golden}{rgb}{0.92, 0.7, 0}
\definecolor{midnight}{rgb}{0, 0, 0.5}
\definecolor{darkblue}{rgb}{0.2, 0, 0.8}

\allowdisplaybreaks[4]

\title{\LARGE Weyl Anomalies of Four Dimensional Conformal Boundaries and Defects}  

\preprint{UUITP-58/21} 
   
\author[a]{Adam Chalabi,}
\author[b]{Christopher P.~Herzog,} 
\author[a]{Andy O'Bannon,}
\author[c]{Brandon Robinson,}
\author[a,d]{Jacopo Sisti}

\affiliation[a]{STAG Research Centre, Physics and Astronomy, University of Southampton, Highfield, Southampton SO17 1BJ, UK}
\affiliation[b]{Mathematics Department, King's College London, The Strand, London, WC2R 2LS, UK}
\affiliation[c]{Instituut voor Theoretische Fysica, K.U. Leuven, Celestijnenlaan 200D, BE-3001 Leuven, Belgium}
\affiliation[d]{Department of Physics and Astronomy, Uppsala University, Box 516, SE-75120 Uppsala, Sweden}
\emailAdd{a.chalabi@soton.ac.uk}
\emailAdd{ christopher.herzog@kcl.ac.uk}
\emailAdd{a.obannon@soton.ac.uk}
\emailAdd{brandon.robinson@kuleuven.be}
\emailAdd{jacopo.sisti@physics.uu.se}

\abstract{Motivated by questions about quantum information and classification of quantum field theories, we consider Conformal Field Theories (CFTs) in spacetime dimension $d\geq 5$ with a conformally-invariant spatial boundary (BCFTs) or $4$-dimensional conformal defect (DCFTs). We determine the boundary or defect contribution to the Weyl anomaly using the standard algorithm, which includes imposing Wess-Zumino consistency and fixing finite counterterms. These boundary/defect contributions are built from the intrinsic and extrinsic curvatures, as well as the pullback of the ambient CFT's Weyl tensor. For a co-dimension one boundary or defect (i.e.\ $d=5$), we reproduce the $9$ parity-even terms found by Astaneh and Solodukhin, and we discover $3$ parity-odd terms. For larger co-dimension, we find $23$ parity-even terms and $6$ parity-odd terms. The coefficient of each term defines a ``central charge'' that characterizes the BCFT or DCFT. We show how several of the parity-even central charges enter physical observables, namely the displacement operator two-point function, the stress-tensor one-point function, and the universal part of the entanglement entropy. We compute several parity-even central charges in tractable examples:  monodromy and conical defects of free, massless scalars and Dirac fermions in $d=6$; probe branes in Anti-de Sitter (AdS) space dual to defects in CFTs with $d \geq 6$; and Takayanagi's AdS/BCFT with $d=5$. We demonstrate that several of our examples obey the boundary/defect $a$-theorem, as expected.
}

\begin{document}
\maketitle
\flushbottom

\section{Introduction}
\label{sec:intro}

What information characterizes a Quantum Field Theory (QFT) uniquely? Can we use this information to classify QFTs, or map the space of QFTs? Can we prove that this information must obey constraints, thus eliminating regions in the space of QFTs? These questions are vitally important for many areas of physics. For example, in condensed matter physics, topological states can be classified by the discrete symmetries of QFTs. In particle physics, constraints on QFTs, such as anomaly matching, are essential in the search for extensions of the Standard Model. In string/M-theory, a classification scheme for QFTs translates to a classification of internal manifolds when compactifying from 10 or 11 dimensions.

Symmetries are essential tools for characterizing and classifying QFTs. In particular, in any local, reflection positive QFT in $d$ Euclidean dimensions, Noether's theorem for the Euclidean group, i.e.\ translations and rotations of $\mathbb{R}^d$, requires the existence of a symmetric, conserved stress tensor, $T^{\mu\nu}= T^{\nu\mu} $ and $\partial_\mu T^{\mu\nu} = 0$, with $\mu,\nu=1,2,\ldots,d$. In this sense, $T^{\mu\nu}$ is \textit{universal}: we can \textit{always} characterize a local Euclidean-invariant QFT, in part, via correlations of $T^{\mu\nu}$ with itself and other operators. 

Conformal symmetry is also an essential tool for characterizing and classifying QFTs. Indeed, Conformal Field Theories (CFTs) are natural starting points for such an analysis because they are fixed points of renormalization group (RG) flows. Moreover, in a local, reflection-positive CFT, conformal symmetry fixes the correlation functions of all local operators in terms of the two- and three-point functions of conformal primary operators, thus dramatically simplifying the problems of characterization and classification for CFTs, and the QFTs connected to them via RG flows.

Conserved currents in $d$-dimensional CFTs are conformal primaries, including in particular $T^{\mu\nu}$.\footnote{Although in $d=2$, $T^{\mu\nu}$ is not a \textit{Virasoro} primary.} Conformal symmetry completely fixes the 2- and 3-point functions of $T^{\mu\nu}$ up to a set of dimensionless coefficients. In even $d$, these coefficients are fixed (in part) by Weyl anomaly coefficients. While conformal symmetry requires tracelessness of the stress tensor
classically,  $T^{\mu}_{~\mu}=0$, for a CFT defined on a non-trivially curved background, quantum effects can break conformal symmetry, leading to $T^{\mu}_{~\mu}\neq0$. This is the Weyl anomaly, which is in fact non-vanishing only in even $d$. In $d=2$, the Weyl anomaly has a single term, proportional to the Ricci scalar of the background manifold. This term's coefficient defines the central charge, $c^{\text{\tiny{(2d)}}}$, which actually fixes all of $T^{\mu\nu}$'s self-correlators~\cite{Polyakov:1981rd}. In $d=4$, and with parity symmetry, the Weyl anomaly has two terms, proportional to the Euler density and Weyl tensor squared~\cite{Duff:1993wm}. Their coefficients define the central charges $a^{\text{\tiny{(4d)}}}$ and $c^{\text{\tiny{(4d)}}}$ respectively.\footnote{The term ``central charge''  is typically reserved for the coefficient of a central extension term of some algebra. Indeed, $c^{\text{\tiny{(2d)}}}$ appears in the coefficient of the central extension term of the Virasoro algebra. However, generically the Weyl anomaly coeffcients of $d>2$ CFTs, like $a^{\text{\tiny{(4d)}}}$ and $c^{\text{\tiny{(4d)}}}$, and those of defects or boundaries, are not coefficients of central extension terms of any algebra. Nevertheless, in analogy with $c^{\text{\tiny{(2d)}}}$, we will follow the standard convention (see for example refs.~\cite{Anselmi:1997am,Intriligator:2003jj}) of calling all Weyl anomaly coefficients ``central charges''. } Reflection positivity also imposes positivity constraints on central charges~\cite{Hofman:2008ar,Hofman:2016awc}, where $c^{\text{\tiny{(4d)}}}$ fixes $T^{\mu\nu}$'s two-point function, and $a^{\text{\tiny{(4d)}}}$ and $c^{\text{\tiny{(4d)}}}$ are two of the three coefficients that fix $T^{\mu\nu}$'s three-point function~\cite{Osborn:1993cr,Erdmenger:1996yc}.

Furthermore, so-called $c$-theorems require some central charges to decrease along RG flows between CFTs. Specifically, in $d=2$ and $d=4$, the $c$-theorem~\cite{Zamolodchikov:1986gt,Cappelli:1990yc,Casini:2004bw,Komargodski:2011xv} and $a$-theorem~\cite{Cardy:1988cwa,Osborn:1989td,Jack:1990eb,Osborn:1991gm,Komargodski:2011vj,Komargodski:2011xv,Casini:2017vbe} state that $c^{\text{\tiny{(2d)}}}$ and $a^{\text{\tiny{(4d)}}}$, respectively, must be smaller in the infrared (IR) CFT than in the ultraviolet (UV) CFT.\footnote{\label{fn:F}
In $d=3$, the $F$-theorem~\cite{Casini:2012ei,Casini:2017vbe} states that the CFT partition function on a sphere, called $F$, must decrease along an RG flow to an IR fixed point.
} 
$c$-theorems rely only on generic principles, like locality, Euclidean invariance, and reflection positivity, making them powerful non-perturbative constraints on QFTs. The decreasing quantities provide a measure of the number of degrees of freedom of a QFT, which we expect to decrease along an RG flow, as massive modes decouple. Most importantly for characterization and classification, $c$-theorems impose irreversibility along RG flows, providing a hierarchical order among QFTs.

Another essential ingredient for characterizing and classifying QFTs are extended operators, a.k.a.\ defects~\cite{Douglas:2010ic}. For example, in gauge theories, Wilson, 't Hooft, and other line operators are crucial for classifying vacua as Higgs, Coulomb, confining, etc., and for distinguishing gauge theories with the same gauge \textit{algebras} but different gauge \textit{groups}~\cite{Aharony:2013hda}. Similarly, higher-dimensional defects can be crucial for classifying vacua, for example by providing order parameters that can detect whether higher-dimensional objects, such as strings, have condensed. Co-dimension $1$ defects, also known as interfaces or domain walls, can provide natural maps between QFTs related by dualities, RG flows, and other transformations. A boundary of a QFT can be considered as a special case of an interface between a QFT and a trivial or topological QFT.

We can define a defect supported on a submanifold in different ways. One approach is to prescribe conditions on the behavior of the ambient fields near the defect, conditions that may allow for coupling with degrees of freedom supported only on the submanifold. Another possible method is to integrate a local operator over a submanifold, an approach used for example in the construction of a Wilson line. We will denote the defect submanifold's dimension as $0<p<d$, and hence the co-dimension as $q\equiv d-p$. These extended objects are in contrast with local operators, which have $p=0$ and $q=d$.

A defect necessarily breaks translational symmetry in directions normal to the submanifold on which it is supported, so that $T^{\mu\nu}$ is no longer conserved. Indeed, now $\partial_{\mu} T^{\mu i} = \delta^{(q)} \, {\cal D}^i$, where $\delta^{(q)}$ is a delta function that localizes to the defect in the $q$ directions normal to the defect, and ${\cal D}^i$ is the displacement operator. The displacement operator is a defect-localized scalar, and is a vector in the normal directions, labeled by the index $i=1,2,\ldots,q$. In correlation functions, the insertion of ${\cal D}^i$ acts as a local geometric deformation of the defect, which ``displaces'' a point on the defect in a normal direction, hence its name. Since it descends from the stress tensor of the ambient theory, the displacement operator is itself \textit{universal} in the defect spectrum: we can always characterize a local QFT with a defect, in part, via correlations of ${\cal D}^i$ with itself and other operators.

A CFT with a defect or boundary that preserves scale invariance and special conformal transformations about points in the submanifold is called a Defect CFT (DCFT) or Boundary CFT (BCFT), respectively. As is the case with ordinary CFTs, DCFTs and BCFTs are natural starting points for characterizing and classifying defects. In particular, they sit at the endpoints of RG flows, including those localized to the defect or boundary, and those of the ambient QFT. Local correlators in a BCFT or DCFT are completely determined by (1) the ambient CFT data, i.e.\ the spectrum of primaries, (2) the spectrum of defect or boundary primaries, whose 2- and 3-point correlators with one another are fixed by the defect or boundary conformal symmetry up to a set of dimensionless coefficients, and (3), the mixed 2-point functions of ambient and defect/boundary primaries. This latter class includes ambient one-point functions, when the defect/boundary primary is the identity.

In a DCFT or BCFT, the conformal symmetry preserved by the defect or boundary requires $T^{\mu}_{~\mu}=0$. Much like CFTs, a DCFT or BCFT in flat space can exhibit a Weyl anomaly in curved space and/or when the defect or boundary is curved. In contrast to the Weyl anomaly in CFTs, now $T^{\mu}_{~\mu}\neq0$ generically consists of contributions from both the ambient CFT, when $d$ is even, and from defect or boundary localized terms. The latter can potentially be non-vanishing for both even and odd $p$ because a submanifold has intrinsic and extrinsic curvatures, which provides a larger basis of conformal invariants that can contribute to the anomaly. To date, defect and boundary contributions to the Weyl anomaly have been determined only for defects/boundaries in $d=2$ CFTs ($p=1$ in $d=2$)~\cite{Polchinski:1998rq}, surface defects ($p=2$ in $d \geq 3$)~\cite{Berenstein:1998ij,Graham:1999pm,Henningson:1999xi,Gustavsson:2003hn,Asnin:2008ak,Schwimmer:2008yh,Cvitan:2015ana,Jensen:2018rxu} and co-dimension $1$ defects in $d=4$~\cite{Herzog:2015ioa,Herzog:2017kkj} and $d=5$~\cite{FarajiAstaneh:2021foi}.

The coefficients of defect/boundary Weyl anomalies define defect/boundary central charges that are crucial for characterization and classification. However, relatively little is known about them, compared to CFT central charges. For instance, they should determine many correlation functions involving $T^{\mu\nu}$ and ${\cal D}^i$, but exactly how has been determined only for a subset of them, in the cases $p=2$ in $d\geq3$, and $q=1$ in $d=4$~\cite{Herzog:2015ioa,Herzog:2017kkj,Herzog:2017xha,Prochazka:2018bpb}. These results subsequently imply some positivity constraints for defect central charges~\cite{Herzog:2017kkj,Jensen:2018rxu}. Defect/boundary $c$-theorem for arbitrary $p$ and $d$ has been proposed ref.~\cite{Kobayashi:2018lil}, for RG flows localized to the defect/boundary. Rigorous proofs of these $c$-theorems have appeared for $p=2$~\cite{Jensen:2015swa,Casini:2018nym} and $p=4$~\cite{Wang:2021mdq}, because in these cases the Weyl anomaly includes an Euler density term intrinsic to the defect/boundary, allowing the proofs of refs.~\cite{Komargodski:2011vj,Komargodski:2011xv} to be suitably adapted.\footnote{When $p=1$, the defect or boundary's contribution to $F$ must decrease under defect/boundary RG flows~\cite{Affleck:1991tk,Friedan:2003yc,Casini:2016fgb,Cuomo:2021rkm}. See footnote~\ref{fn:F}.
} 
In section~\ref{sec:review} we review the state of the art of defect and boundary Weyl anomalies, central charges, and the constraints they obey.

In this paper we determine the contribution to the Weyl anomaly of a defect with $p=4$ in $d \geq 6$. Our motivation comes primarily from $d=6$ CFTs, which occupy privileged positions in the space of QFTs, and have various $p=4$ defects that encode key information.

For CFTs in $d>4$, power counting forbids any interacting local Lagrangian, so na\"ively we expect the only local, reflection-positive $d=6$ CFTs to be free, massless fields, namely scalars, fermions, self-dual three-forms, and combinations thereof. However, string theory has revealed that intrinsically strongly-interacting supersymmetric (SUSY) CFTs (SCFTs) exist in $d=6$. This is in fact the highest $d$ where superconformal symmetry can exist~\cite{Nahm:1977tg}, the only options being $\N=(1,0)$ or $\N=(2,0)$ SUSY. These SCFT's degrees of freedom are tensionless strings, giving rise to a tensor supermultiplet, including a two-form gauge field with self-dual three-form field strength. Despite lacking a Lagrangian description, all evidence to date suggests these SCFTs are local. These special properties suggest $d=6$ SCFTs may be ``parent theories'' that, upon compactification and (super)symmetry breaking, give rise to many (if not all) QFTs in lower $d$~\cite{Heckman:2018jxk}. Moreover, $d=6$ SCFTs have been classified~\cite{Heckman:2013pva,Gaiotto:2014lca,DelZotto:2014hpa,Heckman:2014qba,DelZotto:2014fia,Heckman:2015bfa,Bhardwaj:2015xxa}, making them an especially promising starting point for classifying QFTs.

In fact, these SCFTs arise from M-theory constructions, hence they may also encode important information about the nature of quantum gravity. The most prominent example is the maximally SUSY $d=6$ $\N=(2,0)$ SCFT, which for the gauge algebra $A_{N-1}$ arises as the low-energy worldvolume theory of $N$ coincident M5-branes. This SCFT is especially challenging to study, being an isolated, strongly-interacting fixed point, and having no free parameter besides $N$. To date, most progress has come from holography~\cite{Maldacena:1997re}, the conformal bootstrap~\cite{Beem:2015aoa}, and chiral algebra methods~\cite{Beem:2014kka}. 

The prospect of elucidating the $\N = (2,0)$ theory through its $p=4$ defects in part motivates this work. The $d=5$ superconformal group is not a subgroup of the $d=6$ superconformal group. Hence a $d=6$ SCFT cannot have superconformal boundary conditions. More generally, chiral anomalies may forbid a spatial boundary~\cite{Jensen:2017eof,Thorngren:2020yht}, which could obviate the study of $d=6$ SUSY BCFTs. However, these SCFTs admit various superconformal defects of higher co-dimension. For example, from the M5-brane construction, the $\CN=(2,0)$ SCFT admits a superconformal $p=4$ defect, which arises from the 1/2-BPS intersection with a second stack of M5-branes. These defects not only probe the M5-brane theory, and hence M-theory, but also, upon compactification, determine many properties of QFTs in lower $d$. For instance, wrapping the $\N = (2,0)$ theory on a Riemann surface, depending on how many of the defect's directions wrap the surface, the defect can reduce to defects with $p=2$, $3$, or $4$, the last case being hypermultiplets~\cite{Alday:2009fs}, in class $\mc{S}$ theories.

Among $d=6$ CFTs more generally, with or without SUSY, many other $p=4$ defects are possible. For example, various monodromy and twist defects exist, including in particular the universal twist defect that appears in calculations of entanglement entropy (EE) via the replica trick~\cite{Korepin:2004zz,Calabrese:2004eu,Calabrese:2005zw}. In general, EE depends on the UV cutoff, but in even $d$, EE includes a contribution logarithmic in the cutoff whose coefficient is cutoff-independent, and hence physical. When $d=4$, this universal coefficient has precisely the same form as a $p=2$ defect Weyl anomaly, but now with defect central charges determined by the ambient CFT's central charges~\cite{Schwimmer:2008yh,Solodukhin:2008dh}. However, for EE in $d=6$, so far the relation between the universal log coefficient and a putative $p=4$ defect Weyl anomaly has been determined for a replica twist defect only in special cases, without extrinsic curvature~\cite{Hung:2011xb} or with only extrinsic curvature, i.e.\ a curved defect in flat space~\cite{Safdi:2012sn}.

In short, a $p=4$ defect's contribution to the Weyl anomaly is essential for understanding QFTs and defects with various $d$ and $p$, respectively. To determine the form of a $p=4$ defect's Weyl anomaly, upon allowing both space and the defect's submanifold to be curved, we follow a standard three-step algorithm: 
\begin{enumerate}
\item Find all curvature invariants whose scaling dimension, plus that of $\delta^{(q)}$, is the same as $T^{\mu}_{~\mu}$. These curvature invariants will be built from the defect submanifold's intrinsic Riemann tensor, the second fundamental form, and the pullback of the ambient Weyl tensor. Many such invariants are related by Gauss, Codazzi, Ricci, and other relations, which must be accounted for to identify a linearly-independent basis of invariants.
\item Impose Wess-Zumino (WZ) consistency. This is the statement that the Weyl anomaly must reflect the fact that Weyl transformations are Abelian, i.e.\ two successive Weyl transformations commute. Although trivial in principle, in practice WZ consistency can place non-trivial constraints on curvature invariants, and generically can eliminate some of them.
\item Fix renormalization scheme dependence by adjusting local counterterms to eliminate as many curvature invariants as possible.
\end{enumerate}
The defect central charges are then the coefficients of the remaining WZ-consistent, scheme-independent curvature invariants, up to signs and other conventions.

A corollary of WZ consistency is that the integral of the Weyl anomaly over all of space must be Weyl-invariant. That can happen in two ways, giving rise to two types of terms: A-type and B-type~\cite{Deser:1993yx}. Under a Weyl transformation, A-type terms transform by a total derivative that cannot be eliminated with counterterms, while B-type terms are simply invariant, or vary into total derivatives that can be removed with local counterterms. All known A-type terms are in fact the Euler densities, whose coefficients generically obey $c$-theorems, as mentioned above.

To our knowledge, the only existing results for $p=4$ defect Weyl anomalies appear in refs.~\cite{Hung:2011xb,Safdi:2012sn} for the special case of the entanglement twist field and for specific geometries only, as mentioned above, and in ref.~\cite{FarajiAstaneh:2021foi} for a generic BCFT in $d=5$. In particular, in ref.~\cite{FarajiAstaneh:2021foi}, Astaneh and Solodukhin allowed arbitrary curvature of both the space and the $p=4$ boundary, and imposed parity symmetry. In the boundary contribution to the Weyl anomaly, they found $8$ terms, including those familiar from a $d=4$ CFT, namely the intrinsic Euler density and the square of the pullback of the Weyl tensor. They computed all $8$ central charges in one example, namely a conformally coupled free massless scalar.

Our case of a generic $p=4$ defect in $d\geq 6$ is a significant extension of these previous results, and in particular we generalize them in two non-trivial ways. First, we generalize to co-dimension $>1$, where a larger basis of curvature invariants is available for step 1, leading to many more terms in the final result. Second, we allow for parity symmetry to be broken, which allows for still more curvature invariants. Ultimately, at the end of the algorithm, in the Weyl anomaly of a $p=4$ defect in a CFT with $d \geq 6$, we find $23$ parity-even terms. The number of parity-odd terms depends on the co-dimension, $q$, with $6$ for any $q$, plus an additional $1$ when $q=2$, or an additional $6$ when $q=4$. The parity-even terms include the intrinsic Euler density, which is the only A-type term, and the square of the pullback of the Weyl tensor, while the parity-odd terms include the submanifold Pontryagin density, as expected. Our main result for the anomaly appears in eq.~\eqref{eq:defect-Weyl-anomaly} for the case with any $q$, with the additional parity-odd terms for $q=2$ and $q=4$ in eqs.~\eqref{eq:parity-odd-q=2} and~\eqref{eq:parity-odd-q=4}, respectively.

The status of parity-odd contributions to the Weyl anomaly remains unclear. To our knowledge, only three examples are known. First is a $d=4$ CFT with parity broken, where the Weyl anomaly can include a parity-odd term, the Pontryagin density~\cite{Nakayama:2012gu}. In free field CFTs, the associated parity-odd central charge can be non-zero only for fields in complex representations of the Lorentz group~\cite{Dowker:1976zf,Christensen:1978gi,Christensen:1978md,Nakayama:2012gu}. However, even in the simple example of a free Weyl fermion, no consensus has emerged on whether this central charge is non-zero: see refs.~\cite{Bonora:2014qla,Bonora:2015nqa,Bonora:2015odi,Bastianelli:2016nuf,Bonora:2017gzz,Nakayama:2018dig,Bonora:2018obr,Bastianelli:2018osv,Frob:2019dgf,Bonora:2019dyv,Bastianelli:2019zrq,Nakagawa:2020gqc,Abdallah:2021eii}. The second example is a $p=2$ defect in a $d=4$ CFT with parity broken, where the defect's Weyl anomaly can include $2$ terms odd under parity along the defect~\cite{Cvitan:2015ana,Jensen:2018rxu}. However, in all known examples of this case, the corresponding parity-odd defect central charges vanish~\cite{Jensen:2018rxu}. The third example is a $q=1$ defect/boundary in a $d=4$ CFT, whose B-type terms are parity odd under reflection in the normal direction, and which generically are non-zero~\cite{Herzog:2015ioa,Herzog:2017kkj}. Our results provide a new class of examples useful for studying the many open questions about parity-odd Weyl anomalies.

As mentioned above, in sec.~\ref{sec:review} we review the state of the art of defect/boundary Weyl anomalies. In sec.~\ref{sec:anomalies} we determine the form of a $p=4$ defect's Weyl anomaly, with our main result being eq.~\eqref{eq:defect-Weyl-anomaly}. In sec.~\ref{sec:correlators} we show how some of the $p=4$ defect central charges appear in observables, namely the two-point function of ${\cal D}^i$ and the one-point function of $T^{\mu\nu}$. Using the latter, we show how two of the defect central charges appear in the universal contribution to EE of a spherical region centered on the defect~\cite{Kobayashi:2018lil,Jensen:2018rxu}, and how the Average Null Energy Condition (ANEC) constrains the sign of one defect central charge.

These results provide methods for computing $p=4$ defect/boundary central charges without computing $T^{\mu}_{~\mu}$ directly. Indeed, in sec.~\ref{sec:dcc} we use the results of secs.~\ref{sec:anomalies} and~\ref{sec:correlators} to compute several defect central charges in two classes of examples. First are monodromy defects of free, massless scalars our Dirac fermions in $d=6$, which also provide non-trivial tests of the defect $a$-theorem~\cite{Wang:2021mdq}. We also use the results for monodromy defects in free field theories to find defect central charges for conical defects and orbifolds. Second, we consider defects described holographically by five-dimensional probe branes in $d \geq 7$ Anti-de Sitter (AdS) space, $AdS_{d\geq7}$, where we use recent results from Graham and Reichert~\cite{Graham:2017bew}. This serves as a highly non-trivial check of our result for the defect Weyl anomaly.

In sec.~\ref{sec:bcc}, we compute boundary central charges for two examples of $d=5$ BCFTs in holography. First is five-dimensional probe branes in $AdS_6$, where we compute all $8$ boundary central charges, which provides another non-trivial check of the $p=4$ boundary Weyl anomaly. Second is Takayanagi's AdS/BCFT~\cite{Takayanagi:2011zk,Fujita:2011fp}, where we compute the A-type and several linear combinations of B-type boundary central charges.

In sec.~\ref{sec:summary} we conclude with a summary and discussion of many open questions about defect and boundary Weyl anomalies, for $p=4$ and beyond.

We collect various technical results in several appendices, and in a supplemental Mathematica notebook.

\textbf{Note:} We use Euclidean signature throughout, with only two exceptions: our discussions of the ANEC, which requires Lorentzian signature to define null directions, and our discussions of EE, which requires a Cauchy surface to define a QFT Hilbert space.

\section{Review and Conventions}
\label{sec:review}

In this section, we review some concepts necessary for defect Weyl anomalies. In subsection~\ref{sec:defgeom} we fix our conventions for geometric quantities of the ambient space and the defect, in subsection~\ref{sec:cftweylreview} we review known results for CFT Weyl anomalies, and in subsection~\ref{sec:defweylreview} we review the known results for defect Weyl anomalies. We present no new results in this section. Readers familiar with these topics may skip to section~\ref{sec:anomalies}.

\subsection{Geometry for defects and boundaries}
\label{sec:defgeom}
Let $\CM_d$ be a smooth $d$-dimensional (pseudo-)Riemannian manifold. $\CM_d$ is the background geometry into which we will embed a defect, or introduce a boundary.  We refer to this background as the {\textit{ambient}} space. In this section we take $d \geq 2$, however in all later sections we take $d \geq 5$, unless stated otherwise. Let $x^\mu$ be the coordinates on $\CM_d$, where $\mu = 1,\ldots, d$, and let $g_{\mu\nu}$ be the metric on $\CM_d$. Our ambient metric will have Euclidean signature, except when we discuss the ANEC and EE, which require Lorentzian signature.

Throughout this paper, $D$ will denote the Levi-Civita connection on $\CM_d$, and $\Gamma^\rho_{\mu\nu}$ the Christoffel symbols. The symbol $R$ will generically denote the associated curvature tensors, i.e. $R_{\mu\nu\rho\sigma}$ is the Riemann tensor, $R_{\mu\nu}$ is the Ricci tensor, and $R$ is the Ricci scalar on $\CM_d$. We also need several other tensors related to $R$. The ambient Schouten tensor is
 \begin{align}\label{eq:Schouten-tensor}
 P_{\mu\nu} \equiv \frac{1}{d-2}\Big( R_{\mu\nu} - \frac{1}{2(d-1)}Rg_{\mu\nu}\Big)\,.
\end{align}
Using $P_{\mu\nu}$, we define the Weyl tensor,
\begin{equation}\label{eq:Weyl-tensor}
W_{\mu\nu\rho\tau} \equiv R_{\mu\nu\rho\tau} - P_{\mu\rho}g_{\nu\tau}+ P_{\nu\rho}g_{\mu\tau}- P_{\nu\tau}g_{\mu\rho}+ P_{\mu\tau}g_{\nu\rho}\,,
\end{equation}
and the Cotton tensor,
\begin{equation}\label{eq:Cotton-tensor-1}
C_{\mu\nu\rho} \equiv D_\rho P_{\mu\nu} - D_{\nu} P_{\mu\rho}\,.
\end{equation}
Lastly, the Bach tensor is
\begin{equation}\label{eq:Bach-tensor-1}
B_{\mu\nu}\equiv D^\rho C_{\mu\nu\rho} - P^{\rho\tau}W_{\mu\rho\nu\tau}\,.
\end{equation}
For even $d$, the Euler density of ${\cal M}_d$ is
\beq
\label{eq:eulerdensdef}
E_{d} = \frac{1}{2^{d/2}} \, \delta^{\mu_1\nu_1\ldots \mu_{d/2}\nu_{d/2}}_{\rho_1 \sigma_1 \ldots \rho_{d/2}\sigma_{d/2}} \, R^{\rho_1\sigma_1}{}_{\mu_1\nu_1} \, R^{\rho_2\sigma_2}{}_{\mu_2\nu_2} \ldots R^{\rho_{d/2}\sigma_{d/2}}{}_{\mu_{d/2}\nu_{d/2}},
\eeq
where $\delta^{\mu_1\nu_1\ldots}_{\rho_1\sigma_1\ldots}$ is the generalized Kronecker delta.

The defects that we will study are supported on a $p$-dimensional embedded submanifold, $\Sigma_p \hookrightarrow \CM_d$. For the purposes of this review, we will take $1\leq p\leq d$, but in the following sections, we will set $p=4$, unless stated otherwise. Let $y^{a}$ be coordinates on $\Sigma_p$, where $a = 1,\ldots, p$. We will sometimes also write these coordinates as the vector $\mathbf{y}$. The embedding induces various intrinsic geometric quantities on $\Sigma_p$ that we distinguish from their counterparts on $\CM_d$ with a bar. In particular, let $\ovl{g}_{ab} = e^\mu_a e^\nu_b g_{\m\nu}$ denote the induced metric, where $e^\mu_{a} = \partial_a X^\mu$, and $X^\mu(y^a)$ are the embedding functions. The matrix $e^\mu_a$ acts to pull back ambient tensors onto $\Sigma_p$, e.g. the pullback of the ambient Ricci tensor is $R_{ab} = e^{\mu}_a e^{\nu}_b R_{\mu\nu}$.

On $\Sigma_p$ we denote the induced Levi-Civita connection as $\ovl D$, with its connection coefficients being the Christoffel symbols $\ovl \Gamma^a_{bc}$ built out of $\ovl{g}_{ab}$. We further introduce a covariant derivative that acts on tensors with mixed indices. We will abuse notation and also refer to it as $\ovl{D}_a$, e.g. $\ovl{D}_a w^\mu_b  = \partial_a w^\mu_b + \Gamma^\mu_{\nu a} w^\nu_b - \ovl{\Gamma}^c_{ab}w^\mu_c$ for a mixed tensor $w^{\mu}_b$, where $\Gamma^\mu_{\nu a}=e_a^\rho \Gamma^{\mu}_{\nu \rho}$.  From $\ovl{D}_a$, we define the second fundamental form, $\II^\mu_{ab} \equiv \ovl{D}_a e^{\mu}_b$, and its traceless version, $\oII^\mu_{ab} \equiv \II^{\mu}_{ab} - \frac{1}{p} \ovl{g}_{ab}\II^\mu$, with $\II^\mu \equiv \ovl{g}^{cd}\II^\mu_{cd}$. 

We denote the defect co-dimension as $q \equiv d-p$. Let $x_\perp^i$ denote the normal directions to $\Sigma_p$, where $i= 1,\ldots, q$. The embedding $\Sigma_p\hookrightarrow \CM_d$ splits the ambient space's tangent bundle $T\CM_d \simeq T\Sigma_p \oplus N\Sigma_p$ into a sum of the defect submanifold's tangent bundle, $T\Sigma_p$, and the normal bundle, $N\Sigma_p$. The structure group of the normal bundle depends both on the co-dimension $q$ and the details of the embedding.

The important geometric features of the normal bundle $N\Sigma_p$ derive from the totally antisymmetric normal tensor,
\begin{align}\label{eq:normal}
n_{\mu_1\ldots \mu_q} = \frac{1}{p!} \veps^{a_1\ldots a_p} \veps_{\nu_1\ldots\nu_p \mu_1 \ldots \mu_q} e^{\nu_1}_{a_1} \cdots e^{\nu_p}_{a_p}\,,
\end{align} 
where $\veps_{a_1\ldots a_p}$ and $\veps_{\mu_1\ldots \mu_d}$ are Levi-Civita tensors on $\Sigma_p$ and $\CM_d$ respectively. From $n_{\mu_1\ldots\mu_q}$, we define a projector onto $N\Sigma_p$,
\begin{align}
N_{\mu\nu}\equiv \frac{1}{(q-1)!}n_{\mu \sigma_2\ldots \sigma_q}n_\nu{}^{\sigma_2\ldots \sigma_q}.
\end{align}
We can similarly define a projector onto $T\Sigma_p$, also called the first fundamental form, $h_{\mu\nu} \equiv g_{\mu\nu} - N_{\mu\nu}$. Note that $ h_{\mu\nu} = E^a_\mu E^b_\nu \ovl{g}_{ab} $, where $E^a_\mu$ is a matrix which obeys $E^a_\mu e^\mu_b = \delta^a_b$.

When $q=1$, i.e. for a boundary or interface, $N\Sigma_p$ is a trivial bundle. In that case, the normal projector becomes $N_{\mu\nu} = n_\mu n_\nu$, where $n_\mu$ is the outward pointing unit normal co-vector. Moreover, $h_{\mu\nu}$ reduces to the usual hypersurface metric, $h_{\mu\nu} = g_{\mu\nu} - n_\mu n_\nu$. We define the extrinsic curvature as $K_{\mu\nu} \equiv \frac{1}{2}\mathcal{L}_n h_{\mu\nu}$, where $\mathcal{L}_n$ is the Lie derivative along $n^\mu$, satisfies $n^\mu K_{\mu\nu}=0$, and is related to the second fundamental form by $K_{ab} =-n_\mu \II^\mu_{ab}$, where $K_{ab}=e_a^\mu e_b^\nu K_{\mu\nu}$. We define the traceless version $\oK_{ab}\equiv K_{ab} - \frac{1}{p} \ovl{g}_{ab} K$, with $K \equiv \ovl{g}^{ab}K_{ab}$.

Furthermore, we define two notions of parity. By ``parity along the defect'', we mean simultaneous orientation reversal of the submanifold and the ambient space, such that ${\varepsilon^{a_1\ldots a_p}\to -\varepsilon^{a_1\ldots a_p}}$ and ${\veps^{\mu_1 \ldots \mu_d}\to -\veps^{\mu_1 \ldots \mu_d}}$, such that $n^{\mu_1 \ldots \mu_q}$, as defined by \eq{normal}, is invariant. Intuitively, this corresponds to reversing parity in the directions parallel to the defect. By ``parity in the normal bundle'' we mean orientation reversal of the ambient space alone. Under this transformation, ${\veps^{\mu_1 \ldots \mu_d}\to -\veps^{\mu_1 \ldots \mu_d}}$ while $\varepsilon^{a_1\ldots a_p}$ is invariant, such that ${n^{\mu_1 \ldots \mu_q}\to -n^{\mu_1 \ldots \mu_q}}$. This corresponds to reversing parity in the directions normal to the defect.

In general, a defect will break the conformal symmetry group of the ambient CFT to a subgroup. In flat $d$-dimensional Euclidean space, ${\cal M}_d = \mathbb{R}^d$, the conformal group is $SO(d+1,1)$, whose generators are rotations, translations, dilatations, and special conformal transformations. The maximal subgroup that a $p$-dimensional defect or boundary can preserve is ${SO(p+1,1)}\times SO(q)_N$, where $SO(q)_N$ is the structure group of the normal bundle $N\Sigma_p$. This subgroup is preserved by either a flat or spherical defect that is rotationally symmetric in the normal directions. For example, a flat defect preserves rotations and translations along the defect, dilatations, and special conformal transformations involving inversions about points in the defect, producing the $SO(p+1,1)$ factor, plus rotations in the normal directions, producing the $SO(q)_N$ factor. In what follows, for CFTs in flat space we will always assume the defect or boundary is flat, and so preserves this maximal subgroup, unless stated otherwise.

\subsection{Weyl anomalies of CFTs}
\label{sec:cftweylreview}

To provide context and background for the following sections, in this subsection we review some facts about the Weyl anomaly of CFTs, and in the next subsection we review the current state of the art of defect and boundary Weyl anomalies.

Consider a QFT on an arbitrary background $\CM_d$ with Euclidean metric $g_{\mu\nu}$. The effective action is $\CW = - \log Z$, where $Z$ is the QFT's partition function. In a curved background, the stress tensor is defined as the infinitesimal metric variation of the path integral. In our conventions, the stress tensor one-point function is 
\begin{equation}
\left< T^{\mu\nu} \right> = - \frac{2}{\sqrt{g}} \frac{\delta \mathcal{W}}{\delta g_{\mu\nu}}\,.
\end{equation}
Consider an infinitesimal local Weyl rescaling of $g_{\mu\nu}$ with Weyl parameter $\delta\omega$, i.e. $\delta_\omega g_{\mu\nu} = 2  g_{\mu\nu}\, \delta\omega$. In any correlator, a Weyl variation brings down an insertion of the trace of the stress tensor, $T^\mu{}_\mu$. In paticular, a Weyl variation of the effective action gives
\begin{align}
\delta_\omega \CW = -\int_{\CM_d} d^dx \sqrt{g} \,\delta\omega \,\langle T^\mu{}_\mu\rangle\,.
\end{align}
In addition, a non-trivial $T^\mu{}_\mu$ implies that the effective action includes a contribution that is logarithmically divergent in the UV cutoff $\epsilon$. The coefficient of $\log \epsilon$ in the effective action has the form
\begin{align}
\label{eq:W_log_A}
 \CW|_{\log \epsilon} =    \int d^d x \,  \sqrt{g} \left< T^{\mu}{}_\mu \right>  \,.
\end{align}

In a CFT, the Weyl anomaly $T^\mu{}_\mu\neq 0$ must be built out of external sources, such as $g_{\mu\nu}$. Diffeomorphism invariance and invariance under the $SO(d)$ structure group of the frame bundle imply that $T^\mu{}_\mu$ can only contain scalars (singlets) of the appropriate dimension. If $g_{\mu\nu}$ is the only external source, then these scalars can only be constructed out of curvature tensors and an even number of derivatives. Since curvature tensors contain two derivatives of the metric, no Weyl anomaly is possible when $d$ is odd. When $d$ is even, WZ consistency implies that the anomaly takes the schematic form
\begin{align}\label{eq:general-d-Weyl-anomaly}
T^\mu{}_\mu  = \frac{1}{(4\pi)^{\frac{d}{2}}}\Big( (-)^{\frac{d}{2}-1} a_{\CM}^{\text{\tiny(dim)}}E_{d} + \sum_n c^{\text{\tiny(dim)}}_n I_n\Big)\,,
\end{align}
where we omitted any scheme-dependent contributions. $E_d$ is the $d$-dimensional Euler density and $I_n$ are a set of conformal invariants that are generally expressed as rank-$\frac{d}{2}$ scalar monomials built from contractions of Weyl tensors and their derivatives. Since $E_d$ is a topological density, it transforms as a total derivative under Weyl transformation, whereas the $I_n$ are invariant or vary into total derivatives that can be removed by local counterterms. This leads to the distinguishing nomenclature of ``A-type'' anomalies for the former and ``B-type'' for the latter~\cite{Deser:1993yx}. Since we will discuss central charges of CFTs and defects of various dimensions, we use the superscript ${}^{\text{\tiny{(dim)}}}$ to distinguish the dimension in which the central charges are defined, and the subscripts on $a_{\CM}$ and $a_\Sigma$ to distinguish the ambient CFT and intrinsic defect/boundary A-type central charges, respectively.

The Weyl anomaly of a $d=2$ CFT is
\begin{align}
\label{eq:2d-Weyl-anomaly}
T^\mu{}_\mu = \frac{c^{\text{\tiny{(2d)}}}}{24\pi} \, R\,,
\end{align}
where, $c^{\text{\tiny{(2d)}}}$ is the CFT's central charge, defined from the coefficient of the central extension term in the Virasoro algebra. For a reflection-positive $d=2$ CFT with normalisable vacuum, $c$ is positive semi-definite and obeys the $c$-theorem~\cite{Zamolodchikov:1986gt}: along an RG flow from a UV CFT to an IR CFT, a so-called $c$-function exists, built from correlators of $T^{\mu\nu}$, that decreases monotonically along the flow, and agrees with the values of $c$ of the CFTs at the fixed points. This implies that $c^{\text{\tiny{(2d)}}}_{\text{UV}}\geq c^{\text{\tiny{(2d)}}}_{\text{IR}}$.  Additionally, by WZ consistency, A-type anomalies such as $c^{\text{\tiny{(2d)}}}$ are independent of marginal couplings~\cite{Osborn:1991gm}. The central charge $c^{\text{\tiny{(2d)}}}$ determines many physical observables: all self-correlators of the stress tensor~\cite{Polyakov:1981rd}, the universal contribution to the EE of an interval~\cite{Calabrese:2004eu}, the thermal entropy~\cite{Cardy:1986ie,Affleck:1986bv}, and more. The central charge $c^{\text{\tiny{(2d)}}}$ is thus essential for characterising and classifying $d=2$ CFTs and QFTs.

The Weyl anomaly of a $d=4$ CFT is
\begin{align}
\label{eq:4d-Weyl-anomaly}
T^\mu{}_\mu = \frac{1}{16\pi^2}\Big(- a_\CM^{\text{\tiny(4d)}} E_4 + c^{\text{\tiny(4d)}} W_{\mu\nu\rho\sigma}W^{\mu\nu\rho\sigma} + \tilde{c}^{\text{\tiny(4d)}}\,\varepsilon^{\mu\nu\rho\lambda} R_{\mu\nu\sigma\omega}R^{\sigma\omega}{}_{\rho\lambda}\Big),
\end{align}
where the final term in eq.~\eqref{eq:4d-Weyl-anomaly} is the $d=4$ Pontryagin density. In eq.~\eqref{eq:4d-Weyl-anomaly}, WZ consistency allows for a scheme-dependent term $\propto \Box R$~\cite{Bonora:1985cq}, whose coefficient we have set to zero using a local counterterm. For reflection-positive, local $d=4$ CFTs, the A-type central charge $a_\CM^{\text{\tiny(4d)}}$ obeys the $a$-theorem: $a^{\text{\tiny{(4d)}}}_{\text{UV}}\geq a^{\text{\tiny{(4d)}}}_{\text{IR}}$~\cite{Cardy:1988cwa,Osborn:1989td,Komargodski:2011vj,Komargodski:2011xv}. Like all A-type CFT central charges, $a^{\text{\tiny{(4d)}}}$ is independent of marginal couplings. Explicit examples are known in which the parity-even B-type central charge, $c^{\text{\tiny(4d)}}$, can decrease or increase along an RG flow~\cite{Cappelli:1990yc,Anselmi:1997am}. In $d=4$ CFTs holographically dual to Einstein gravity in $AdS_5$, $a^{\text{\tiny(4d)}}=c^{\text{\tiny(4d)}}$~\cite{Henningson:1998gx,Henningson:1998ey}. Unlike $d=2$ CFTs, no single central charge determines all of $T^{\mu\nu}$'s self-correlators. Instead, the B-type anomaly $c^{\text{\tiny{(4d)}}}$ fixes $T^{\mu\nu}$'s two-point function, while $a_\CM^{\text{\tiny(4d)}}$ and $c^{\text{\tiny(4d)}}$ are two of the three numbers that fix $T^{\mu\nu}$'s three-point function~\cite{Osborn:1993cr,Erdmenger:1996yc}. Reflection positivity then requires $c^{\text{\tiny(4d)}}\geq0$. If $T^{\mu\nu}$ is the unique conserved spin-2 operator, then the conformal bootstrap and other CFT ``first principles'' bound the ratio $a_\CM^{\text{\tiny(4d)}}/c^{\text{\tiny(4d)}}$~\cite{Hofman:2008ar,Hofman:2016awc}, so that in particular $a_\CM^{\text{\tiny(4d)}}\geq0$. Somewhat like a $d=2$ CFT, the $d=4$ central charges determine the universal contribution to EE~\cite{Solodukhin:2008dh}. In contrast to $a_\CM^{\text{\tiny(4d)}}$ and $c_\CM^{\text{\tiny(4d)}}$, little is known about the parity-odd B-type central charge, $\tilde{c}^{\text{\tiny(4d)}}$: for the state of the art, see refs.~\cite{Bonora:2014qla,Bonora:2015nqa,Bonora:2015odi,Bastianelli:2016nuf,Bonora:2017gzz,Nakayama:2018dig,Bonora:2018obr,Bastianelli:2018osv,Frob:2019dgf,Bonora:2019dyv,Bastianelli:2019zrq,Nakagawa:2020gqc,Abdallah:2021eii}.

The Weyl anomaly of a $d=6$ CFT is
\begin{align}
\label{eq:6d-Weyl-anomaly}
T^\mu{}_{~\mu} = \frac{1}{(4\pi)^3}\left(a_\CM^{\text{\tiny(6d)}} E_6 + c_1^{\text{\tiny(6d)}} I_1 + c_2^{\text{\tiny(6d)}} I_2 + c_3^{\text{\tiny(6d)}} I_3\right),
\end{align}
where the B-type Weyl invariants are
\begin{subequations}
\begin{eqnarray}
I_1 & = & W_{\mu\lambda\rho \nu} W^{\lambda \sigma \tau \rho} {{W_\sigma}^{\mu\nu}}_\tau, \\
I_2 & = & {W_{\mu\nu}}^{\lambda \rho} {W_{\lambda \rho}}^{\sigma \tau} {W_{\sigma \tau}}^{\mu\nu} \,, \\
I_3 & = & W_{\mu\nu\lambda\rho} \left(D^2 \, \delta^{\nu}_{~\sigma} - \frac{6}{5} R \, \delta^{\nu}_{~\sigma} + 4 R^{\nu}_{~\sigma} \right) W^{\sigma\nu\lambda\rho},
\end{eqnarray}
\end{subequations}
where we have set total derivative terms to zero using local counterterms~\cite{Bastianelli:2000hi}. For $d=6$ SCFT tensor branch RG flows, a $c$-theorem has been proven for $a_\CM^{\text{\tiny(6d)}}$~\cite{Cordova:2015fha}. However, whether $a_\CM^{\text{\tiny(6d)}}$ obeys a $c$-theorem more generally remains an open question: for the state of the art, see refs.~\cite{Elvang:2012st,Grinstein:2014xba,Heckman:2015axa,Gracey:2015fia,Stergiou:2016uqq,Heckman:2021nwg} and references therein. $\N=(1,0)$ SUSY imposes a linear relation on the B-type central charges, so that only two are independent, while $\N=(2,0)$ SUSY imposes a second linear relation, so that only one is independent~\cite{Beccaria:2015ypa}. For CFTs in $d=6$ holographically dual to Einstein-Hilbert gravity, all four central charges are linearly related, so that only one is independent~\cite{Henningson:1998gx,Henningson:1998ey}. The central charge $a_\CM^{\text{\tiny(6d)}}$ appears in $T^{\mu\nu}$'s $4$-point function, $c_1^{\text{\tiny(6d)}}$ and $c_2^{\text{\tiny(6d)}}$ are related to the two free parameters in $T^{\mu\nu}$'s $3$-point function, and $c_3^{\text{\tiny(6d)}}$ fixes $T^{\mu\nu}$'s $2$-point function~\cite{Osborn:1993cr,Erdmenger:1996yc,Bastianelli:1999ab}. Reflection positivity then requires $c_3^{\text{\tiny(6d)}}\geq 0$. As in lower $d$, the $d=6$ central charges determine the universal contribution to EE~\cite{Hung:2011xb,Safdi:2012sn}.

\subsection{Weyl anomalies of defects and boundaries}
\label{sec:defweylreview}

Now consider a $p$-dimensional defect or boundary of a $d$-dimensional CFT. In particular, consider a first-order variation of the effective action, ${\cal W}$, with respect to the metric, $g_{\mu\nu}$, and the defect or boundary's embedding functions, $X^{\mu}(y^a)$. This variation picks up contributions from terms that are localized to the defect or boundary submanifold, $\Sigma$,
\begin{align}
\delta \CW   = -\frac{1}{2}\int_{\CM_d} d^d x \sqrt{g}\,\delta g_{\mu\nu}\langle \left. T^{\mu\nu}\right|_{\mathcal{M}_d}\rangle - \frac{1}{2} \int_{\Sigma_p} d^py \sqrt{\ovl{g}}\left(\delta g_{\mu\nu}\langle \left.  T^{\mu\nu}\right|_{\Sigma_p}  \rangle  + 2 \delta X^i(y^a) \langle \CD_i\rangle\right),
\end{align}
where $\left. T^{\mu\nu}\right|_{\mathcal{M}_d}$ and $\left. T^{\mu\nu}\right |_{\Sigma_p}$ denote the contributions to the stress tensor from the ambient CFT and the boundary/defect, respectively. $\delta X^i (y^a)$ is the variation of the embedding functions in the directions transverse to $\Sigma$, and $\CD_i$ is the displacement operator defined from the broken Ward identities for translations normal to the defect,
\begin{align}\label{eq:displacement-operator-def}
D_\mu T^{\mu i} = \delta^{(q)}(x_\perp) \CD^i\,.
\end{align}
We emphasize that generically the defect/boundary does not have its own intrinsically defined, conserved stress tensor. However, the components of the ambient stress tensor along $\Sigma_p$ are conserved everywhere on $\CM_d$, including $\Sigma_p$, i.e. $D_\mu T^{\mu a} = 0$.

If we instead consider an infinitesimal Weyl rescaling without affecting the defect/boundary's embedding, then the Weyl anomaly picks up a contribution localized to $\Sigma_p$,
\begin{align}
T^\mu{}_\mu = T^\mu{}_\mu\big|_{\CM_d} + \delta^{(q)}(x_\perp) T^\mu{}_\mu\big|_{\Sigma_p}.
\end{align}
Here we indicate by $T^\mu{}_\mu\big|_{\CM_d} $ the contributions to the Weyl anomaly purely from the ambient CFT, and by $T^\mu{}_\mu\big|_{\Sigma_p}$ we denote the defect/boundary Weyl anomaly, which will be the primary focus of the following sections. Importantly, $T^\mu{}_\mu\big|_{\Sigma_p}$ is built out of structures that involve (derivatives of) the metric $g_{\mu\nu}$ and the embedding functions $X^\mu(y^a)$. This leads to a much richer basis for conformal invariants, and hence novel defect/boundary anomalies.

As mentioned in sec.~\ref{sec:intro}, the defect/boundary Weyl anomaly has been determined in four cases: $p=1$ in $d=2$~\cite{Polchinski:1998rq}, $p=2$ in $d \geq 3$~\cite{Berenstein:1998ij,Graham:1999pm,Henningson:1999xi,Gustavsson:2003hn,Asnin:2008ak,Schwimmer:2008yh,Cvitan:2015ana,Jensen:2018rxu}, $p=3$ in $d=4$~\cite{Herzog:2015ioa,Herzog:2017kkj}, and $p=4$ in $d=5$~\cite{FarajiAstaneh:2021foi}. In the rest of this section we will review the first three cases, and in the next section we will review the fourth case.

The Weyl anomaly of a $p=1$ defect in a $d=2$ CFT is~\cite{Polchinski:1998rq}
\begin{align}\label{eq:1d-defect-Weyl-anomaly}
T^\mu{}_\mu|_{\Sigma_1} = \frac{c^{\text{\tiny{(2d)}}}}{12\pi}\,K.
\end{align}
A defect with odd $p$ has no intrinsic Euler density. However, if the ambient CFT has even $d$, and the defect has co-dimension $q=1$, then the restriction of the ambient Euler density to the defect will appear in the defect's Weyl anomaly. Moreover, WZ consistency fixes the associated coefficient in terms of the ambient CFT's A-type central charge~\cite{Polchinski:1998rq}. We see both of these features in eq.~\eqref{eq:1d-defect-Weyl-anomaly}, and we will see them again for $p=3$ in $d=4$, eq.~\eqref{eq:3d-boundary-Weyl-anomaly}. For $p=1$ in a $d>2$ CFT, the defect Weyl anomaly vanishes.

The Weyl anomaly of a $p=2$ defect in a $d \geq 3$ CFT is~\cite{Berenstein:1998ij,Graham:1999pm,Henningson:1999xi,Gustavsson:2003hn,Asnin:2008ak,Schwimmer:2008yh,Cvitan:2015ana,Jensen:2018rxu}
\begin{align}\label{eq:2d-defect-Weyl-anomaly}
T^\mu{}_\mu|_{\Sigma_2} = \frac{1}{24\pi}( \bc \ovl{R} + \dc{1} \oII^2 + \dc{2} W^{ab}{}_{ab})+\frac{\delta_{q,2}}{2\pi}\e^{ab}n_{ij}(\tdc{1}\oII^{i}_{ac}\oII^{j}_{b}{}^c +\tdc{2}W_{a}{}^i{}_b{}^j)\,,
\end{align}
where $\delta_{q,2}$ is a Kronecker delta. Within the first set of parentheses, the first term is an A-type central charge, while the remaining two are B-type. The term $\dc{2}W^{ab}{}_{ab}$ does not exist for $q=1$ because the ambient Weyl tensor $W$ vanishes identically when $d= 3$. In that term, the trace over the indices is performed with the induced metric $\ovl{g}_{ab}$, hence $W^{ab}{}_{ab}$ is generically non-vanishing when $d\geq 4$. In the second set of parentheses, both terms exist only when $q=2$, are B-type, and are odd under parity along the defect and separately under parity in the normal bundle. The indices $i, j , \ldots$ are valued in the normal bundle. We often use them instead of $\mu, \nu \ldots$ to denote projection onto the normal bundle. Note that even if $d$ is odd, and hence the ambient CFT has no Weyl anomaly, the defect Weyl anomaly in~\eq{2d-defect-Weyl-anomaly} still exists.

Like their CFT counterparts, the defect/boundary central charges should appear in a number of observables, besides $T^{\mu}_{~\mu}$ itself. We will review what is known to date. The central charge $\dc{1}$ controls $\CD^i$'s two-point function. For example, in $d=4$ flat space~\cite{Lewkowycz:2014jia,Bianchi:2015liz}
\begin{align}
\label{eq:p2D2}
\langle \CD^i ({\bf{y}})\CD^j({\bf{0}})\rangle = \frac{4}{3\pi^2}\frac{\dc{1}}{|{\bf{y}}|^6}\,.
\end{align}
Reflection positivity then requires $\dc{1} \geq 0$. When $q>1$, the central charge $\dc{2}$ controls $T^{\mu\nu}$'s one-point function, which is allowed to be non-zero by the conformal symmetry preserved by the defect: in flat space, the non-zero components are~\cite{Kapustin:2005py,Billo:2016cpy}
\begin{align}\label{eq:stress-tensor-one-pt-fn}
\langle T^{ab} \rangle = - h \frac{(q-1)\delta^{ab}}{|x_\perp|^d}\, ,\qquad \langle T^{ij} \rangle =h\frac{(d-q+1)\delta^{ij} - d \frac{x^i_\perp x_\perp^j}{|x_\perp|^2}}{|x_\perp|^d}\,.
\end{align}
In other words, the symmetries completely determine the form of $T^{\mu\nu}$'s one-point function, up to the single number, $h$. As shown in refs.~\cite{Lewkowycz:2014jia,Bianchi:2015liz,Jensen:2018rxu}, $h$ is determined by $\dc{2}$,
\begin{align}
\label{eq:2d1pt}
h = -\frac{1}{6\pi(d-1){\text{vol}}(\mathds{S}^{d-3})} \,\dc{2}\,,
\end{align}
where $\text{vol}(\mathds{S}^d)$ is the volume of a unit $d$-sphere. If we Wick-rotate to Lorentzian signature, then applying the ANEC to eq.~\eqref{eq:stress-tensor-one-pt-fn} implies $\dc{2}\leq 0$~\cite{Jensen:2018rxu}.\footnote{The proofs of the ANEC in refs.~\cite{Faulkner:2016mzt,Hartman:2016lgu,Kravchuk:2018htv} have not yet been extended to include defects. Nevertheless, in unitary DCFTs we expect the ANEC to hold for defects, because physically a defect should not change the fact that the total energy measured by a null observer should be non-negative.} A linear combination of $\bc$ and $\dc{2}$ determines the defect contribution to the universal part of the EE for a spherical region $A$ of radius $L$ centered on the defect, with UV cutoff $\epsilon$~\cite{Kobayashi:2018lil,Jensen:2018rxu},
\begin{align}
\label{eq:2dee}
S_{A,\Sigma_2} = \frac{1}{3}\left( \bc + \frac{d-3}{d-1}\dc{2} \right)\log \left(\frac{L}{\e}\right)\,.
\end{align}

The A-type defect central charge, $\bc$, shares a few key features with the $d=2$ CFT central charge, $c^{\text{\tiny{(2d)}}}$. For instance, under a defect RG flow, $\bc$ obeys a $c$-theorem: $a_{\Sigma,\text{UV}}^{\text{\tiny{(2d)}}}\geq a_{\Sigma,\text{IR}}^{\text{\tiny{(2d)}}}$~\cite{Jensen:2015swa,Casini:2018nym}. Furthermore, WZ consistency implies that $\bc$ is independent of \textit{defect} marginal couplings. However, the similarities seem to end there. For example, $\bc$ can depend on marginal couplings present in the ambient CFT~\cite{Herzog:2019bom,Herzog:2019rke,Bianchi:2019umv}. Also, even in reflection-positive theories, $\bc$ is not necessarily positive semi-definite. In particular, a free, massless scalar with Dirichlet boundary conditions in $d=3$ has $\bc<0$~\cite{Nozaki:2012qd,Jensen:2015swa,Fursaev:2016inw}. This raises the question of in what sense $\bc$ is counting defect/boundary degrees of freedom. More generally, what bounds $\bc$ obeys, if any, remains an open question.

In a $d=4$ SCFT, a $p=2$ defect that preserves at least $\N=(2,0)$ two-dimensional SUSY has $d_1=d_2$, which is conjectured to extend to $p=2$ defects in $d>4$ SCFTs as well~\cite{Bianchi:2019sxz}. Furthermore, for such defects in a SCFT with $d\geq 3$, $\bc$ is fixed by an 't Hooft anomaly~\cite{Wang:2020xkc}, hence the IR defect R-symmetry can be identified by extremizing a trial $\bc$, similar to $a$-maximization~\cite{Intriligator:2003jj} or $c$-extremization~\cite{Benini:2012cz,Benini:2013cda}.

The Weyl anomaly of the $p=3$ boundary in a $d=4$ CFT is~\cite{Herzog:2015ioa,Herzog:2017kkj}
\begin{align}\label{eq:3d-boundary-Weyl-anomaly}
T^\mu{}_\mu\big|_{\Sigma_3} = \frac{1}{16\pi^2}\Big(a_{\CM}^{\text{\tiny{(4d)}}} E_4|_{\partial\CM} + b_1^{\text{\tiny (3d)}} \oK^3 + b_2^{\text{\tiny (3d)}}\oK^{ab}W^{c}{}_{acb}\Big).
\end{align}
As anticipated below eq.~\eqref{eq:1d-defect-Weyl-anomaly}, the first term in the parentheses in eq.~\eqref{eq:3d-boundary-Weyl-anomaly} is the restriction of the bulk A-type anomaly to the boundary,
\begin{align}\label{eq:4d-3d-boundary-Euler}
E_4|_{\partial\CM} = \delta^{abc}_{def}\Big(2K^d{}_aR^{ef}{}_{bc} + \frac{8}{3}K^d{}_aK^e{}_bK^f{}_c\Big),
\end{align}
with a coefficient determined entirely by the ambient CFT's A-type central charge, $a_\CM^{\text{\tiny{(4d)}}}$, as required by WZ consistency of the ambient CFT's Weyl anomaly in the presence of the boundary. The boundary thus has only two new central charges, $b_1^{\text{\tiny (3d)}}$ and $b_2^{\text{\tiny (3d)}}$, which are both B-type and parity odd in the normal bundle. In free field CFTs, heat kernel methods fix $b_2^{\text{\tiny (3d)}}$ in terms of the ambient CFT's B-type central charge, $c^{\text{\tiny(4d)}}$, specifically, $b_2^{\text{\tiny (3d)}} = -8c^{\text{\tiny(4d)}}$, independent of the boundary conditions~\cite{Dowker:1989ue,Fursaev:2015wpa,Solodukhin:2015eca}. However, ref.~\cite{Herzog:2017xha} showed that the same is not true in interacting CFTs. The other boundary central charge, $b_1^{\text{\tiny (3d)}}$, does depend on boundary conditions, even for free fields~\cite{Melmed:1988hm,Moss:1988yf}. 
 
Similarly to the $p=2$ defect/boundary, the B-type central charges $b_1^{\text{\tiny (3d)}}$ and $b_2^{\text{\tiny (3d)}}$ determine correlation functions of the boundary displacement operator, $\CD$. In particular, $b_2^{\text{\tiny{(3d)}}}$ control's ${\cal D}$'s two-point function~\cite{Herzog:2017xha,Herzog:2017kkj}: in flat space,
 \begin{align}
 \langle \CD({\bf{y}})\CD({\bf{0}})\rangle =-\frac{15}{2\pi^4}\frac{ b_2^{\text{\tiny{(3d)}}}}{|{\bf{y}}|^8}\,.
 \end{align}
 Reflection positivity then requires $b_2^{\text{\tiny{(3d)}}}\leq0$. Further, $b_1^{\text\tiny{(3d)}}$ controls $\CD$'s three-point function~\cite{Herzog:2017kkj}: in flat space
\begin{align}
\label{eq:d4p3DDD}
 \langle \CD({\bf{y}}_1)\CD({\bf{y}}_2)\CD({\bf{0}})\rangle =-\frac{35}{2\pi^6}\frac{ b_1^{\text{\tiny{(3d)}}}}{ |{\bf{y}}_1|^4|{\bf{y}}_2|^4|{\bf{y}}_1- {\bf{y}}_2|^4}\,.
\end{align}

\section{Defect Weyl anomaly}
\label{sec:anomalies}

In this section we present our results for the Weyl anomaly of a $p=4$ conformal defect of arbitrary co-dimension $q\geq1$. For co-dimension $q \geq 2$ our results are novel. For $q=1$, which includes the case of a conformal boundary of a $d=5$ BCFT, the parity-even part of the Weyl anomaly was reported in ref.~\cite{FarajiAstaneh:2021foi}. We reproduce their results, and find three new terms that break parity along the boundary.

We determine the anomaly through the three-step algorithm mentioned in sec.~\ref{sec:intro}, which we outline in detail in sec.~\ref{sec:algorithm}. Applying the algorithm is computationally involved, so we do not present all the details here. Many details can be found in appendix~\ref{app:basis_terms}, and we refer readers wishing to reproduce our results to our supplemental Mathematica notebook. To illustrate the algorithm, in~\App{WZ-example} we review the simple example of a $p=2$ defect in a $d=4$ CFT, reproducing the known result in eq.~\eqref{eq:2d-defect-Weyl-anomaly}. Our result for the $p=4$ defect Weyl anomaly for general $q$ is presented in sec.~\ref{sec:defect_Weyl_anomaly}, and specifically in eq.~\eqref{eq:defect-Weyl-anomaly}, which is the main result of this paper. In sec.~\ref{sec:boundary_Weyl_anomaly} we set $q=1$ and present the full anomaly of a $p=4$ conformal boundary or interface, including the parity-odd contributions.

\subsection{Algorithm for defect Weyl anomalies} \label{sec:algorithm}

Starting from the effective action $\CW$, the most general form of the defect Weyl anomaly $\delta_\omega \CW = \int_{\Sigma_p} (\ldots) \delta\omega$ can be determined algorithmically by following three steps:\footnote{For a co-dimension $q=1$ defect embedded in an even dimensional CFT there is an additional step. As explained below \eq{4d-3d-boundary-Euler}, the Euler density $E_d$ in the presence of a boundary or interface contains a boundary term. It ensures that the ambient part of the Weyl anomaly is Wess-Zumino consistent, and its coefficient is fixed in terms of the ambient CFT's A-type central charge. These boundary terms are known for all even $d$~\cite{Herzog:2015ioa}. In the following sections, however, we will only consider conformal defects of dimension $p=4$ for which no such extra boundary contribution can arise from the bulk $d=5$ CFT.}

\begin{enumerate}

\item \textbf{Find a basis of terms for the anomaly $\delta_\omega \CW$.} This amounts to enumerating all possible terms that may appear in $\delta_\omega \CW$ to linear order in the Weyl transformation parameter $\delta\omega$. We insist that the terms appearing in $\delta_\omega \CW$ are local, diffeomorphism and Lorentz invariant, both in the ambient space and on the defect. Moreover, we require that these terms have vanishing mass-dimension, since Weyl transformations amount to a rescaling of lengths by a dimensionless scalar. Accounting for the measure of integration over the $p$-dimensional defect requires that each term contains $p$ derivatives acting on the metric $g_{\mu\nu}$, the pullbacks $e_a^\mu$, or the Weyl variation parameter $\delta\omega$. These conditions put strong constraints on the terms that can appear in the anomaly, and imply that they must be finite in number. However, not all of these terms are linearly independent. The second Bianchi identity can be combined with the Gauss-Codazzi-Ricci equations, summarized in~\App{relations}, to produce geometric relations. These can be further adorned with additional derivatives. These linear relations between curvature invariants allow some terms to be expressed as a sum of others. Imposing these relations on the anomaly basis ensures that the remaining terms are linearly independent. Each of the linearly independent basis elements are then included in the integrand of $\delta_\omega \CW$, with free coefficients.

\item \textbf{Impose Wess-Zumino (WZ) consistency.} Since Weyl transformations are Abelian, the WZ consistency condition forces the anti-symmetrization of two independent Weyl variations of $\CW$ to vanish, i.e., $[\delta_{\omega_1} , \delta_{\omega_2} ] \CW = 0$. Solving WZ consistency sets to zero some of the coefficients of the linearly independent anomaly basis. Moreover, it can fix some coefficients in terms of others, which leads to linear combinations of terms with a single unfixed coefficient.

\item \textbf{Determine which terms in the Weyl anomaly are scheme-independent.} To do so, we introduce in the effective action $\CW$ all possible local, diffeomorphism and Lorentz invariant counterterms with $p$ derivatives acting on $g_{\mu\nu}$ and $e_a^\mu$. Including each of these terms with their own free coefficient builds up the counterterm action, $\CW_{CT}$. Taking an infinitesimal Weyl transformation of $\CW_{CT}$, one can shift the anomaly $\delta_\omega \CW \mapsto \delta_\omega \CW+ \delta_\omega \CW_{CT}$. Since the coefficients in $\CW_{CT}$ are free, one can then tune them so as to cancel terms appearing in $\delta_\omega \CW$. The WZ consistent terms that cannot be (partially) removed by any choice of counterterms are the scheme-independent contribution to the Weyl anomaly.

\end{enumerate}

This algorithm can be computationally intensive. Typically, the algorithm grows more difficult to implement as the defect's dimension, $p$, increases. In particular, the number of diffeomorphism and Lorentz invariant terms with $p$ derivatives increases sharply with $p$. Moreover, ensuring that the terms are linearly independent becomes increasingly involved, because geometric relations can be adorned with derivatives in numerous ways. 

In~\App{WZ-example}, we illustrate the algorithm with the simple example of a $p=2$ defect in a $d=4$ ambient CFT. For such a surface defect, there are 10 terms in step 1. After step 2 there are 7 terms, of which 5 are scheme-independent and remain in the end. By contrast, for a defect of dimension $p=4$, those numbers are larger by an order of magnitude, as we explain in the following subsections and in app.~\ref{app:basis_terms}. We employ the xAct package for Mathematica~\cite{xAct} to facilitate dealing with these large numbers of terms. In our supplemental Mathematica notebook, we also derive many intricate geometric relations that ensure that our basis is linearly independent.

\subsection{Defect Weyl anomaly for $d\geq 6$}
\label{sec:defect_Weyl_anomaly}

Following the above algorithm for arbitrary $q$, in step 1 we find a 101-dimensional basis of terms. We report this basis of terms in appendix~\ref{app:basis-any-q}. Step 2, WZ consistency, selects linear combinations of these terms with 61 unfixed coefficients. In step 3, we find that of these 61 contributions, 29 are scheme independent, and thus make up the Weyl anomaly of a $p=4$ conformal defect.

These 29 terms comprise the expected A-type anomaly of the induced connection on the defect, $\overline{E}_4$, as well as 28 B-type terms. The Weyl variation of $\overline{E}_4$ is a total derivative that cannot be removed by a counterterm, whereas the Weyl variations of the B-type terms are either exactly zero or a total derivative that can be removed by a counterterm\footnote{We have checked this statement explicitly for the non-trivial conformal invariants $\CI$ in \eq{boundary-Weyl-anomaly} and $\CJ_1$ in \eq{defect-Weyl-anomaly}.  However, we have not confirmed that this is true for $\CJ_2$.   }.  Of the 28 B-type terms, 22 are of even parity. The remaining 6 break parity along the defect because they contain a defect Levi-Civita tensor $\varepsilon^{abcd}$.

All of the terms mentioned above are admissible in any co-dimension $q$. The case $q=1$, however, is special as the symmetry properties of curvature tensors cause many terms to vanish identically. Moreover, terms that are distinct for general $q$ may reduce to the same term when $q=1$. The 22 parity-even terms in general $q$ reduce to 8 different terms when $q=1$, and of the 6 parity-odd terms, only 3 are non-zero when $q=1$.

In addition to the terms that exist for any co-dimension $q\geq 2$, for special values of $q$ certain additional terms may appear that contain the totally anti-symmetric tensor in the normal bundle, $n^{\mu_1 \ldots \mu_q}$. As explained in sec.~\ref{sec:defgeom}, we refer to these terms as being parity-odd in the normal bundle. We find that for $q=2$, there is 1 additional such term, and for $q=4$ there are 8. For $q=3$ and $q \geq 5$, the index structure is too restrictive, and effectively rules out the existence of such terms. 

We now present the full Weyl anomaly of a $p=4$ defect.  For clarity, and since we will only be considering $p=4$ defects and boundaries, from this point forward we will drop the ${}^{\text{\tiny{(4d)}}}$ superscript label on all anomaly coefficients. After implementing the algorithm outlined above, and using the same normalisation as in~\eq{4d-Weyl-anomaly}, we arrive at
\begin{align}
\begin{split}\label{eq:defect-Weyl-anomaly}
\hspace{-1cm}\left. T^\mu{}_{\mu}\right|_{\Sigma_4} =\frac{1}{(4\pi)^2}\Big(&-\bcc\overline{E}_4 +\dcc{1}\CJ_1 +\dcc{2}\CJ_2+ \dcc{3}W_{abcd}W^{abcd} +\dcc{4}(W_{ab}{}^{ab})^2  \\
&+ \dcc{5}W_{aibj}W^{aibj}+\dcc{6}W^b{}_{iab}W_c{}^{iac}+\dcc{7}W_{ijkl}W^{ijkl}+\dcc{8}W_{aijk}W^{aijk} \\
&+ \dcc{9}W_{abjk}W^{abjk} +\dcc{10}W_{iabc}W^{iabc}+ \dcc{11}W^c{}_{acb}W_d{}^{adb} +\dcc{12}W^a{}_{iaj}W_b^{ibj}\\
& +\dcc{13}W_{ab}{}^{ab}\oII^i_{cd}\oII^{cd}_i+\dcc{14}W^{a}{}_{bij}\oII^i_{ac}\oII^{jbc}+ \dcc{15}W^a{}_{ibj}\oII^i{}_{ac}\oII^{jbc}\\
&+\dcc{16}W^{abcd}\oII^i_{ac}\oII_{ibd}+ \dcc{17}W_a{}^{bac}\oII^i_{bd}\oII_{ic}{}^d+\dcc{18}W^c{}_{icj}\oII^i_{ab}\oII^{jab}\\
&+\dcc{19}\Tr~\oII^i\oII_i\oII^j\oII_j+\dcc{20}\Tr~\oII^i\oII^j\oII_i\oII_j + \dcc{21}(\Tr~\oII^i\oII_i)^2
  + \dcc{22}(\Tr~ \oII^i\oII^j)(\Tr~\oII_i\oII_j)\\
&+\tdcc{1}W_{abcd}W^{ab}{}_{ef}\varepsilon^{cdef} +\tdcc{2}W_{ijab}W^{ij}{}_{cd}\varepsilon^{abcd}+ \tdcc{3}\overline{D}_a\oII^i_{bf}\overline{D}_c\oII_{id}{}^f \varepsilon^{abcd}\\
&+\tdcc{4}W_{abcd}\oII^{ia}{}_e\oII_{if}^b\varepsilon^{cdef}+ \tdcc{5}W_{ijab}\oII^{ie}_c\oII^j_{de}\varepsilon^{abcd}+\tdcc{6}\oII^i_a{}^e\oII^j_{be}\oII_{ic}{}^f\oII_{jdf}\varepsilon^{abcd} \Big)\,.
\end{split}\hspace{-2cm}
\end{align}

The only A-type term is the first term, involving the intrinsic Euler density, whose coefficient defines $\bcc$. The next $22$ terms, with coefficients $(\dcc{1},\ldots,\dcc{22})$, are B-type and parity even. The final $6$ terms, with coefficients $(\tdcc{1},\ldots,\tdcc{6})$, are B-type and parity odd along the defect.

In what follows, for B-type central charges, $d$ or $\tilde{d}$ with subscripts denote defect central charges, as in eq.~\eqref{eq:defect-Weyl-anomaly} above, and $b$ or $\tilde{b}$ with subscripts denote boundary central charges, as in eq.~\eqref{eq:boundary-Weyl-anomaly} below. A handy mnemonic device is then ``$d$ for defect and $b$ for boundary''.

As we noted above, solving WZ consistency can sometimes relate the coefficients of the linearly independent terms in our basis to one another, which leads to non-trivial conformal invariants built out of a linear combination of basis elements. For a $p=4$ defect, there are indeed two such WZ consistent non-trivial conformal invariants, $\CJ_1$ and $\CJ_2$, which appear in the first line of eq.~\eqref{eq:defect-Weyl-anomaly}, with coefficients $\dcc{1}$ and $\dcc{2}$, and take the form
\begin{align}\label{eq:J1}
\begin{split}
\CJ_1 &= \frac{1}{d-1}R \oII^i_{ab}\oII^{ab}_i-\frac{1}{d-2}N^{\mu\nu}R_{\mu\nu} \oII^i_{ab}\oII^{ab}_i-\frac{2}{d-2}R^a{}_b\oII^{i}_{ac}\oII_{i}^{bc}-\frac{1}{2}W^c{}_{acb}\II_i\oII^{iab}\\
&\qquad+\frac{4}{9}W^c{}_{ica}\overline{D}^b\oII^i_{ab}+\oII^{iab}D_i W^c{}_{acb}-\frac{1}{2}\II^i\Tr~\oII_i\oII^j\oII_j+\frac{1}{16}\II^i\II_i\Tr~\oII^j\oII_j\\
&\qquad+\frac{2}{9}\overline{D}^b\oII^i_{ab} \overline{D}^c\oII_{ic}{}^a\,,
\end{split}
\end{align}
\begin{align}\label{eq:J2}
\begin{split}
\CJ_2 =&~\frac{d-4}{d-2}W_{ab}{}^{ab}N^{\mu\nu}R_{\mu\nu}-\frac{d-4}{d-1} RW_{ab}{}^{ab}+\frac{4(d-5)}{3(d-2)}R_{ab}W_c{}^{acb}\\
&-\frac{5(d-4)}{48}W_{ab}{}^{ab}\II^i\II_i+\frac{2(d-5)}{3}W^c{}_{ica}\overline{D}^b\oII^i_{ab}+\frac{4(d+1)}{9}\oII^{iab}D_i W^c{}_{acb}\\
&-\frac{1}{3}W_{ic}{}^{ac}\overline{D}_a\II^i-\frac{2(d-5)}{3}\II^i\Tr~\oII_i\oII^j\oII_j+\frac{(d-10)}{12}\II^iD_i W_{ab}{}^{ab}+\frac{1}{3}D^iD_iW_{ab}{}^{ab}\,,
\end{split}
\end{align}
where for example $D^i D_i W_{ab}{}^{ab} = N^{\mu\nu} h^{\rho\tau} h^{\sigma\kappa} D_\mu D_\nu W_{\rho\sigma\tau\kappa}$.

Also as noted above, even though our basis elements for the defect Weyl anomaly are linearly independent, the basis is not unique. To illustrate this, consider the square of the intrinsic Weyl tensor $\ovl{W}_{abcd}\ovl{W}^{abcd}$. The Gauss \eq{gauss} and its index contractions imply
\begin{align}
\begin{split}\label{eq:intrinsic-Weyl-squared}
\overline{W}_{abcd}\overline{W}^{abcd} =~& W_{abcd}W^{abcd}+\frac{1}{3} (W_{ab}{}^{ab})^2-2W^c{}_{acb}W_d{}^{adb}-\frac{2}{3}W_{ab}{}^{ab}\oII^i_{cd}\oII^{cd}_i\\
&+4W^{abcd}\oII^i_{ac}\oII_{ibd}+4W_a{}^{bac}\oII^i_{bd}\oII_{ic}{}^d-2\Tr~\oII^i\oII_i\oII^j\oII_j-2\Tr~\oII^i\oII^j\oII_i\oII_j\\
&+\frac{1}{3}(\Tr~\oII^i\oII_i)^2+2(\Tr~ \oII^i\oII^j)(\Tr~\oII_i\oII_j)\,.
\end{split}
\end{align}
Thus, in eq.~\eqref{eq:defect-Weyl-anomaly}, replacing any of the terms that appear on the right-hand side in eq.~\eqref{eq:intrinsic-Weyl-squared} with $\overline{W}_{abcd}\overline{W}^{abcd}$ yields an equally admissible basis. One could also consider replacing the intrinsic Euler density $\ovl{E}_4$ with the scheme-independent part of the defect's intrinsic four-dimensional $Q$-curvature~\cite{Branson:1991fd}, which can be written as\footnote{The full form of Branson's $Q$-curvature in $d=4$ contains a total derivative,
\begin{align*}
Q = \frac{1}{6}(R^2 - 3R_{\mu\nu}R^{\mu\nu} - \Box R)\,.
\end{align*}
The $\Box R$ term linearises the Weyl variation, and ensures that $\delta Q$ is a total derivative linear in $\delta\omega$, with all terms at higher order in $\delta\omega$ vanishing identically. Since the Weyl transformation of $Q$ is a total derivative, $\int Q$ is Weyl invariant. In \eq{intrinsic-Q}, we consider the $Q$-curvature of a $p=4$ submanifold obtained by replacing $R\to \ovl R$ and $\Box \to \ovl \Box$. The $\ovl\Box \ovl R$ term plays no role in our discussion as it can be removed by a local counterterm on the defect.} 
\begin{equation}\label{eq:intrinsic-Q}
\int \ovl Q \,\delta \omega = \frac{1}{4} \int \left(\overline{E}_4-\overline{W}_{abcd}\overline{W}^{abcd} \right)\delta\omega\,,
\end{equation}
where we have put a bar on $Q$ to emphasise that it is constructed with intrinsic curvatures of the submanifold. In~\eq{defect-Weyl-anomaly}, the effect of replacing $\ovl{E}_4$ with $Q$-curvature and using~\eq{intrinsic-Weyl-squared} amounts to scaling the A-type coefficient by a factor of $4$ and shifting several B-type anomaly coefficients. Specifically, replacing $\ovl E_4$ with $\ovl Q$ in~\eq{defect-Weyl-anomaly} shifts
\begin{align}
\begin{split}\label{eq:Q-curvature-shifts}
 &\dcc{3} \to \dcc{3}+\bcc\,,\quad \dcc{4}\to \dcc{4} + \frac{1}{3} \bcc\,,\quad \dcc{11}\to \dcc{11} -2 \bcc\,,\quad \dcc{13}\to \dcc{13} -\frac{2}{3} \bcc\,,\\
 & \dcc{16}\to \dcc{16} + 4 \bcc\,,\quad \dcc{17}\to \dcc{17} + 4 \bcc\,,\quad \dcc{19}\to \dcc{19} -2 \bcc\,,\quad \dcc{20}\to \dcc{20} -2 \bcc\,,\\
 &\dcc{21}\to \dcc{21} + \frac{1}{3} \bcc\,,\quad  \dcc{22}\to \dcc{22} + 2 \bcc\,.
\end{split} 
\end{align}
Note that a number of coefficients are unaffected by the change to $Q$-curvature. Further, several linear combinations of $\dcc{n}$ in~\eq{Q-curvature-shifts} are invariant under the change to $Q$-curvature, many of which we will find manifest in the AdS/BCFT computation in \sn{AdS-BCFT}.

In eq.~\eqref{eq:defect-Weyl-anomaly}, we emphasise the presence of the non-trivial parity-odd anomalies, with coefficients $(\tdcc{1},\ldots,\tdcc{6})$. The term whose coefficient is $\tdcc{3}$ can be written as a linear combination of a total derivative plus the terms whose coefficients are $\tdcc{4}$ and $\tdcc{5}$. The anomalous variation of $\CW$, however, includes an additional factor of $\delta\omega$. The $\tdcc{3}$ term is not a total derivative in $\delta\CW$ because the derivative does not act on $\delta\omega$. If one were to try to absorb the $\tdcc{3}$ term into a shift of the $\tdcc{4}$ and $\tdcc{5}$ terms via partial integration, one would be left with a parity-odd Weyl invariant in $\delta_\omega \CW$ with derivatives acting on the Weyl variation parameter $\delta\omega$. Concretely, that term is $\CD_{41}$ in eq.~\eqref{eq:D4d}. This term is also WZ consistent and cannot be removed by a counterterm.  So, we find it convenient to write the anomaly with the $\tdcc{3}$ term instead of $\CD_{41}$.

Just as in the case for the parity even anomalies, the basis of independent terms for the parity odd defect anomalies is not unique.  To illustrate this, we could use the form of the intrinsic Pontrjagin density, which can be expressed as
\begin{equation}
\overline{R}_{abcd}\overline{R}^{ab}{}_{ef}\varepsilon^{cdef} = W_{abcd}W^{ab}{}_{ef}\varepsilon^{cdef}+4W_{abcd}\oII^{ia}{}_e\oII_{if}^b\varepsilon^{cdef}-4 \oII^i_a{}^e\oII^j_{be}\oII_{ic}{}^f\oII_{jdf}\varepsilon^{abcd}\,,
\end{equation}
to rewrite the parity odd part of \eq{defect-Weyl-anomaly}.  The net effect would again to be to shift and possibly rescale $\tdcc{1}$, $\tdcc{4}$, and $\tdcc{6}$, depending on the term in \eq{defect-Weyl-anomaly} that we replace.

As mentioned in sec.~\ref{sec:intro}, additional parity-odd terms may appear in the defect Weyl anomaly, for special values of the co-dimension $q$. More specifically, using our definitions of parity in sec.~\ref{sec:defgeom}, these terms are parity-odd in the normal bundle, and their existence depends on $q$ by construction, because the totally antisymmetric normal tensor has $q$ indices. The requirement that each term has $p=4$ derivatives, and the symmetry properties of the curvature tensors are so restrictive, that only when $q=2$ and $q=4$ can~\eq{defect-Weyl-anomaly} pick up such parity odd contributions. When $q=2$, in step 1 we find 41 parity-odd terms in the normal bundle. We list them in~\App{basis-q=2}. However, only one of these is WZ consistent and scheme-independent:
\begin{equation}\label{eq:parity-odd-q=2}
\left. T^\mu{}_{\mu}\right|_{\Sigma_4} \supset \frac{\delta_{q,2}}{(4\pi)^2}\; \tdcc{7} W_{ajbk}\oII^{ac}_i\oII_c{}^{bj}n^{ik}\,.
\end{equation}
When $q=4$, in step 1 the basis of parity-odd terms in the normal bundle is 6-dimensional. All 6 terms are WZ consistent and scheme-independent. They are
\begin{align}\label{eq:parity-odd-q=4}
\begin{split}
\left. T^\mu{}_{\mu}\right|_{\Sigma_4} \supset \frac{\delta_{q,4}}{(4\pi)^2}&\Big(\tdcc{8} \e^{abcd}n^{ijk\ell} W_{abij}W_{cdk\ell}+\tdcc{9} n^{ijk\ell}W_{abij}W^{ab}{}_{k\ell}\\
&+\tdcc{10} n^{ijk\ell}W_{mirj}W^m{}_k{}^r{}_\ell+\tdcc{11} n^{ijk\ell}W_{aimj}W^a{}_k{}^m{}_\ell\\
&+\tdcc{12}n_{ijk\ell}\e_{abcd} W^{abij} \oII_f{}^{ck} \oII^{fd\ell}+\tdcc{13} n_{ijk\ell}W^{abij} \oII_{ac}^{k} \oII_b{}^{c\ell}\Big)\,.
\end{split}
\end{align}

To date, little to nothing is known about the $p=4$ defect central charges we have found in eqs.~\eqref{eq:defect-Weyl-anomaly},~\eqref{eq:parity-odd-q=2}, and~\eqref{eq:parity-odd-q=4}. The one exception is the A-type central charge, $\bcc$. In ref.~\cite{Wang:2021mdq}, Wang showed that $\bcc$ obeys an $a$-theorem, for RG flows localized to the defect, and furthermore, if the defect preserves at least $\N=1$ four-dimensional SUSY, then $\bcc$ obeys a defect version of $a$-maximization.

\subsection{Boundary Weyl anomaly for $d=5$}\label{sec:boundary_Weyl_anomaly}

Here we employ the algorithm outlined above to construct the most general expression for the boundary Weyl anomaly in $d=5$.  In doing so, we will recover the expression for the parity even anomalies found in ref.~\cite{FarajiAstaneh:2021foi} (see also ref.~\cite{Blitz:2021qbp} and references therein).\footnote{The dictionary between our central charges and those in ref.~\cite{FarajiAstaneh:2021foi} is (Here $\to$ There): $-\bcc \to a/5760$, ${\dbcc{1}\to c_8/5760}$, $\dbcc{2}\to \left(c_1-\frac{1}{3}c_8\right)/5760$, $\dbcc{3}\to (c_2+c_8)/5760$, $\dbcc{i} \to c_{i-1}/5760$ for $i = 4,\ldots, 6$, ${\dbcc{7}\to (c_6+c_8)/5760}$, and $\dbcc{8}\to c_7/5760$.   Note the different sign convention for the A-type anomaly, where here the $\bcc$-theorem goes the usual way, $\bcc{}_{\text{,\tiny{UV}}} \geq \bcc{}_{\text{,\tiny{IR}}}$. }  We also identify three previously unknown parity odd anomalies, whose coefficients we denote $(\tdbcc{1},\tdbcc{2},\tdbcc{3})$. The full anomaly is
\begin{align}
\begin{split}\label{eq:boundary-Weyl-anomaly}
\left. T^\mu{}_\mu\right|_{\Sigma_4}= \frac{1}{(4\pi)^2}\Big(&-\bcc \overline{E}_4+\dbcc{1}\CI +\dbcc{2}(\Tr~\oK^2)^2+\dbcc{3}\Tr~\oK^4+\dbcc{4}W_{abcd}W^{abcd} \\
&+ \dbcc{5} W_{anbn}W^{a}{}_n{}^b{}_n +\dbcc{6}W_{abcd}\oK^{ac}\oK^{bd}+\dbcc{7}W_{anbn}\oK^a{}_c\oK^{cb} \\
&  +\dbcc{8}W_{nabc}W_n{}^{abc}+ \tdbcc{1}\overline{D}_a \oK_b{}^e\overline{D}_c\oK_{de}\varepsilon^{abcd}\\
& +\tdbcc{2}W_{ab}{}^{ef}W_{cdef}\varepsilon^{abcd}+\tdbcc{3}W_{abcd} \oK^a{}_e \oK^b{}_f \varepsilon^{cdef}\Big)\,,
\end{split}
\end{align}
where $W_{anbn}=e_a^\mu n^\nu e_b^\rho n^\sigma W_{\mu\nu\rho\sigma}$, and similarly for $W_{nabc}$. The conformal invariant $\CI$ is 
\begin{align}\label{eq:boundary-I}
\begin{split}
\hspace{-1cm}\CI &= -\frac{2}{3}R_{ab}\oK^a{}_c\oK^{cb} +\frac{1}{4}R~\Tr~\oK^2 -\frac{1}{3}R_{nn}\Tr~\oK^2 + \frac{1}{2}W_{anbn}K\oK^{ab}+ \frac{1}{16}K^2~\Tr~\oK^2   \\
&\qquad+ \oK^{ab}D_n W_{anbn}- \frac{1}{2}K~\Tr~\oK^3 +\frac{2}{9}\overline{D}^a\oK_{ab}\overline{D}_c\oK^{bc}\,,
\end{split}
\end{align}
where $D_n W_{anbn}=n^\tau e_a^\mu n^\nu e_b^\rho n^\sigma D_\tau W_{\mu\nu\rho\sigma}$.

One can also deduce the form of \eq{boundary-Weyl-anomaly} from \eq{defect-Weyl-anomaly} by setting $q\to1$.  Most of the terms in \eq{defect-Weyl-anomaly} do not have analogous $q=1$ structures: they vanish due to various symmetry properties of the curvature tensors.  Further, some of the $q>1$ structures have identical $q=1$ analogues, and so a linear combination of their coefficients end up determining the $q=1$ anomaly coefficients.  The exact dictionary for mapping the B-type central charges is as follows, where $\to$ indicates taking $q \to 1$:
\begin{align}\label{eq:defect-boundary-central-charge-map}
\begin{split}
&\dcc{1}\to\dbcc{1}\,,\quad \dcc{21}+\dcc{22}\to \dbcc{2}\,,\quad\dcc{19}+\dcc{20}\to \dbcc{3}\,,\quad \dcc{3}\to \dbcc{4}\,,\quad \dcc{5}+\dcc{11}\to\dbcc{5}\,,\\&
 \dcc{16}\to\dbcc{6}\,,\quad \dcc{15}-\dcc{17}\to\dbcc{7}\,,\quad \dcc{10}\to\dbcc{8}\,,\quad \tdcc{3}\to\tdbcc{1}\,,\quad \tdcc{1}\to\tdbcc{2}\,,\quad\tdcc{4}\to\tdbcc{3}\,,
\end{split}
\end{align}
with all other $q>1$ terms vanishing when $q=1$.

\section{Defect central charges from observables}
\label{sec:correlators}
In this section, we connect some of the coefficients that appear in the defect Weyl anomaly in~\eq{defect-Weyl-anomaly} to various physical quantities. In subsection~\ref{sec:correlators-DD} we will find a relation between the coefficient of the two-point function of the displacement operator, $\langle \CD\CD\rangle$, and the defect central charge $\dcc{1}$. In a reflection-positive DCFT, we will then argue that $\dcc{1}$ is a negative semi-definite $c$-number, by reflection positivity of $\langle \CD\CD\rangle$. In subsection~\ref{sec:correlators-T} we will relate the coefficient, $h$, in the 1-point function of the stress tensor in the presence of a co-dimension $q>1$ defect, $\langle T^{\mu\nu}\rangle$ in eq.~\eqref{eq:stress-tensor-one-pt-fn}, to the defect central charge $\dcc{2}$.  Assuming that the defect ANEC is true, we then argue that $\dcc{2}$ must be negative semi-definite. Further, we will be able to show that since $h\propto -\dcc{2}$, the defect contribution to the universal part of EE for a spherical region centered on the defect, as computed in ref.~\cite{Kobayashi:2018lil}, contains a linear combination of $\bcc$ and $\dcc{2}$, similar to the $p=2$ result in eq.~\eqref{eq:2dee}~\cite{Jensen:2018rxu}. Finally, in subsection~\ref{sec:boundary-stress-tensor}, for $d=5$ BCFTs we compute $\langle T^{\mu\nu}\rangle$ for a flat boundary of an ambient space with non-trivial curvature, and relate the coefficients of the first two leading divergent terms to linear combinations of the boundary central charges in~\eq{boundary-Weyl-anomaly}.

\subsection{Displacement operator two-point function}
\label{sec:correlators-DD}

In this subsection, we connect the anomalous scale dependence in the displacement two-point function to a particular term in the 
Weyl anomaly. For a flat defect, conformal invariance on the defect determines the $2$- and $3$-point functions of defect primaries up to constants.  
In particular, the displacement $2$-point function takes the form
\begin{align}
\label{eq:dd2pt}
\langle \CD^i({\bf{y}}_1) \CD^j({\bf 0}) \rangle = \delta^{ij} \frac{c_{\CD\CD}}{|{\bf y}|^{2p+2}} \,,
\end{align}
with constant $c_{\CD\CD}$. There are a variety of techniques for isolating the scale dependence of the correlator in eq.~\eqref{eq:dd2pt}, and hence matching to the Weyl anomaly in eq.~\eqref{eq:defect-Weyl-anomaly} or eq.~\eqref{eq:boundary-Weyl-anomaly}. In subsection~\ref{sec:shapepert} we use some of them to fix $c_{\CD\CD}$ in terms of the defect central charge $d_1$ in eq.~\eqref{eq:defect-Weyl-anomaly}, or boundary central charge $b_1$ in eq.~\eqref{eq:boundary-Weyl-anomaly}. In subsection~\ref{sec:check} we check our result in the case of the free scalar BCFT in $d=5$. In subsection~\ref{sec:dispcomments} we comment on other correlators of $\CD^i$ and their possible relation to defect/boundary central charges.

\subsubsection{Relating $c_{\CD\CD}$ to $d_1$ and $b_1$}
\label{sec:shapepert}

Here we consider an infinitesimal shape perturbation of the flat defect, $\delta X^i({\bf y})$. In this subsection we will keep $p$ generic, and at the end set $p=4$. Up to second order in the shape perturbation, the variation of the effective action reads
\begin{equation}
\label{eq:var_W_disp}
\delta_X \mathcal{W} = - \frac{1}{2} \int_{\Sigma_p} d^p y_1 \, d^p y_2\,\langle \CD_i({\bf y}_1)\CD_j({{\bf y}_2})\rangle \delta X^i({\bf y}_1) \delta X^j({\bf y}_2) + \mathcal{O}(\delta X^3)\,, 
\end{equation} 
where by translational invariance along the flat defect, $\langle \mathcal{D}_i \rangle = 0$. To isolate the logarithmic divergence we first define ${\bf s} \equiv {\bf y}_1-{\bf y}_2$ and substitute \eq{dd2pt} into \eq{var_W_disp} obtaining
\begin{equation}
\begin{split}
\delta_X \mathcal{W} & =- \frac{c_{\mathcal{D}\mathcal{D}}}{2}  \int_{\Sigma_p} d^p y \int_{\Sigma_p}  d^p s \, \frac{1}{{\bf s}^{2p+2}}  \delta X^i({\bf y}) \delta X^i({\bf y}- {\bf s}) + \mathcal{O}(\delta X^3) \\
& \supset - \frac{c_{\mathcal{D}\mathcal{D}}}{2} \int_{\Sigma_p} d^p y \int_{\Sigma_p}  d^p s \, \frac{1}{{\bf s}^{2p+2}} \frac{s^{a_1} \dots s^{a_{p+2}}}{(p+2)!} \delta X^i({\bf y}) \partial_{a_1} \dots \partial_{a_{p+2}} \delta X^i({\bf y})
\,,
\end{split}
\end{equation}
where we are summing over the repeated $i$ indices. In the second line, we Taylor expanded $\delta X^i({\bf y}- {\bf s})$ up to order $p+2$  in $\bf{s}$, and only kept the highest order to isolate the logarithmic part. For even $p$, the integral over ${\bf s}$ can be computed using the identity
\begin{equation}
\int d^p s \, \frac{1}{{\bf s}^{2p+2}} s^{a_1} \dots s^{a_{p+2}} = 
\text{vol}(\mathds{S}^{p-1})\frac{(p-2)!!}{(2p)!!} \int_\epsilon^L d s \, \frac{1}{s} \left[\delta^{a_1 a_2}\dots \delta^{a_{p+1}a_{p+2}} + \text{perm}\right],
\end{equation}
where $\epsilon$ and $L$ are UV and IR cutoffs, respectively, and perm stands for permutations of indices excluding pairwise exchange on each $\delta^{ab}$. For odd $p$, the integral over ${\bf s}$ vanishes identically, so we will assume that $p$ is even for the remainder of this computation.  Using the fact that the number of permutations is $(p+1)!!$, we obtain for the coefficient of $\log \left( \frac{\epsilon}{L} \right)$ 
\begin{equation}
\label{eq:d_W_f}
\left. \delta_X \mathcal{W} \right|_{\log \e} =  \left(-1\right)^{p/2+1}\frac{2^{-(p+2)}\pi^{\frac{p}{2}} }{p!\Gamma (\frac{p}{2}+2)} c_{\mathcal{D}\mathcal{D}}   \int_{\Sigma_p} d^p y \,\partial^{a_1} \dots \partial^{a_{p/2+1}} \delta X^{i} \partial_{a_1} \dots \partial_{a_{p/2+1}} \delta X^{i} \,,
\end{equation}
where we have used $\text{vol}(\mathds{S}^{p-1})=2\pi^{\frac{p}{2}} / \Gamma\left(\frac{p}{2}\right)$. Let us rewrite the above equation in terms of the second fundamental form. At leading order we have $ \delta\II^i_{ab} = \partial_a \partial_b \delta X^{i}$. We thus find
\begin{equation}
\label{eq:d_W_f_2}
\left. \delta_X \mathcal{W} \right|_{\log \e} =  \left(-1\right)^{p/2+1}\frac{2^{-(p+2)}\pi^{\frac{p}{2}} }{p!\Gamma (\frac{p}{2}+2)} c_{\mathcal{D}\mathcal{D}}  \int_{\Sigma_p} d^p y \,\partial^{a_1} \dots \partial^{a_{p/2-1}} \delta\II^{icd} \partial_{a_1} \dots \partial_{a_{p/2-1}} \delta\II^{i}_{cd}\, .
\end{equation}

Taking $p=4$, we find for the logarithmically divergent part of $\delta_X \CW$,
\begin{equation}
\left. \delta_X \mathcal{W} \right|_{\log\e} = \frac{1}{2}\frac{\pi ^2}{4608} c_{\mathcal{D}\mathcal{D}}  \int_{\Sigma_p} d^p y \, \delta\II^{icd} \ovl\Box \delta\II^{i}_{cd}\, .
\end{equation}
The numerical prefactor can equivalently be obtained in the following two ways. The first is to Fourier transform~\eq{dd2pt} along the defect. Setting $p=4$, and introducing a UV cut-off $\epsilon$ as a regulator, we find
\begin{align}
\int_{\Sigma_4}d^4y \langle \CD^i({\bf y})\CD^j({\bf{0}})\rangle e^{i {\bf k} \cdot {\bf{y}}} &= \delta^{ij} 
\frac{4\pi^2 c_{\CD\CD}}{|{\bf k}|}\int_\e^\infty  \frac{dr}{r^8}J_1(|{\bf k}| r)\\
& = \delta^{ij} \pi^2c_{\CD\CD}\Big(\frac{1}{3\e^6} -\frac{|{\bf k}|^2}{16\e^4} + \frac{|{\bf k}|^4}{192\e^2} +\frac{|{\bf k}|^6}{4608} \log \left(\e|{\bf k}|\right) +\ldots \Big)\,,
\nonumber
\end{align}
with $\mathbf{k}$ and $r$ the momentum and radial coordinate along the defect, respectively. The power law divergences in $\epsilon$ can be cancelled by counterterms, while the coefficient of the logarithm is the regulator-independent scale anomaly. The second, equivalent, approach is differential regularization~\cite{Freedman:1991tk}. We replace the large inverse powers of distance
with derivatives and an energy scale $\mu$:
\begin{align}
\langle \CD^i({\bf y}) \CD^j({\bf 0})\rangle = -\delta^{ij} \frac{c_{\CD\CD}}{36864}\ovl{\Box}^4\frac{\log\mu^2|{\bf y}|^2}{|{\bf y}|^2}\,.
\end{align}
Taking a logarithmic derivative with respect to the scale $\mu$ then yields
\begin{align}\label{eq:DD-scaling}
\mu\partial_\mu\langle\CD^i({\bf y}) \CD^j({\bf 0})\rangle = -\delta^{ij} \frac{c_{\CD\CD}}{18432}\ovl{\Box}^4\frac{1}{|{\bf y}|^2} =\delta^{ij} \frac{c_{\CD\CD}\pi^2}{4608} \ovl{\Box}^3 \delta^{(4)}({\bf{y}})\,,
\end{align}
which of course agrees with the methods above.

The next step is to relate $c_{\CD \CD}$ to the coefficients in the defect or boundary Weyl anomaly,~\eq{defect-Weyl-anomaly} or~\eqref{eq:boundary-Weyl-anomaly}, respectively. By \eq{W_log_A}, the logarithmic divergence in the effective action needs to match the anomaly, so we compute to second order the shape deformation of \eq{defect-Weyl-anomaly} around the configuration of a flat defect embedded in a flat ambient space. Among all of the terms in the defect anomaly, only $\CJ_1$ contains an appropriate structure to contribute at second order in the variation, which reads
\begin{equation}
\label{eq:d_A_f}
\begin{split}
\delta_X \int_{\Sigma_4} d^4y \langle T^\mu{}_\mu\rangle & =  \frac{\dcc{1}  }{72 \pi^2} \int_{\Sigma_4} d^4 y \, \partial^b \delta\oII^i_{ab} \partial^c \delta\oII_{i c}^{\,a} + \mathcal{O} \left( \delta\II^3 \right) \\
& =    \frac{\dcc{1} }{72 \pi^2} \int_{\Sigma_4} d^4 y  \, \frac{9}{16}\partial^{a} \partial^{b} \partial^{c}  \delta X^{i} \partial_{a} \partial_{b} \partial_{c} \delta X^{i} + \mathcal{O}\left( \delta X^3 \right) .
\end{split}
\end{equation}
Comparing \eq{d_A_f} to~\eqref{eq:d_W_f} then gives our main result of this subsection,
\begin{equation}
\label{eq:CD-d1-2}
c_{\CD\CD} = - \frac{72}{\pi^4} \, \dcc{1} \,. 
\end{equation}
Performing an identical computation in the boundary case using~\eq{boundary-Weyl-anomaly}, we find
\begin{align}
\label{cDDd1rel}
c_{\CD\CD} = -\frac{72}{\pi^4}\, \dbcc{1}\,.
\end{align}

As a check of our methods, let us consider $p=2$. In that case,~\eq{d_W_f_2} reduces to
\begin{equation}
\label{eq:d_W_f_2_p2}
\left. \delta_X \mathcal{W} \right|_{\log \e} = \frac{\pi}{64} c_{\mathcal{D}\mathcal{D}}   \int_{\Sigma_2} d^2 y \, \delta\II^{icd}  \delta\II^{i}_{cd}\,, 
\end{equation}
and by direct comparison to the defect Weyl anomaly in~\eq{2d-defect-Weyl-anomaly} we obtain $d_1^{(\text{\tiny 2d})} = 3 \pi^4 c_{\mathcal{D}\mathcal{D}}/4$, reproducing the known result for $d=4$ in eq.~\eqref{eq:p2D2}. In fact, our calculation shows that eq.~\eqref{eq:p2D2} is valid for any $d$.

\subsubsection{Check of the result for the free scalar BCFT in $d=5$}
\label{sec:check}

We can check our result for the boundary case, eq.~\eqref{cDDd1rel}, using a free, massless scalar in $d=5$, in the presence of a boundary. On the one hand, $\dbcc{1} = -\frac{1}{256}$ was computed in ref.~\cite{FarajiAstaneh:2021foi} using heat kernel methods. This answer is in fact independent of the boundary conditions, Robin or Dirichlet. On the other hand, we can compute the displacement operator two-point function in flat space using Wick's theorem. The displacement operator in the $q=1$ case can be identified with the boundary limit of the $T^{nn}$ component of the stress tensor. For a conformally coupled scalar, the improved stress tensor takes the form
\[
T^{\mu\nu} = (\partial^\mu \phi) (\partial^\nu \phi) - \frac{1}{2} \delta^{\mu\nu} (\partial_\rho \phi) (\partial^\rho \phi) - \frac{d-2}{4(d-1)}(\partial^\mu \partial^\nu - \delta^{\mu\nu} \Box) \phi^2\,.
\] 
Using the scalar's equation of motion, the displacement operator follows from the boundary limit of $T^{nn}$:
\begin{align}
\CD_D &= \frac{1}{2} (\partial_n \phi)(\partial_n \phi) \ , \; \; \;
\CD_N = -\frac{1}{2} (\partial_a \phi) (\partial^a \phi) + \frac{d-2}{4(d-1)} \partial_a \partial^a \phi^2 \ ,
\end{align}
with subscript $D$ for Dirichlet and $N$ for Neumann.  (Note the general Robin boundary condition reduces to Neumann in the flat space limit.)
  If we normalize the two point function of the scalar to take the form
\[
\langle \phi(x_\perp, {\bf y}) \phi(0, {\bf 0}) \rangle = \kappa \left( \frac{1}{( x_\perp^2 + {\bf y}^2)^{(d-2)/2}} \pm \frac{1}{( x_\perp^2 + {\bf y}^2)^{(d-2)/2}}  \right),
\]
with a plus sign for Neumann and minus sign for Dirichlet, then Wick's Theorem yields
\begin{align}
\langle \CD({\bf y}) \CD({\bf 0}) \rangle = \frac{2 (d-2)^2\kappa^2}{|{\bf y}|^{2d}} \ ,
\end{align}
in both cases. The standard normalization, $\kappa^{-1} = (d-2) \mbox{vol}({\mathds S}^{d-1})$, with $d=5$ then gives $c_{\CD \CD} = \frac{9}{32 \pi^4}$. Plugging this into eq.~\eqref{cDDd1rel} then gives us $\dbcc{1}=-\frac{1}{256}$, in agreement with the heat kernel result, as expected.

\subsubsection{Comments on other correlators of the displacement operator}
\label{sec:dispcomments}

Having successfully established a relationship between $c_{\CD \CD}$ and the Weyl anomaly for $p=4$ defects, it is natural to wonder if additional relationships can be established between the Weyl anomaly and other displacement operator correlation functions. One natural candidate is the displacement operator three-point function,
\begin{equation}
\langle \CD({\bf y}_1) \CD({\bf y}_2) \CD({\bf y}_3) \rangle = \frac{ c_{\CD \CD \CD}}{|{\bf y}_1 - {\bf y}_2 |^5 |{\bf y}_2 - {\bf y}_3|^5 |{\bf y}_1 - {\bf y}_3|^5} \,,
\end{equation}
with constant $c_{\CD \CD \CD}$. In the $d=4$ case with $p=3$ dimensional boundary, such a correlation function determines the coefficient of the $\Tr~\oK^3$ term in the Weyl anomaly~\cite{Herzog:2017kkj}, as shown in eq.~\eqref{eq:d4p3DDD}. However, in our $p=4$ case, an odd number of derivatives is required to produce a $\delta({\bf y}_1 - {\bf y}_2) \delta({\bf y}_1 - {\bf y}_3)$ type contact term with the correct dimensionality. No such term exists in the Weyl anomaly \eq{defect-Weyl-anomaly}, precluding the displacement $3$-point function from determining a piece of the Weyl anomaly.
 
Another obvious correlation function to investigate is $\langle T^{\mu\nu}(x) \CD({\bf y}) \rangle$, which is also determined by conformal symmetry, up to a constant~\cite{Billo:2016cpy}. In fact, the constant is $c_{\CD \CD}$, making it unlikely that further information can be obtained, while the tensor structures involved make the analysis more complicated. After some work in the $q=1$ case, we were able to check that the scale anomaly in $\langle T^{\mu\nu}(x) \CD({\bf y}) \rangle$ is consistent with the coefficients of the $ \oK^{ab}D_n W_{anbn}$ and $\overline{D}^a\oK_{ab}\overline{D}_c\oK^{bc}$ terms in \eq{boundary-I}, as we show in~\App{TD}.
 
Other correlation functions involving the stress tensor and displacement operator tend to involve undetermined functions of a cross ratio,
as well as sums over tensor structures. We have not found useful candidates to explore, except for $\langle T^{\mu\nu} \rangle$, which we turn to next.

\subsection{Stress-tensor one-point function}
\label{sec:correlators-T}

In this subsection, we compute the one-point function of the stress tensor, $\langle T^{\mu\nu}\rangle$, in two cases. First, in subsection~\ref{sec:defect-stress-tensor}, we consider $\langle T^{\mu\nu}\rangle$ for a flat $p=4$ defect in a $d\geq 6$ ambient CFT. Our results will be similar to those for a $p=2$ defect in $d>3$, reviewed in sec.~\ref{sec:defweylreview}. By applying the method of ref.~\cite{Lewkowycz:2014jia} (see also ref.~\cite{Bianchi:2015liz}), we will show that the coefficient $h$ in~\eq{stress-tensor-one-pt-fn} and the defect Weyl anomaly coefficient $\dcc{2}$ in eq.~\eqref{eq:defect-Weyl-anomaly} are related as $h \propto -\dcc{2}$. Using this result, we will then Wick-rotate to Lorentzian signature and assume the ANEC applies in the presence of a defect to argue that $\dcc{2}\leq0$. By combining the relation between $h$ and $\dcc{2}$ with a result of ref.~\cite{Kobayashi:2018lil}, we will also show that the defect contribution to the universal part of the EE of a region centered on the defect is a linear combination of $\bcc$ and $\dcc{2}$. Second, in subsection~\ref{sec:boundary-stress-tensor}, we will consider $\langle T^{\mu\nu}\rangle$ in a $d=5$ BCFT with a curved boundary. In that case the near-boundary expansion of $\langle T^{\mu\nu}\rangle$ has a number of free coefficients~\cite{PhysRevD.20.3063,Miao:2017aba}, which we determine in terms of some boundary central charges from eq.~\eqref{eq:boundary-Weyl-anomaly}.

\subsubsection{Relating $h$ to $d_2$ for a flat defect in flat space}
\label{sec:defect-stress-tensor}

In the context of computing R\'enyi entropies, where the defect is the co-dimension $q=2$ twist defect, the authors of ref.~\cite{Lewkowycz:2014jia} consider the relation between $\langle T_{\mu\nu}\rangle$ and the anomaly term $\Box W^{ij}{}_{ij}$ when $d=6$.\footnote{See also ref.~\cite{Hung:2011xb}, which considered EE in a $d=6$ dimensional CFT, but which has an error in the analysis of the contribution to the anomaly containing the $\Box W^{ij}{}_{ij}$ term. Specifically, in the third line of eq.~(2.28) of ref.~\cite{Hung:2011xb}, the variation of the term $I_3 = W_{\mu \rho \sigma\tau}(\Box \delta^\mu{}_\nu + 4 R^\mu{}_\nu - \frac{6}{5} R\delta^{\mu}{}_\nu)W^{\nu \rho \sigma\tau}$ in the ambient Weyl anomaly does not result in a conformally invariant term along the entangling surface. Indeed, a further Weyl variation vanishes up to terms that are total derivatives in the {\emph{normal}} directions, but which cannot be dropped in the integral over the entangling surface's directions.} Here we repeat their analysis, now keeping the co-dimension of the defect arbitrary, provided $q>1$. We will focus on $\CJ_2$ in~\eq{defect-Weyl-anomaly}, the only term in the anomaly that contains $\Box W^{ij}{}_{ij}$, 
\begin{equation}
\label{eq:cw_invariant}
\begin{split}
 T^{\mu}{}_{\mu}  \supset \frac{\dcc{2}}{16 \pi^2} \, \frac{1}{3}D^i D_i W^{ab}{}_{ab} \, \delta^{(q)}(x_{\perp}) + \mathcal{O}(R^2)\,,
\end{split}
\end{equation}
where we used tracelessness of $W_{\mu\nu\rho\sigma}$ to swap freely between $W^{ij}{}_{ij}$ and $W^{ab}{}_{ab}$. In \eq{cw_invariant}, the $\mathcal{O}(R^2)$ stands for terms at least quadratic in the curvatures, which will not be important to establish that $h\propto -\dcc{2}$.

The starting point for the computation is the ambient CFT on $\CM_d = \mathbb{R}^d$ with metric $\delta_{\mu\nu}$ and a flat defect wrapping $\Sigma = \mathbb{R}^4 \hookrightarrow \mathbb{R}^d$.  We then perturb the flat ambient metric so that $\delta_{\mu\nu} \rightarrow g_{\mu\nu} = \delta_{\mu\nu} +  \delta g_{\mu\nu}$. 
To first order in the metric perturbation, the effective action changes as
\begin{equation}
\label{eq:delta_W_T}
\delta_g \CW =- \frac{1}{2} \int d^d x \, \langle T^{\mu\nu} \rangle \, \delta g_{\mu\nu}\,.
\end{equation}
Crucially, since we have assumed that the perturbation is about both a flat background and flat defect, we can employ the form of the stress tensor given in~\eq{stress-tensor-one-pt-fn}.

Since in curved space the Weyl anomaly is generically non-trivial, we expect that \eq{delta_W_T} will contain a logarithmic divergence of the form in~\eq{W_log_A}. In particular, consistency between~\eq{delta_W_T} and~\eq{W_log_A} implies
\begin{equation}
\label{eq:log_rel}
\delta_g  \int d^d x \, \sqrt{g} \, \langle T^{\mu}{}_{\mu} \rangle    = -  \frac{1}{2} \left.\int d^d x \, \langle T^{\mu\nu} \rangle \, \delta g_{\mu\nu} \right|_{\log\e} \,.
\end{equation}
In the anomaly~\eq{defect-Weyl-anomaly}, $\CJ_2$ is the only conformal invariant that contains terms at most linear in both the Weyl tensor and $\oII^i_{ab}$. The term in~\eq{cw_invariant} is the only one that contributes to the first-order perturbation about the flat configuration.

Let us analyze the expressions on either side of \eq{log_rel} separately, starting with the left-hand side. The first order variation of the integral of \eq{cw_invariant} over the defect's submanifold $\Sigma_4$ may be written as
\begin{equation}
\label{eq:var_I_C}
\begin{split}
\delta_g \int_{\Sigma_4}d^4 y \sqrt{\gamma}\, \langle \left.T^\mu{}_\mu\right|_{\Sigma_4}\rangle = &  \frac{\dcc{2}}{16\pi^2} \int_{\Sigma_4} d^{4} y \, \frac{1}{3} \partial^k \partial_k  \delta W^{ab}{}_{ab}\,,
\end{split}
\end{equation}
where we dropped terms that are higher order in curvature, or subleading in the perturbation. This short computation gives us all of the information that we will need about the left-hand side of \eq{log_rel}.

The evaluation of the right-hand side of \eq{log_rel} is bit more involved. Using the form of $\langle T^\mu{}_\mu\rangle$ in~\eq{stress-tensor-one-pt-fn}, we can write the right-hand side of~\eq{log_rel} as
\begin{equation}\label{eq:int-T-one-pt-fn-variation-1}
 \hspace{-0.5cm}\int d^d x \, \langle T^{\mu\nu} \rangle \, \delta g_{\mu\nu}  =  h \int d^d x \,\left[ \frac{(d-q+1)}{|x_\perp|^d}  \delta^{ij}\delta g_{ij} -\frac{(q-1) }{|x_\perp|^d} \delta^{ab}\delta g_{ab} - \frac{d\, x_\perp^j  x_\perp^i }{|x_\perp|^{d+2}} \delta g_{ij}  \right].
\end{equation}
The perturbed metric $g_{\mu\nu} = \delta_{\mu\nu} + \delta g_{\mu\nu}$ admits an expansion near the defect of the form in eq. (2.16) of ref.~\cite{Alvarez-Gaume:1981exa}. We observe that log-divergent contributions in \eq{int-T-one-pt-fn-variation-1} can only arise from terms with near-defect behavior like $1/|x_\perp|^d$. Since $d=q+4$, we will need the fourth order in the perturbed metric's near-defect expansion, which has
\begin{equation}\label{eq:near-defect-metric-perturbation}
\delta g_{ij} \supset -\frac{1}{20} \left. \partial_r \partial_s   \delta R_{ikjl} \right|_{\Sigma_4} \, x_{\perp}^r x_{\perp}^s x_{\perp}^k x_{\perp}^l\,,
\end{equation}
where $\left.\delta R_{ikjl}\right|_{\Sigma_4} $ is the Riemann tensor of the metric perturbed around flat space,  $g_{\mu\nu}$, and evaluated on the defect. Note that terms of the form $\mathcal{O}(R^2)$ in the near defect expansion of the metric vanish for a first order perturbation around flat space. Since the orthogonal and transverse directions are independent, for computational ease we can simply match terms that have only transverse components. Plugging \eq{near-defect-metric-perturbation} into \eq{int-T-one-pt-fn-variation-1} and using the anti-symmetry of the Riemann tensor to eliminate the $x_{\perp}^i x_{\perp}^j\delta g_{ij}$ term, we then adopt cylindrical coordinates $(\rho,\,\theta_i)$ around the defect, located at $\rho =0$, to write
\begin{equation}
\begin{split}\label{eq:int-T-one-pt-fn-variation-2}
\int d^d x \, \langle T^{\mu\nu} \rangle \, \delta g_{\mu\nu} 
\supset& - \frac{h}{4}\int_{\Sigma_4} d^4 y \, \left.\partial_r \partial_s \, \delta R_{ikil}\right|_{\Sigma_4} \int_\epsilon^L d\rho \, \frac{1}{\rho} \, \int d \Omega_{q-1} \, \hat x^r \hat x^s \hat x^k \hat x^l\,,    
\end{split}
\end{equation}
where we defined $\hat x^i \equiv x^i_{\perp}/|x_\perp|$.  We also introduced a UV cutoff $\epsilon$ and an IR cutoff $L$ to regulate the $\rho$-integral around the location of the defect. The angular integral can easily be computed using the following relation,
\begin{equation}
\int d\Omega_{q-1} \, \hat x^{i_1} \dots \hat x^{i_n} =\frac{(q+n-2)!!}{(q-2)!!}  \text{vol} (\mathds S^{q-1})(\delta^{i_1 i_2}\dots \delta^{i_{n-1}i_n}+ \text{perm})\,,
\end{equation} 
where perm denotes pairwise permutations in the $i_n$ indices. Performing the angular integral in~\eq{int-T-one-pt-fn-variation-2} then allows us to compute the $\rho$-integral easily, which produces a logarithmic divergence in $\epsilon$. The totally transverse part of the coefficient of $\log\epsilon$ is
\begin{equation}
\label{eq:log_T_int_interm}
\hspace{-0.5cm}\left.    \int d^d x \, \langle T^{\mu\nu} \rangle \, \delta g_{\mu\nu} \right|_{\log\e,\,\perp} =  \frac{h}{4}\frac{ \text{vol}(\mathds{S}^{q-1})}{q(q+2)}\int_{\Sigma_4} d^4 y \, \left[ \partial^2 \delta R_{ikik}+2 \partial_k \partial_l \delta R_{ikil} \right] .
\end{equation}
Moreover, one can easily show that at first-order in the perturbation
\begin{equation}
2\partial_k \partial_l \delta R_{ikil} =  \partial^2 \delta R_{ikik}\,,
\end{equation}
so that eq.~\eqref{eq:log_T_int_interm} becomes
\begin{equation}
\label{eq:log_T_int}
\left.    \int d^d x \, \langle T^{\mu\nu} \rangle \, \delta g_{\mu\nu} \right|_{\log\e,\,\perp} =  \frac{h}{2\, q(q+2)}\text{vol}(\mathds{S}^{q-1})\int_{\Sigma_4} d^4 y \,  \partial^2 \delta R_{ikik} \,.
\end{equation}

In order to compare \eq{log_T_int} to \eq{var_I_C}, we use the explicit form of the Weyl tensor in \eq{Weyl-tensor} to find that at first order in the metric perturbation,
\begin{equation}
\label{eq:C_ij_transv}
\left.\delta W^{ab}{}_{ab}\right|_{\perp} =  \frac{12}{6+5 q + q^2} \, \delta R_{ijij} \,,
\end{equation}
where we have used the fact that we are perturbing around flat space, and we have only kept terms that solely have transverse components. Finally, by plugging~\eq{C_ij_transv} into~\eq{var_I_C}, and comparing to~\eq{log_T_int} through the relation~\eq{log_rel}, we arrive at the generic relation between the coefficient $h$ that fixes $\langle T^\mu{}_\mu\rangle$ and the defect Weyl anomaly coefficient $\dcc{2}$, for a co-dimension $q$ conformal defect in a $d=q+4$ CFT:
\begin{equation}
\label{eq:h_d_2_rel}
 h = - \frac{ \, \Gamma \left(\frac{q}{2}+1\right)}{\pi ^{\frac{q}{2}+2}\,(q+3)} \dcc{2}\,.
\end{equation}
For the special case of $q=2$, which will be useful for the monodromy defects considered in section~\ref{sec:monodromy},~\eq{h_d_2_rel} becomes
\begin{equation}\label{eq:h-d2-q=2}
h = - \frac{1}{5 \pi^3} \dcc{2}\,.
\end{equation}
Upon taking into account the different conventions, this agrees with ref.~\cite{Lewkowycz:2014jia}.

\begin{figure}[t]
\centering
\includegraphics[width=0.5\textwidth]{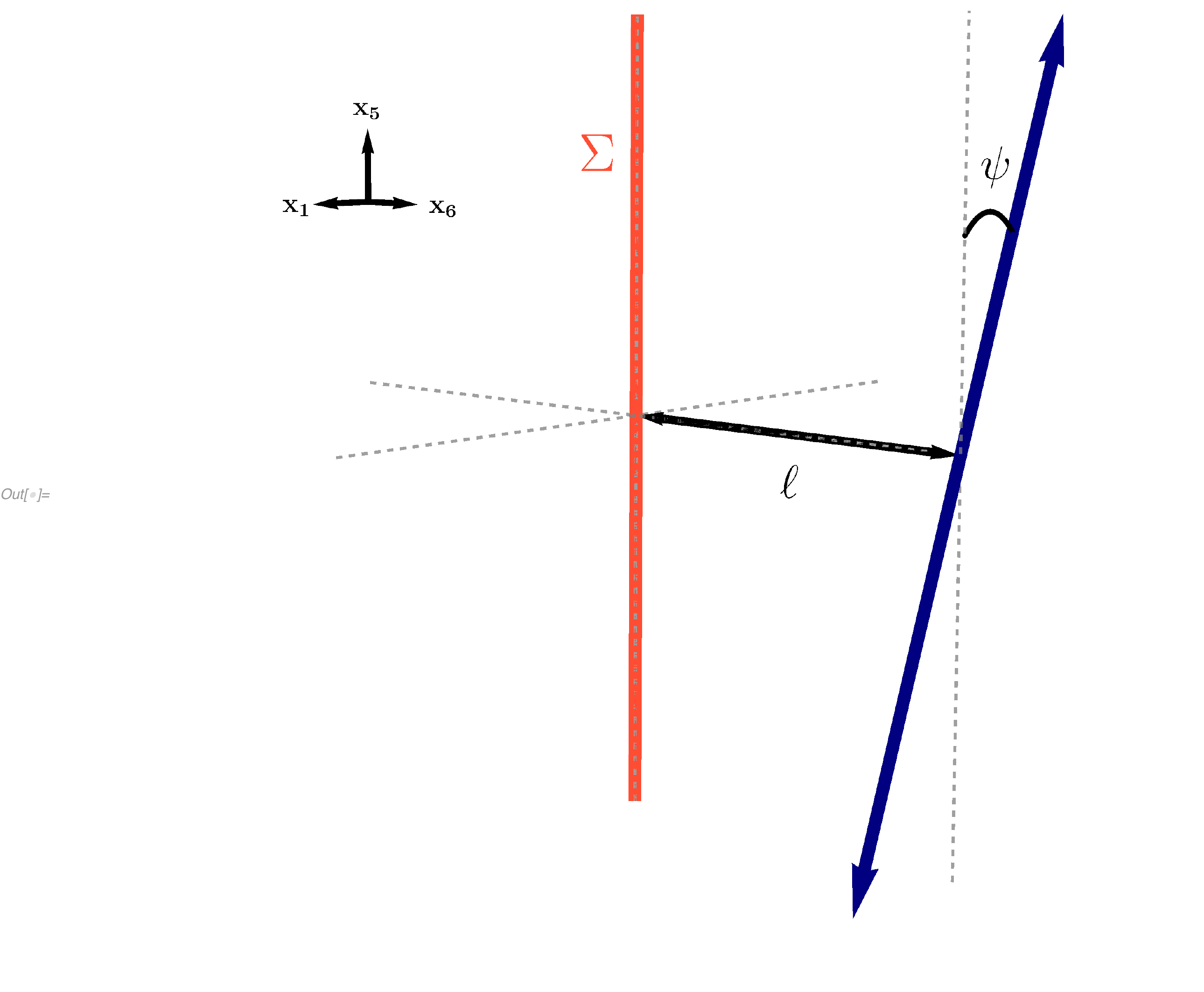}
\caption{Configuration for the null geodesic $v^\mu$ described in \eq{null_param}. The null geodesic (blue) passes by the defect (orange, labelled $\Sigma$) at an angle $\psi$ in the $x_1-x_5$ plane and at a distance $\ell$ away in the $x_6$ direction.}\label{fig:ANEC}
\end{figure}

Following the example of a $p=2$ defect in $d > 3$, reviewed in sec.~\ref{sec:cftweylreview}, we can use our result in eq.~\eqref{eq:h_d_2_rel} to show that $\dcc{2} \leq 0$, if we assume that the ANEC holds in the presence of the defect. Wick-rotating to Lorentzian signature, the ANEC states
\begin{align}
\int_{-\infty}^\infty du \langle T_{\mu\nu}\rangle v^\mu v^\nu \geq 0\,,
\end{align}
where $v^\mu$ is the tangent vector to a null geodesic with affine parameter $u$. Take $\Sigma_4$ to be a flat, static defect in $d\geq6$-dimensional Minkowski space. We consider a null geodesic passing at a minimal distance $\ell$ away from $\Sigma_4$ and oriented at an angle $\psi$ out of the plane, as in~\fig{ANEC}. We take the following family of null geodesics parametrized as
\begin{equation}
\label{eq:null_param}
t=\ell u\,, \qquad x_1 = \ell u \cos \psi\,,\qquad  x_5 = \ell u \sin \psi\,, \qquad x_6 = \ell\,,
\end{equation}
while all the other components are set to zero. Here $t, x_1$ are coordinates parallel to the defect and $x_5, x_6$ are orthogonal. By plugging~\eq{null_param} and \eq{stress-tensor-one-pt-fn} into the ANEC and using \eq{h_d_2_rel}, we obtain
\begin{align}
\begin{split}
\hspace{-0.5cm}\int_{-\infty}^\infty du \langle T_{\mu\nu}\rangle v^\mu v^\nu&= -\dcc{2}\frac{   q \, \Gamma \left(\frac{d+1}{2}\right) \Gamma \left(\frac{q}{2}\right) }{(q+3) \Gamma \left(\frac{d}{2}\right) \pi ^{\frac{q+3}{2}} \ell^{d-2}} \left|\sin \psi \right| \geq0\,,
\end{split}\end{align}
and hence $\dcc{2}\leq0$, as advertised.

Also following the example of a $p=2$ defect in $d > 3$, we can use our result in eq.~\eqref{eq:h_d_2_rel}, combined with a result of ref.~\cite{Kobayashi:2018lil}, to show that $\dcc{2}$ contributes to the universal part of the EE of a spherical region centered on the defect. Wick-rotating to Lorentzian signature and fixing the time, we consider a compact, spherical entangling region $A$ of radius $L$ that is co-original with $\Sigma_4$, such that the intersection $\partial A \cap \Sigma_4$ is an equatorial $\mathds{S}^2$. The general expression for the universal part of the defect EE for a $p$-dimensional conformal defect is~\cite{Kobayashi:2018lil}
\begin{equation}
\label{eq:S_A_gen}
\left . S_{A,\Sigma_p} \right |_{\textrm{univ.}}= -F_{\Sigma_p}- \frac{2 (d-p-1) \pi^{\frac{d}{2}+1}}{\sin\left( \frac{p}{2}\pi \right)\Gamma\left(\frac{p}{2}+1\right)\Gamma\left( \frac{d-p}{2} \right)} h\,,
\end{equation}
where
\begin{align}
F_{\Sigma_p} = -\log \frac{ Z_{\Sigma_p}[L]}{Z_{0}[L]}\,,
\end{align}
 is the defect free energy derived from the Euclidean partition function on an $\mathds{S}^d$ of radius $L$ in the presence of a defect, $Z_{\Sigma_p}[L]$, normalized by the partition function on the same background without a defect, $Z_0[L]$. The second term in eq.~\eqref{eq:S_A_gen} follows from the defect Killing energy for the time translation Killing vector. Importantly, the pole for even $p$, due to choosing a dimensional regularization scheme, maps to a logarithmic divergence in the UV-cutoff $\epsilon$ in a short distance expansion around the intersection $\partial A \cap \Sigma$. In particular, for $p=4$, the universal part of the defect EE is 
\begin{equation}
S_{A,\Sigma_4} = -4\left[\bcc - \frac{(d-5) \pi ^{d/2}}{2\,  \Gamma \left(\frac{d}{2}-2\right) } \, h \right]\, \log\left(\frac{L}{\e}\right),
\end{equation}
where we used that $F_{\Sigma_4}= 4 a_\Sigma \log \left(\frac{L}{\epsilon}\right)$. Using our result in~\eq{h_d_2_rel}, we thus find, for the universal part of the EE,
\begin{equation}\label{eq:defect-EE}
S_{A,\Sigma_4} =  -4\left[ \, \bcc + \frac{1}{4}\frac{ (d-5) (d-4)}{d-1} \, d^{{\text{\tiny(4d)}}}_2 \right]\log \left(\frac{L}{\e}\right).
\end{equation}
This result highlights the key fact that the universal part of the defect EE is not necessarily monotonic under defect RG flows. That is, in spite of the $c$-theorem for $\bcc$ proven in ref.~\cite{Wang:2021mdq}, and since no $c$-theorems are known for B-type anomalies, the presence of $\dcc{2}$ means eq.~\eqref{eq:defect-EE} is not necessarily monotonic under defect RG flows. Indeed, the $p=2$ result for EE in eq.~\eqref{eq:2dee} was explicitly shown in ref.~\cite{Rodgers:2018mvq} not to be monotonic for holographic examples of certain defect RG flows. We conclude the discussion about the defect EE by mentioning the possibility that additional central charges may appear in the coefficient of the logarithmic divergence if one considers entangling regions with a generic shape intersecting the defect, as discussed in refs.~\cite{Fursaev:2013mxa,FarajiAstaneh:2017hqv} in the $d=4$ case with a boundary. We leave the study of this more general case to future work.

\subsubsection{Boundary Weyl anomalies and $\langle T^{\mu\nu}\rangle$ with curved boundaries}
\label{sec:boundary-stress-tensor}

In this subsection, we consider a $d=5$ dimensional ambient CFT on a curved background $\CM_5$ with a boundary, $\partial\CM_5 \neq \emptyset$.  Since we assume that $\CM_5$ is not flat, the stress-tensor picks up a non-trivial one-point function in the near-boundary expansion. We will thus find a relation between some of the boundary central charges in~\eq{boundary-Weyl-anomaly} and the coefficients in the leading divergences of $\langle T_{\mu\nu}\rangle$.

Generically, when a CFT is defined on a background with a curved boundary, the near-boundary expansion of $\langle T_{\mu\nu} \rangle$ has divergences of the form~\cite{PhysRevD.20.3063,Miao:2017aba}
\begin{equation}
	\label{eq:T_exp}
	\langle T_{\mu\nu} \rangle = \frac{T_{\mu\nu}^{(d)}}{x_\perp^d} + \frac{T_{\mu\nu}^{(d-1)}}{x_\perp^{d-1}} + \frac{T_{\mu\nu}^{(d-2)}}{x_\perp^{d-2}} + \dots,
\end{equation}
where $x_\perp$ is the geodesic distance from the boundary located at $x_\perp =0$. The first three divergences can be computed simply by requiring that $T_{\mu\nu}$ is conserved and traceless~\cite{PhysRevD.20.3063}.  

The residual conformal symmetry at the boundary is enough to constrain the leading $1/x_\perp^d$ divergence to vanish identically, $T_{\mu\nu}^{(d)} = 0$.  The subleading divergences, however, have much richer structures determined by the Weyl and extrinsic curvatures:
\begin{equation}
	\label{eq:T_coe1}
(4\pi)^2 T_{\mu\nu}^{(d-1)} = A_T \oK_{\mu\nu}\,,
\end{equation}
and\footnote{In ref.~\cite{Miao:2017aba} the authors allowed for a term of the form $\bar R_{\mu\nu} - 1/4 \, \bar  R \, \bar  g_{\mu\nu}$, which is argued to be inconsistent with conformal symmetry in ref.~\cite{PhysRevD.20.3063}. In addition, in ref.~\cite{Miao:2017aba}, they found that this term is absent both when $d=3$ and $d=4$. The presence or absence of this term does not affect the calculation here.}
\begin{equation}
	\label{eq:T_coe2}
	\begin{split}
	(4\pi)^2	T_{\mu\nu}^{(d-2)} =&  \frac{A_T}{d-2} \left( n_\mu n_\nu - \frac{h_{\mu\nu}}{d-1}\right) \tr \oK^2 -2 \frac{A_T}{d-1} n_{(\mu} h^\rho_{\nu)} D_\rho K - \frac{2 A_T}{d-2} n_{(\mu} h_{\nu )}^\rho R_{\rho n}  \\
		& -2 A_T \oK_{(\mu}^\rho K_{\nu) \rho} + \beta_1 W_{\mu\rho \nu \sigma}n^\rho n^\sigma + \beta_2 K \oK_{\mu\nu} + \beta_3 \left( (K^2)_{\mu\nu} - \frac{h_{\mu\nu}}{d-1} \tr K^2 \right) ,
	\end{split}
\end{equation}
where parentheses indicate symmetrization over the enclosed indices. In ref.~\cite{Miao:2017aba}, the authors related the coefficients $A_T,\,\beta_1,\,\beta_2$ and $\beta_3$ to the boundary central charges for dimensions $d=3$ and $4$. Below, we will find such relations for the case $d=5$. 

To relate the near-boundary data $(A_T,\beta_1,\beta_2,\beta_3)$ to the coefficients in \eq{boundary-Weyl-anomaly}, we again employ~\eq{log_rel}. We begin by evaluating the left-hand side of \eq{log_rel}, which requires a first-order metric variation of the integrated Weyl anomaly. For ease of computation, and to facilitate matching with the right-hand side, we write the background metric in Gaussian normal coordinates,
\begin{equation}\label{eq:Gauss-normal-metric}
ds^2 = dx_\perp^2 + g_{ab}(x_\perp,y)dy^a dy^b.
\end{equation}
We next write the near-boundary expansion of the components $g_{ab}(x_\perp,y)$ in \eq{Gauss-normal-metric} up to third order in the geodesic distance $x_\perp$ from the boundary,
\begin{equation}
	\label{eq:bdy_metric}
	\begin{split}
		ds^2 = \, &  dx_\perp^2 + \Big[\bar g_{ab}-2 K_{ab}x_\perp + (K^2_{ab}- R_{anbn})x_\perp^2 +  \\
		&  - \frac{1}{3} \left( \partial_n R_{anbn} - R_{ancn}K^c{}_b - R_{cnbn}K^c{}_a  \right) x_\perp^3 +  \mathcal{O}(x_\perp^4) \Big] dy^a dy^b.
	\end{split}
\end{equation}
On the right-hand side, the variations $\delta K_{ab}$ and $\delta \bar g_{ab}$ around the background \eq{bdy_metric} would contribute to the logarithmic divergence also at the orders $\mathcal{O}\left(1/x_\perp^{2}\right)$ and $\mathcal{O}\left(1/x_\perp\right)$, respectively, corresponding to $T_{\mu\nu}^{(2)}$ and $T_{\mu\nu}^{(1)}$. However, since the form of the one-point function of the stress tensor is only known up to order $\mathcal{O}\left(1/x_\perp^3\right)$, we need to restrict to metric perturbations that obey $\delta K_{ab}= \delta \bar g_{ab}=0$, allowing only the variations $\delta R_{anbn}$ and $\delta \partial_n R_{anbn}$ to be non-trivial. To simplify the computation further, we assume without loss of generality that the boundary metric is flat, $\bar g_{ab}=\delta_{ab}$. With these assumptions, the left-hand side, i.e. the variation of the anomaly, gives
\begin{equation}
\label{eq:var_A_bdy}
\begin{split}
\hspace{-0.6cm}\delta_g \int d^5x \,\sqrt{g}\,\langle T^\mu{}_\mu\rangle =\frac{1}{(4\pi)^2} &\int_{\partial\CM_5} d^4 y \, \left\{- b_1 \frac{2}{3} \oK^{ab} \delta \partial_n R_{anbn} + \left[ \frac{2}{3}(b_6+b_7- 2 b_1)K^{ac}K^b{}_c  \right.\right. \\
&	\left.\left. + \frac{1}{12}\left( b_6+ b_7  \right) K^2 \delta^{ab}  +  \frac{1}{6}\left( 3 b_1 -2 b_6 -2 b_7 \right) K K^{ab} \right.\right. \\
&	\left.\left. - \frac{1}{6} \left(  b_1 + b_6 + b_7 \right) \tr K^2 \delta^{ab}  + \frac{4}{3}(2 b_4+b_5) W^{anbn} \right] \delta R_{anbn}  \right\}.
\end{split}
\end{equation}
On the right-hand side of \eq{log_rel}, we need the log-divergent part of the integral of \eq{T_exp} in the near-boundary expansion of the metric in \eq{bdy_metric}.  A straightforward computation as in the previous subsection yields the same structure as in \eq{var_A_bdy}, with the identifications
\begin{equation}
\label{eq:T_anom_rel}
\begin{split}
& A_T =   4 b_1, \quad \beta_1 = - \frac{8}{3} \left( 2 b_4 + b_5 \right), \quad \beta_2 =  \frac{2}{3} \left(b_6 + b_7 \right) + \frac{13}{3} b_1   \quad \beta_3 = - \frac{4}{3} \left(2 b_1 + b_6+ b_7 \right) .
\end{split}
\end{equation} 

Our results in~\eq{T_anom_rel} will play a crucial role in the holographic computations in~\sn{AdS-BCFT}. One interesting point to note about the relations in eq.~\eqref{eq:T_anom_rel} is that they are all invariant under the change of basis that replaces the intrinsic Euler density with $Q$-curvature. That is, the relations in eq.~\eqref{eq:T_anom_rel} are invariant under the shifts in eq.~\eqref{eq:Q-curvature-shifts}, after using the map from defect to boundary central charges in eq.~\eqref{eq:defect-boundary-central-charge-map}. Specifically, under these shifts $\dbcc{1}$ and $\dbcc{8}$ are invariant, while all other boundary central charges are shifted non-trivially by multiples of $\bcc$. This raises the question of whether the invariance of~\eq{T_anom_rel} under this change of basis is universal to all orders, or if it is spoiled at fourth-order by contributions due to $\delta K_{ab}$ and $\delta \bar g_{ab}$. We leave this question for future research.

\section{Defect central charges in $d\geq 6$}
\label{sec:dcc}

In this section we use the results of sec.~\ref{sec:correlators} to compute some defect central charges in examples of $p=4$ conformal defects in $d\geq 6$ CFTs. In subsection~\ref{sec:monodromy}, we consider monodromy defects in $d=6$ free field theories. Specifically, we consider free, massless complex scalars and free, massless Dirac fermions, and compute the defect central charges $a_\Sigma$, $d_1$ and $d_2$. We further consider the closely related $n$-fold cover and orbifold defects. In subsection~\ref{sec:probe-brane}, we consider the bottom-up holographic example of an AdS${}_5$ probe brane in AdS${}_{d+1}$ with $d\geq 6$. In this case we compute all defect central charges in terms of the brane tension.

\subsection{Monodromy defects in $d=6$ free field theories}
\label{sec:monodromy}

In this subsection, we compute the defect central charges $\bcc$, $\dcc{1}$, and $\dcc{2}$ for monodromy defects in $d=6$ free field CFTs. Generally, one can think of monodromy defects as surface ($q=2$) defects on which a topological $q=1$ defect that implements a flavor symmetry rotation can end. The authors of ref.~\cite{Giombi:2021uae} used holographic techniques for monodromy defects in the large $N$ limit of the interacting $O(N)$ symmetric scalar theory, mapped to $\mathds{S}^1$ times hyperbolic space, to compute correlation functions such as $\langle T_{\mu\nu}|_{\Sigma}\rangle$. Separately, in ref.~\cite{Bianchi:2021snj} monodromy defects were analyzed using defect Operator Product Expansion (OPE) and free field techniques in $\mathbb{R}^d$. In particular, in $d=4$ ref.~\cite{Bianchi:2021snj} related the one-point function of the $U(1)$ flavor current, $\langle J_\mu \rangle$, as well as $\langle \CD^i \CD^j\rangle$ and $\langle T_{\mu\nu}|_{\Sigma_2}\rangle$, to the defect central charges $\bc$, $\dc{1}$, and $\dc{2}$, respectively. In a similar fashion, we will use the form of the defect Weyl anomaly in~\eq{defect-Weyl-anomaly} and the relations of its coefficients to correlation functions in eqs.~\eqref{cDDd1rel} and~\eqref{eq:h-d2-q=2}, as well as $\langle J_\mu\rangle$, to compute $a_\Sigma$, $d_1$, and $d_2$ for monodromy defects in $d=6$ theories of single free, massless complex scalars and Dirac fermions.

To begin, our ambient geometry is flat space in Euclidean signature, $\CM_6 = \mathbb{R}^{6}$, with metric $ds^2 = g_{\mu\nu}dx^\mu dx^\nu$, which we write as
\begin{align}\label{eq:monodromy-geometry}
ds^2  &=\gamma_{ab} dy^a dy^b + d\rho^2 + \rho^2d\theta^2 = d\tau^2 + d\vec{y}^2 +d\rho^2 + \rho^2d\theta^2\,.
\end{align}
Here we have adopted polar coordinates in the $\{\rho,\,\theta\}$-plane normal to the defect, which we take to be embedded along $y^a = \{\tau,\,\vec{y}\}$, at $\rho =0$. In the computation of $\langle \CD^i \CD^j\rangle$ below, we will find it more convenient to use the complex coordinate $z = \rho e^{i\theta}$ in the normal directions.

On this background we will place a free field CFT and turn on a non-trivial monodromy around the surface $\rho=0$.  That is, we will start with the Euclidean action for a free, massless, conformally coupled complex scalar $\vphi$,
\begin{align}
\label{eq:free-scalar-action}
I_{\tiny{\text{scalar}}} = \int d^6x\sqrt{g} \left(\hat{D}^\mu \vphi (\hat{D}_\mu \vphi)^\dagger + \frac{1}{5}R |\vphi|^2\right),
\end{align}
where, anticipating the introduction of a background gauge field for the $U(1)$ flavor symmetry, $\hat{D}_\mu$ denotes a gauge covariant derivative, i.e. $\hat{D}_\mu \vphi = \partial_\mu \vphi -i A_\mu\vphi$, with $\vphi$ having unit flavor charge. We will also consider a free, massless Dirac fermion $\psi$, with action
\begin{align}
\label{eq:free-fermion-action}
I_{\tiny{\text{fermion}}}= - \int d^6x |e| \bar\psi \hat{\Dslash} \psi\,,
\end{align}
where $|e|$ is the determinant of the components of frame fields, $e^A$,
\begin{align}
e^0 = d\tau\,,\quad e^1 = d\rho\,,\quad e^2 = \rho \, d\theta\,,\quad e^\b = dy^\b\,,
\end{align}
and $\b = 3,\ldots, d$ runs over the indices for the defect's spatial directions. We also denote the gauge covariant Dirac operator $\hat{\Dslash}= \Gamma^\mu(\partial_\mu + \Omega_\mu -iA_\mu)$, with $\Omega_\mu = \frac{1}{8} \omega_\mu{}^{AB}[\Gamma_A,\Gamma_B]$, $\Gamma_\mu = e^A{}_\mu \Gamma_A$ denoting the $d=6$ gamma matrices obeying the usual Clifford algebra $\{\Gamma_\mu,\,\Gamma_\nu\} =2 g_{\mu\nu}\mathds{1}_6$, and $\omega^{AB}$ being the spin connection.

We construct a monodromy defect by turning on a constant background gauge field for the $U(1)$ flavor symmetry, where for free Dirac fermions we explicitly choose the non-anomalous vector $U(1)_V$ flavor symmetry,
\begin{align}
\label{eq:Adef}
A &= \a \, d\theta\,,
\end{align}
and minimally coupling to $\vphi$ or $\psi$ with unit charge. In what follows, we set $\a\in[0,1)$ for both free scalars and free Dirac fermions. Turning on $A$ is equivalent to prescribing a non-trivial monodromy, $\vphi \to e^{-i\a\theta}\vphi$, and similarly for $\psi$, if we gauge away the singular (at $\rho=0$) background gauge field. The conserved current $J_\mu$ sourced by $A_\mu$ has a non-trivial one-point function, whose form is fixed up to a function of $\alpha$,
\begin{align}
\langle J^\theta (x)\rangle = \frac{C_J(\a)}{\rho^d}\,.
\end{align}
If we consider a spherical monodromy defect, then ref.~\cite{Bianchi:2021snj} showed that, for $d$ even, $C_J(\alpha)$ is related to the defect A-type anomaly coefficient by
\begin{align}
\label{eq:b-CJ-integral}
\frac{d\CA}{d\a} =(-1)^{\frac{d}{2}}\frac{4\pi^{\frac{d}{2}}}{\Gamma(\frac{d}{2})} C_J(\a)\,,
\end{align}
where $\CA = \int \sqrt{g}\langle T^\mu{}_\mu\rangle$ is the integrated Weyl anomaly.
On the sphere the only part of $\CA$ that is non-vanishing is the integrated Euler density. In our case, eq.~\eqref{eq:b-CJ-integral} implies
 \begin{align}
 \label{eq:b-CJ-6d-integral}
\bcc(\a)= \frac{\pi^3}{2} \int d\a \, C_J(\a)\,.
\end{align}
For a monodromy defect, eq.~\eqref{eq:b-CJ-6d-integral} thus gives us another observable, in addition to those in sec.~\ref{sec:correlators}, to obtain a defect central charge, namely $\bcc$.

Ref.~\cite{Bianchi:2021snj} provides the details for the solutions of the free fields' equations of motion, using a mode expansion 
that exploits the defect's cylindrical symmetry, followed by the computation of the propagators, and then the 
calculation of $\langle J_\mu \rangle$, $\langle \CD^i \CD^j\rangle$, and $\langle T_{\mu\nu}|_{\Sigma}\rangle$.

Importantly, however, the choice of $\alpha$ does not uniquely specify the monodromy defect. There is a possibility of having two different defect operators with the same transverse spin that both appear in the defect OPE of a free field. The free field equations of motion do put strong constraints on the defect OPE~\cite{Bianchi:2021snj,Lauria:2020emq}. For co-dimension $2$ defects, a free scalar can only couple to defect scalars  $\hat{O}^\pm_s$ of dimension $\Delta_\pm(s) = \frac{d-2}{2} \pm |s|$, while a free fermion couples to defect fermions of dimension $\Delta_\pm(s) = \frac{d-1}{2} \pm |s|$, where $s$ is the transverse spin of the defect operator. While the plus sign is possible for any $s$, defect unitarity restricts the choice of the minus sign to cases where $|s| < 1$. The monodromy boundary condition means further that $s \in {\mathbb Z} - \alpha$. Further information about spin $s = -\alpha$ and $s=1-\alpha$ defect operators must be specified to define the monodromy defect. In the case of the scalar, we use the two free constants $\xi,\,\tilde\xi \in [0,1]$ to parametrize the admixture of the $\hat{O}^-_{-\alpha}$ and $\hat{O}^- _{1-\alpha}$ operators in the defect OPE of the bulk scalar. We also note that these two primary operators correspond to the defect limit of singular modes with a mild divergence, i.e. modes with divergences softer than $\rho^{-1}$ as $\rho\to0$. 

The analysis of the Dirac fermions has an additional wrinkle. In this case, consistent solutions to the equations of motion in the presence of a monodromy defect required at least one of two singular modes, which in the conventions of ref.~\cite{Bianchi:2021snj} are contained in some components of the defect spinors $\hat \psi_{\alpha}$ and $\hat \psi_{-\alpha}$. It is enough then to introduce only one constant $\xi \in [0,1]$ to fix the defect OPE of the bulk fermion.

\subsubsection{$\langle T_{\mu\nu}\rangle$ for monodromy defects in $d=6$}

Using the relations in eqs.~\eqref{eq:stress-tensor-one-pt-fn} and~\eqref{eq:h-d2-q=2}, we can use free field methods to derive $\dcc{2}$ from the one-point function of the stress tensor for both free scalars and free Dirac fermions. Starting with the theory of free scalars in~\eq{free-scalar-action} and the definition of the stress-tensor,
\begin{align}
T_{\mu\nu} \equiv \frac{2}{\sqrt{g}}\frac{\delta I_{\tiny{\text{scalar}}}}{\delta g^{\mu\nu}}\,,
\end{align}
we find
\begin{align}
T_{\mu\nu} = \hat{D}_\mu\vphi (\hat{D}_\nu\vphi)^\dagger + (\hat{D}_\mu\vphi)^\dagger \hat{D}_\nu\vphi - \frac{2}{5}\left[D_\mu D_\nu + \frac{1}{4}g_{\mu\nu}\Box - R_{\mu\nu}\right]|\vphi|^2 \,.
\end{align}
To fix the coefficient of the one-point function of the stress tensor we can analyze the $\rho\rho$-component, which takes the form~\cite{Bianchi:2021snj}
\begin{align}\begin{split}\label{eq:scalar-T-one-point-fn}
\langle T_{\rho\rho}\rangle &=- \frac{\alpha(1-\alpha^2)(2-\alpha)}{360\pi^3}\left(\alpha(1-\alpha) + \frac{6\alpha^2\xi}{2-\alpha}+\frac{6(1-\a)^2\tilde{\xi}}{1+\a}\right)\frac{1}{\rho^6}\,,\\
\Rightarrow \quad h &= \frac{\alpha(1-\alpha^2)(2-\alpha)}{360\pi^3}\left(\alpha(1-\a) +\frac{6\a^2\xi}{2-\a}+\frac{6(1-\a)^2\tilde{\xi}}{1+\a}\right).
\end{split}\end{align}
Using~\eq{h-d2-q=2}, we can then easily read off the value of the defect central charge,
\begin{align}\label{eq:monodromy-scalar-d2}
\dcc{2} = -\frac{\a(1-\a^2)(2-\a)}{72}\left(\a(1-\a) +\frac{6\a^2\xi}{2-\a} + \frac{6(1-\a)^2\tilde{\xi}}{1+\a}\right).
\end{align}

For free Dirac fermions, we start from~\eq{free-fermion-action} and vary with respect to the frame fields to find,
\begin{align}
T_{\mu\nu} = \frac{1}{2}\Big(\bar\psi \Gamma_{(\mu} \partial_{\nu)}\psi -\partial_{(\mu}\bar\psi\Gamma_{\nu)}\psi -\bar\psi (\Omega_{(\mu}\Gamma_{\nu)} -i A_{(\mu}\Gamma_{\nu)})\psi\Big)\,,
\end{align}
where parenthesis denotes symmetrization across indices. In this case, the $\rho\rho$-component of the one-point function of the stress tensor takes the form~\cite{Bianchi:2021snj}
\begin{align}\begin{split}
\label{eq:fermion-T-one-point-fn}
\langle T_{\rho\rho}\rangle &=- \frac{\alpha(1-\alpha^2)(2-\alpha)}{90\pi^3}\big(\a(\a +2) +3(1-2\a)\xi\big)\frac{1}{\rho^6}\,,\\
\Rightarrow \quad h &=  \frac{\alpha(1-\alpha^2)(2-\alpha)}{90\pi^3}\big(\a(\a +2) +3(1-2\a)\xi\big)\,.
\end{split}\end{align}
Using~\eq{h-d2-q=2}, we read off the value of the defect central charge,
\begin{align}
\label{eq:monodromy-fermion-d2}
\dcc{2} = -\frac{\a(1-\a^2)(2-\a)}{18}\big(\a(\a+2) +3(1-2\a)\xi \big)\,.
\end{align}

Note that $\dcc{2}$ for monodromy defects in theories of both free scalars and free Dirac fermions in eqs.~\eqref{eq:monodromy-scalar-d2} and \eqref{eq:monodromy-fermion-d2}, respectively, are negative semi-definite. That is, for any value of $\xi,\,\tilde{\xi}\in[0,1]$, or $\xi\in[0,1]$ and $\a\in[0,1)$, $\dcc{2}\leq0$, in accord with the defect ANEC argument in subsection~\ref{sec:defect-stress-tensor}.

\subsubsection{$\langle \CD \CD\rangle$ for monodromy defects in $d=6$}

\begin{figure}[tb]
\centering
\includegraphics[width = \textwidth]{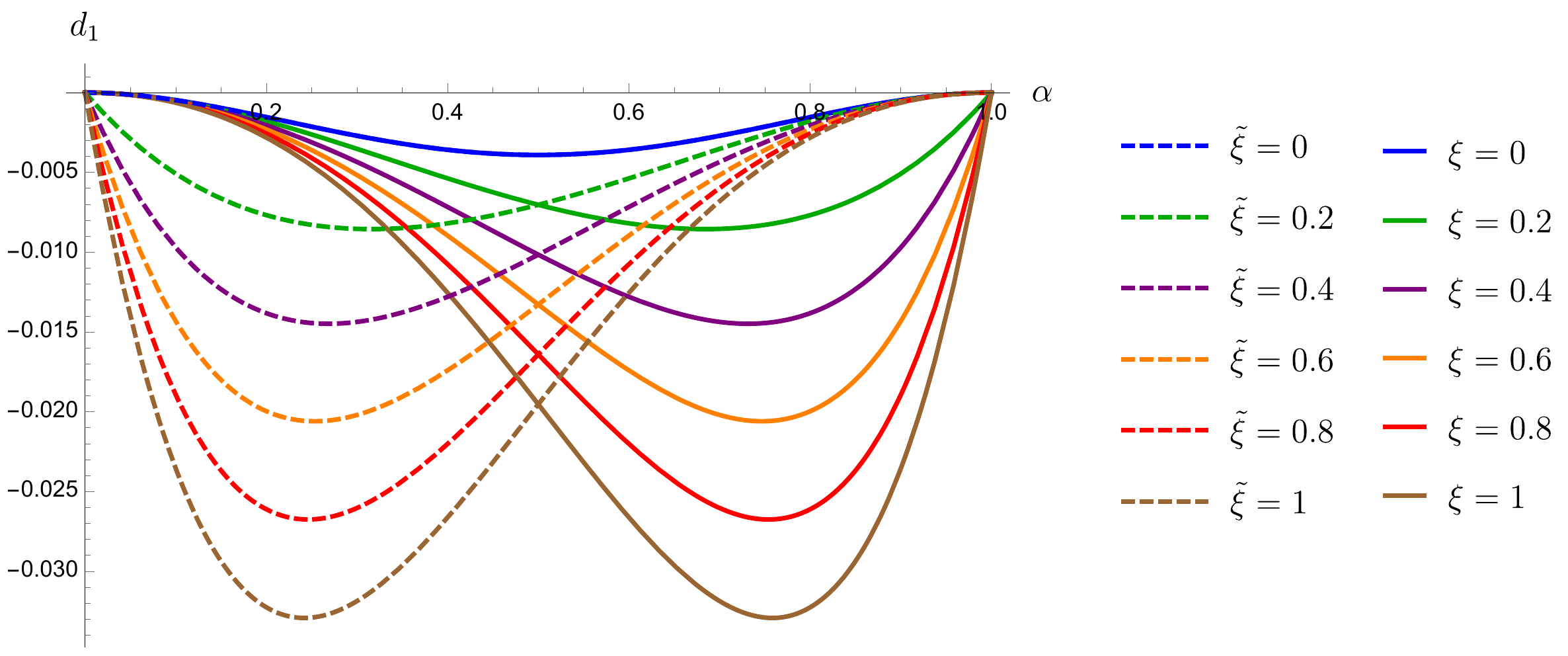}
\caption{\label{fig:d1-scalar-monodromy}Behavior of $\dcc{1}$ as a function of $\a$ for a monodromy defect in a theory of a single conformally coupled complex scalar in $d=6$, from eq.~\eqref{eq:monodromy-scalar-d1}. The solid lines have fixed $\tilde{\xi} =0$ and varying $\xi$, while the dashed lines have fixed $\xi=0$ and varying $\tilde{\xi}$.}
\end{figure}
By virtue of the simple nature of the construction of the monodromy defect, the stress-tensor of a field coupled to the $A$ in eq.~\eqref{eq:Adef} obeys
\begin{align}
D^\mu T_{\mu\nu} = J^\mu F_{\mu\nu}\,,
\end{align}
where $F_{\mu\nu}$ is the curvature of $A$, which has non-vanishing $\theta\rho$-components proportional to $\delta(\rho)$.  Comparing to \eq{displacement-operator-def}, we can see that the current $J$ is proportional to the displacement operator $\CD$. Explicitly, adopting complex coordinates $z = \rho e^{i\theta}$ and $\bar z = \rho e^{-i\theta}$, the relation between the $\CD$ and $J$, when $\xi = \tilde\xi =0$, is
\begin{align}\label{eq:monodromy-dispalcement}
\CD_z =-2\pi i \a J_z|_{z,\bar{z}=0}\,,\qquad \CD_{\bar{z}} = 2\pi i \a J_{\bar{z}}|_{z,\bar{z}=0}\,.
\end{align}
For generic values of $\xi,\tilde\xi \neq0$ we need the defect OPE, which was computed in ref.~\cite{Bianchi:2021snj}.

For the theory of free scalars in~\eq{free-scalar-action}, a computation of the defect OPE coefficients that determine $c_{\CD\CD}$ appears in sec.~3.2.4 of ref.~\cite{Bianchi:2021snj}. Sparing the details, the displacement operator two-point function for a single free complex scalar in the presence of a monodromy defect in $d=6$ takes the form
\begin{align}\label{eq:scalar-D-two-point-fn}
\langle \CD_z(y) \CD_{\bar z}(0) \rangle  = \frac{2\alpha(1-\alpha^2)(2-\alpha)}{\pi^4}\left(\alpha(1-\alpha) + \frac{6\alpha^2\xi}{2-\alpha}+\frac{6(1-\a)^2\tilde{\xi}}{1+\a}\right)\frac{1}{|y^a|^{10}}\,.
\end{align}
Using the relation between $c_{\CD\CD}$ and $\dcc{1}$ in eq.~\eqref{cDDd1rel}, we find that a generic monodromy defect in a theory of free scalars in $d=6$ is given by
\begin{align}\label{eq:monodromy-scalar-d1}
\dcc{1} = -\frac{\a(1-\a^2)(2-\a)}{36}\Big(\a(1-\a) +\frac{6\a^2 \xi}{2-\a} + \frac{6(1-\a)^2\tilde{\xi}}{1+\a}\Big)\,.
\end{align}
In~\fig{d1-scalar-monodromy} we plot $\dcc{1}$ as a function of $\alpha$ for the monodromy defect in the theory of free complex scalars, for various values of $\xi,\,\tilde{\xi}$.

\begin{figure}[tb]
\centering
\includegraphics[width =0.8\textwidth]{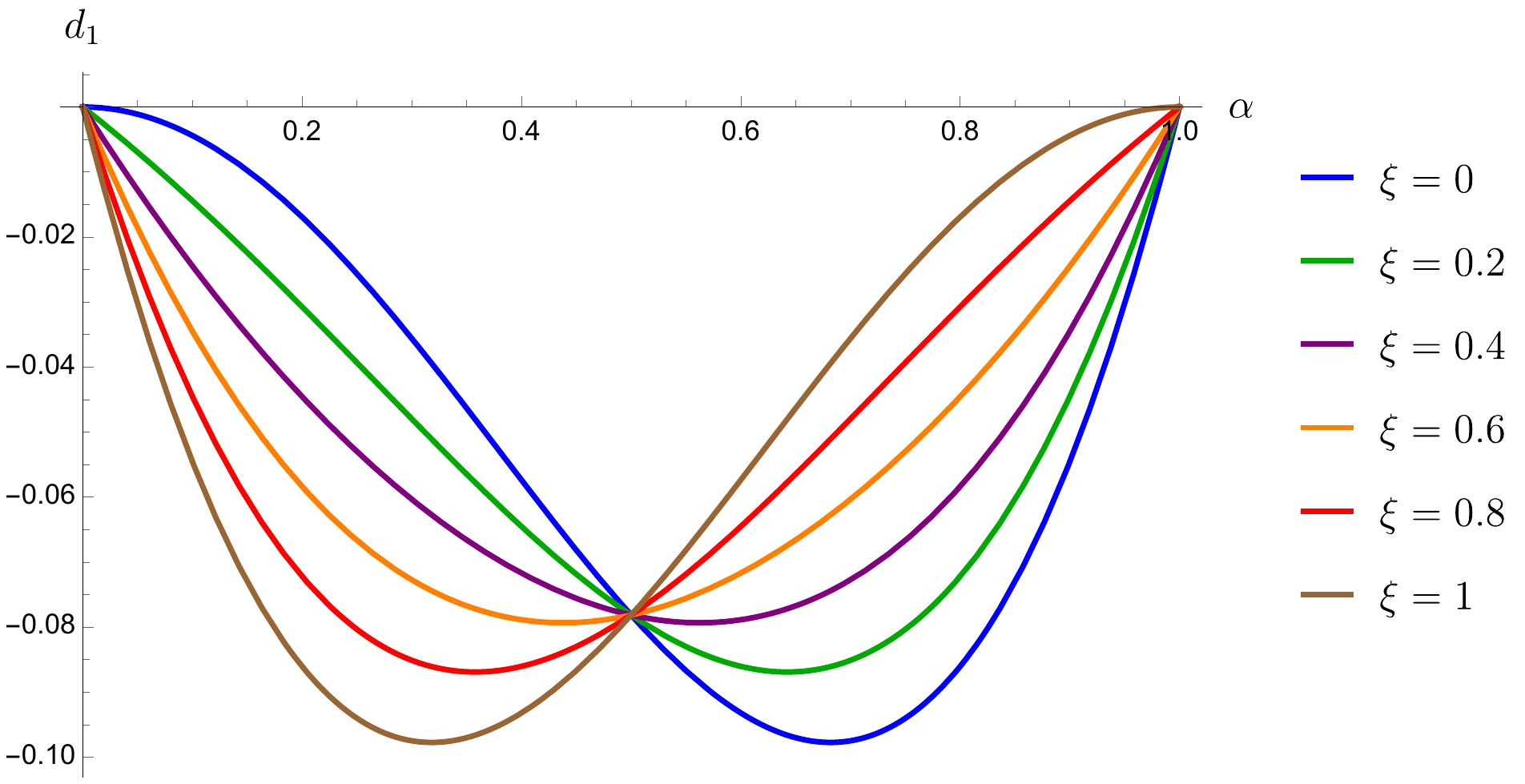}
\caption{\label{fig:d1-fermion-monodromy}Behavior of $\dcc{1}$ as a function of $\a$ for a monodromy defect in a theory of a single free Dirac fermion in $d=6$ for various values of $\xi$, from eq.~\eqref{eq:monodromy-fermion-d1}. Note that when $\a=1/2$, $\dcc{1}$ is independent of $\xi$.}
\end{figure}

For the theory of free Dirac fermions in $d=6$ in \eq{free-fermion-action}, a similar analysis using the defect OPE for $\CD$ was carried out in Sec.~4.2.3 of ref.~\cite{Bianchi:2021snj}. Again sparing the details, the displacement operator two-point function for free Dirac fermions in the presence of a monodromy defect with singular modes turned on, with generic $0<\xi<1$, takes the form
\begin{align}\label{eq:fermion-D-two-point-fn}
\langle \CD_z(y) \CD_{\bar z}(0) \rangle  = \frac{8\alpha(1-\alpha^2)(2-\alpha)}{\pi^4}\big(\alpha(2+\alpha) + 3(1-2\a)\xi\big)\frac{1}{|y^a|^{10}}\,.
\end{align}
Using the same relation between $c_{\CD\CD}$ and $\dcc{1}$ in eq.~\eqref{cDDd1rel}, we find 
\begin{align}
\label{eq:monodromy-fermion-d1}
\dcc{1} = -\frac{\a(1-\a^2)(2-\a)}{9}\big(\a(2+\a) + 3(1-2\a)\xi\big)\,.
\end{align}
We plot this $\dcc{1}$ in~\fig{d1-fermion-monodromy}.

Monodromy defects in free field theories obey the following relation between $h$ and $c_{\CD\CD}$, originally conjectured for superconformal defects in ref.~\cite{Bianchi:2015liz} 
\begin{align}
\label{d1d2rel}
c_{\CD\CD} =  \frac{2^{d} d}{\pi^{\frac{d-3}{2}} }\Gamma\Big(\frac{d+1}{2}\Big) h\qquad \Rightarrow \qquad \dcc{1}  =  2 \dcc{2}\,.
\end{align}
Comparing eqs.~\eqref{eq:monodromy-scalar-d2} and~\eqref{eq:monodromy-scalar-d1} for free scalars, and eqs.~\eqref{eq:monodromy-fermion-d2} and~\eqref{eq:monodromy-fermion-d1} for free Dirac fermions, we find precisely $\dcc{1}= 2\dcc{2}$ in both cases. Note also that in both cases, $\dcc{1}\leq 0$ for all $\a\in[0,1)$ and $\xi,\tilde{\xi}\in[0,1]$, as expected from reflection positivity for $\langle \CD\CD\rangle$.

\subsubsection{$\langle J\rangle$ for monodromy defects in $d=6$}

\begin{figure}[tb]
\centering
\includegraphics[width = \textwidth]{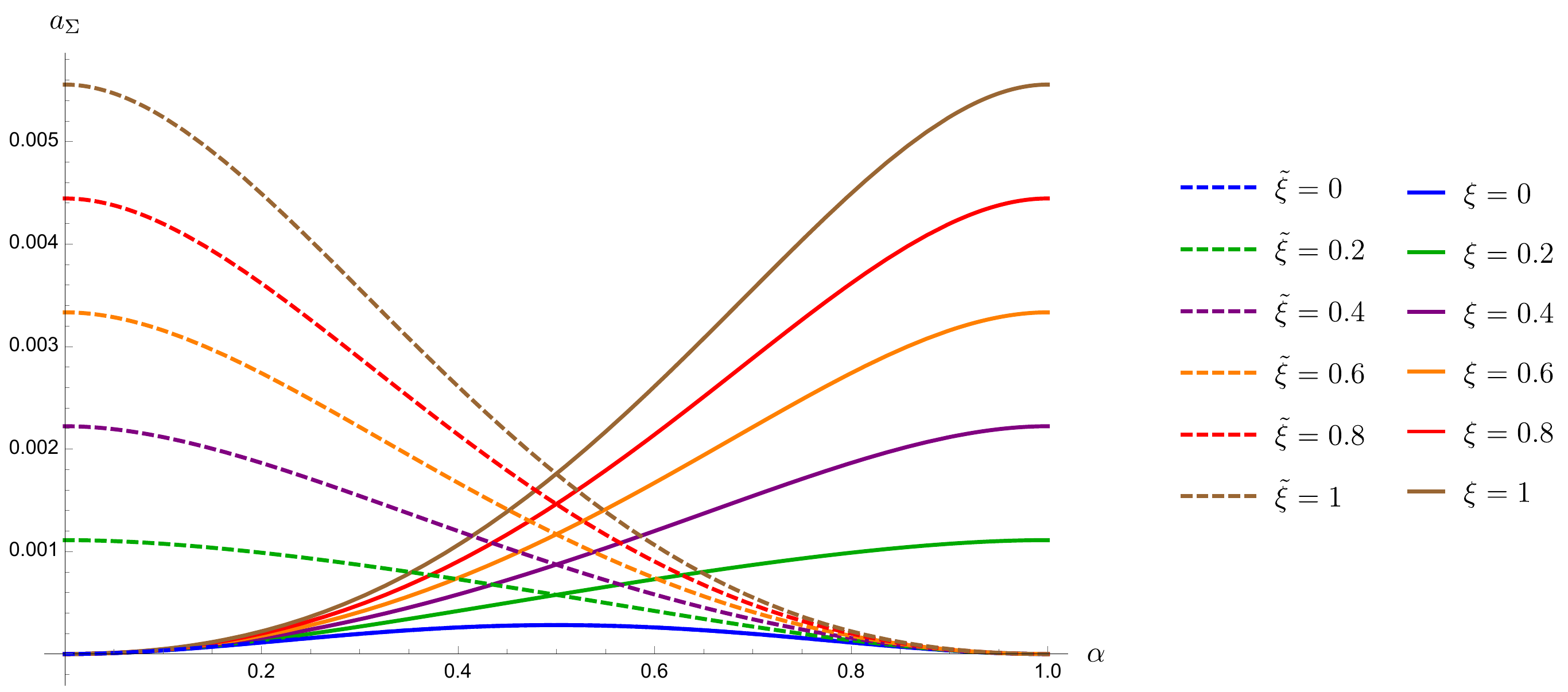}
\caption{\label{fig:a4-scalar-monodromy} Behavior of $\bcc$ as a function of $\a$ for a monodromy defect in a theory of a single conformally coupled complex scalar in $d=6$. The solid lines have fixed $\tilde{\xi} =0$ with varying $\xi$, and the dashed lines have fixed $\xi=0$ with varying $\tilde{\xi}$.}
\end{figure}

Starting with the theory of free scalars in \eq{free-scalar-action}, the conserved $U(1)$ current is given by
\begin{align}
J_\mu = \frac{1}{\sqrt{g}}\frac{\delta I_{\tiny{\text{scalar}}}}{\delta A^\mu} = -i\vphi  D_\mu \vphi^\dagger + i D_\mu \vphi \vphi^\dagger + 2A_\mu |\vphi|^2\,.
\end{align}
A brief computation of the non-vanishing components of $J_\mu$ using the mode expansion for $\vphi$ with arbitrary $\xi,\,\tilde\xi$ yields \cite{Bianchi:2021snj}
\begin{align}\label{eq:scalar-J-one-point-fn}
\langle J_\theta\rangle = \frac{\alpha (1-\alpha^2)(2-\alpha)}{120\pi^3}\Big(1-2\alpha +\frac{10\alpha\xi}{2-\alpha}-\frac{10(1-\alpha)\tilde{\xi}}{1+\alpha}\Big)\frac{1}{\rho^4} \,.
\end{align}
From~\eq{scalar-J-one-point-fn} we can compute $a_\Sigma$ using~\eq{b-CJ-6d-integral}. The integral generates an undetermined, $\a$-independent constant, $\mf{c}$, that can be fixed by imposing the boundary conditions that $a_\Sigma= 0$ for $\a =0$ with $\tilde\xi =0$, and for $\a=1$ with $\xi=0$. Computing the integral with the appropriate boundary conditions, $\mf{c}(\xi,\tilde\xi)  = \tilde\xi / 15$, then gives 
\begin{align}\label{eq:monodromy-scalar-b}
\bcc(\a,\xi,\tilde\xi) = \frac{\a^2}{720}(1-\a)^2(3+\a-\a^2 ) +\frac{\a^3}{360}(5-3\a^2)\xi -\frac{(1-\a)^3}{360}(3\a^2-6\a-2)\tilde\xi \,.
\end{align}
We plot this $\bcc$ in~\fig{a4-scalar-monodromy}.

With an explicit expression for $\bcc$, we can test the $p=4$ defect $c$-theorem of ref.~\cite{Wang:2021mdq} by studying a simple defect RG flow. Suppose only the $\Delta = 2-|\a|$ mode, $\hat{O}^-_{-\a}$, is present in the defect OPE, i.e. $\tilde\xi =0$. We can then build a defect-relevant quadratic deformation
\begin{align}\label{eq:defect-RG-flow-1}
\lambda \int d^4y \, \hat{O}^-_{-\a} (y)\hat{O}^{\dagger-}_{-\a}(y) \,,
\end{align}
with relevant defect coupling constant $\lambda$. As shown in ref.~\cite{Bianchi:2021snj} for general $d$, starting from any value of $\xi$ in the UV and deforming the monodromy defect by the defect relevant operator in eq.~\eqref{eq:defect-RG-flow-1}, the defect theory flows at fixed $\a$ to an IR fixed point with $\xi=0$. Looking at the behavior of $\bcc$ in~\eq{monodromy-scalar-b} under this flow, we see that $\bcc{}_{,\tiny{\text{UV}}}(\a,\xi,0) \geq \bcc{}_{,\tiny{\text{IR}}}(\a,0,0)$ for all values of $\xi\in [0,1]$, consistent with the $c$-theorem of ref.~\cite{Wang:2021mdq}.

\begin{figure}[tb]
\centering
\includegraphics[width =0.8\textwidth]{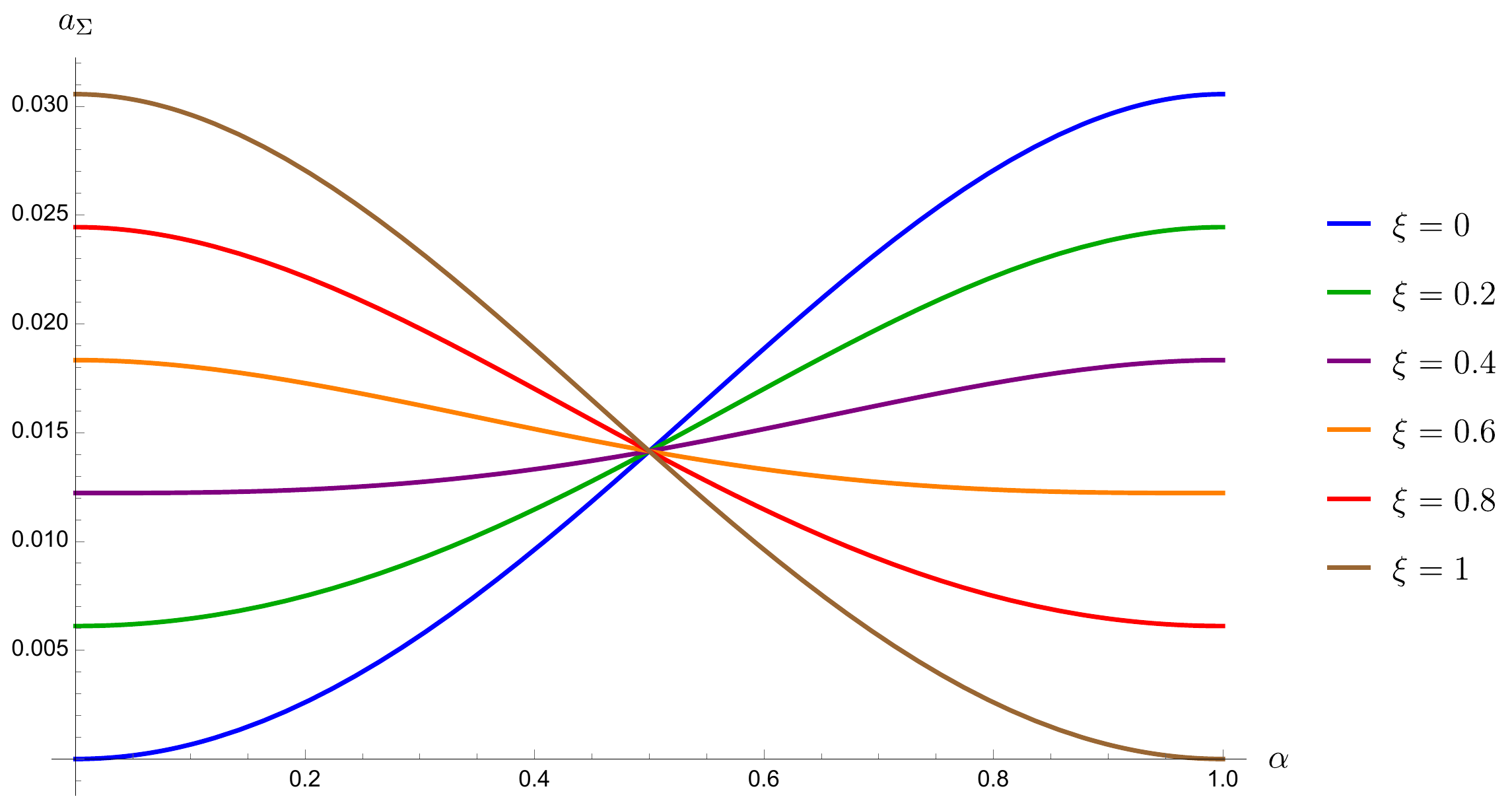} 
\caption{\label{fig:a4-fermion-monodromy} Behavior of $\bcc$ as a function of $\a$ for a monodromy defect in a theory of a single free Dirac fermion in $d=6$ for various values of $\xi$. Note that when $\a=1/2$, $\bcc$ is independent of $\xi$.}
\end{figure}

Repeating the same analysis for the theory of free Dirac fermions in $d=6$ in~\eq{free-fermion-action}, the conserved vector $U(1)$ current is
\begin{align}
J_\mu = i \bar\psi\Gamma_\mu \psi\,.
\end{align}
Using the mode expansion of $\psi$ with arbitrary $\xi$, the non-vanishing components of $J_\mu$ are found to be of the form~\cite{Bianchi:2021snj}
\begin{align}\label{eq:fermion-J-one-point-fn}
\langle J_\theta\rangle = \frac{\alpha(1-\alpha^2)(2-\alpha)(2+\alpha - 5\xi)}{15\pi^3}\frac{1}{\rho^4} \,.
\end{align}
In order to fix the undetermined constant in~\eq{b-CJ-6d-integral}, we fix the boundary conditions such that $\bcc=0$ when $\a = 0$ and $\xi =0$, and when $\a = 1$ and $\xi =1$. Computing the integral with $C_J(\a)$ in~\eq{fermion-J-one-point-fn}, we set $\mf{c}(\xi) = - 11\xi/360$, which gives 
\begin{align}
\label{eq:monodromy-fermion-b}
\bcc(\a,\xi) = \frac{\a^2}{360} (2\a^4 -15\a^2 +24) +\frac{(1-2\a) }{360}(6\a^4 -12 \a^3 -16\a^2+22\a+11)\xi \,.
\end{align}
We plot this $\bcc$ in~\fig{a4-fermion-monodromy}.

The defect RG flows for monodromy defects in a theory of free Dirac fermions are similar to the scalar case.  We again try to construct relevant boundary deformations from the extra operators, either $(\hat{\psi}_{-\alpha})^2$ or $(\hat{\psi}_{1-\alpha})^2$.  These composite operators have scaling dimension
$\Delta = 5-2 \alpha$ and $\Delta = 3 + 2 \alpha$ respectively.  Thus the first is relevant in the range $\alpha > \frac{1}{2}$ and the second in the range
$\alpha < \frac{1}{2}$. If $\a>\frac{1}{2}$, then $(\hat{\psi}_{-\alpha})^2$ triggers a flow from a UV fixed point with arbitrary $\xi$ to an IR fixed point with $\xi=1$, while if $\a<\frac{1}{2}$, then $(\hat{\psi}_{1-\alpha})^2$ triggers an RG flow from a UV fixed point with arbitrary $\xi$ to an IR fixed point with $\xi =0$. Looking at the behavior of $\bcc(\a,\xi)$ in~\eq{monodromy-fermion-b}, we again find that $\bcc{}_{\tiny{,\text{UV}}}(\a,\xi)\geq \bcc{}_{,\tiny{\text{IR}}}(\a,1)$ for $\a\geq \frac{1}{2}$, and $\bcc{}_{,\tiny{\text{UV}}} (\a,\xi)\geq \bcc{}_{,\tiny{\text{IR}}}(\a,0)$ for $\a\leq \frac{1}{2}$. 
 In the limiting case $\a=\frac{1}{2}$, both $(\hat \psi_{+\frac{1}{2}})^2$ and $(\hat \psi_{-\frac{1}{2}})^2$ are exactly marginal deformations, and do not trigger an RG flow. Because A-type defect central charges are independent of defect marginal couplings~\cite{Osborn:1991gm}, we expect $\bcc(\frac{1}{2}, \xi)$ to be independent of $\xi$. Indeed, $\bcc(\frac{1}{2},\xi) = \frac{163}{11520}$ for all $\xi\in[0,1]$, as seen in fig.~\ref{fig:a4-fermion-monodromy}.

As a final note regarding monodromy defects, we can use the computation of $\bcc$ and $\dcc{2}$ to make a prediction for the defect contribution to the universal part of EE of a spherical region centered on the defect. From \eq{defect-EE}, we find that for monodromy defects in free scalar field theories in $d=6$,\footnote{We are grateful to J.~S.~Dowker for alerting us to a typo in an earlier version of the manuscript.}${}^{,}$\footnote{Ref.~\cite{Dowker:2021nju} computes the universal contribution of a monodromy defect to EE, and suggests that their result is in conflict with our \eq{EE_scalar}. However, we do not find any discrepancy between our results. Indeed, we find perfect agreement. The defect EE crucially depends on the relative position of the entangling surface and the defect. Ref.~\cite{Dowker:2021nju} considers a spherical monodromy defect on $\mathds{S}^d$, and computes the defect contribution to the universal part of the EE of a region \textit{coincident} with the defect. This set-up differs from the one in the present paper where the entangling surface \textit{intersects} the defect in an equatorial $\mathds{S}^{d-4}$, as explained above \eq{S_A_gen}. The results of these two computations do not agree in general. Rather, in changing the orientation of the defect relative to the entangling surface from the intersecting case to the coincident one, the contribution from the stress tensor one-point function acquires an overall factor of $(d+1-q)/(1-q) = -(d-1)$ for co-dimension $q=2$. Accounting for this additional factor, one recovers the result of ref.~\cite{Dowker:2021nju} from the one presented here.}
\begin{align}\label{eq:EE_scalar}
S_{A,\Sigma_4}= -\left[\frac{\alpha ^2}{180}  (1-\alpha )^2+\frac{\alpha ^3 \xi }{45}+\frac{\tilde{\xi}}{45} (1-\alpha )^3 \right] \log \left(\frac{L}{\e}\right),
\end{align}
while for a monodromy defect in a theory of free Dirac fermions in $d=6$,
\begin{align}
S_{A,\Sigma_4} = -\left[\frac{\alpha ^2}{90} \left(16-5 \alpha ^2\right) +\frac{\xi}{90} \left(20 \alpha ^3-30 \alpha ^2-12 \alpha +11\right) \right]\log \left(\frac{L}{\e}\right)\,.
\end{align}
For the defect RG flows described above, this contribution to EE is monotonic, that is, the UV values, with $0\leq\xi,\tilde{\xi}$ for scalars and $0\leq\xi\leq 1$ for fermions, are greater than the IR values. However, whether the same is true for more general defects, and in particular whether this is required by some physical principle, remain open questions.

\subsubsection{$n$-fold covers and orbifolds}
\label{sec:orbifolds}

The monodromy defects of the previous subsection are convenient building blocks for constructing other $p=4$ defects in free-field CFTs, namely $n$-fold covers and orbifolds. In this subsection we will use the results of the previous subsection to compute $\bcc$, $d_1$, and $d_2$ for $n$-fold covers and orbifolds in $d=6$ free CFTs. We will start with the scalar field because the boundary conditions are easier to implement. Once we have those results, we will consider the free fermion.

For the scalar field, we can construct an $n$-fold cover of the plane $(x_1, x_2)$ using a vector of scalar fields $\vec \phi = (\phi_1, \cdots , \phi_n)$ with funny boundary conditions along the cut $x_2 = 0$ and $x_1>0$.
The boundary conditions are implemented using the following $n \times n$ shift matrix:
\begin{equation}
T_S = \left(
\begin{array}{cccccc}
0 & 1  & 0 &  \cdots & & 0\\ 
 & 0 & 1 & & \cdots & 0 \\
\vdots  &&&\ddots &  & \vdots  \\
0  & & & \cdots & 0 & 1 \\
1 & 0 &&  \cdots  & 0  & 0 
\end{array}
\right) \ .
\end{equation}
 We impose continuity and smoothness of the normal derivative along the cut $x_2 = 0$ and $x_1>0$: $\vec \phi = T_S \vec \phi$ and $\partial_2 \vec \phi = T_S \partial_2 \vec \phi$. Since the original Lagrangian is quadratic, it is convenient to switch to a basis where $T_S$ is diagonal. In that new basis, we instead have $n$ scalar fields, each of which experiences a monodromy upon going around the branch point, $x_1 = x_2 = 0$: $\tilde \phi_j \to  \exp(\frac{2 \pi i j}{n}) \tilde \phi_j$ for $j = 0, \ldots, n-1$. The $n$ phases $\exp(\frac{2 \pi i j}{n})$ are the eigenvalues of $T_S$. We have chosen the range of $j$ so that the monodromy parameter $\alpha \in [0, 1)$. We also fix $\xi = \tilde \xi = 0$, the physical intuition being that these values correspond to IR fixed points and should be ``more stable''. While our results will be sensitive to this selection of $\alpha$, our choice is consistent with some cross checks we perform below.

As a result of this map from the monodromy defects to the $n$-fold cover, we can express each defect central charge for the $n$-fold cover as a sum over the central charge of the monodromy defects. We immediately have $\dcc{1} = 2 \dcc{2}$ from eq.~\eqref{d1d2rel}, because this is obeyed by the monodromy defects themselves. For $\bcc$ and $d_1$ we perform the requisite sums over $j$ to find
\begin{subequations}
\begin{eqnarray}
\label{scalarb}
\bcc &=& \frac{(n^2-1)(9n^4 + 9n^2+2)}{60480 n^5},\\
& & \nonumber \\
\label{scalard1}
d_1 & = & -\frac{(n^2-1)(31 n^4 + 31 n^2+10)}{15120  n^5}.
\end{eqnarray}
\end{subequations}
Note that $d_1$ is negative for integer $n>1$, as expected for a reflection-positive theory. The final bit of magic is to recognize that the ${\mathbb Z}_m$ orbifold, i.e. $\mathbb{R}^4\times(\mathbb{R}^2/{\mathbb Z}_m)$, can be obtained upon the identification $n = 1/m$. As the orbifold theories have $m>1$, and thus $n<1$, remarkably  $\dcc{2} >0$, in seeming violation of the ANEC. Furthermore, $\dcc{1} > 0$, in violation of reflection positivity for the displacement operator.  

As a cross check, we can compare our result for $m=2$ with an equivalent calculation in ref.~\cite{Herzog:2020bqw} for the orbifold ${\mathbb R}^{p} \times ( {\mathbb R}^q/{\mathbb Z}_2)$, using the method of images. In that case, the one-point function for the stress tensor had\footnote{  We are setting the parameter ${\mathcal N}=1$ in ref.~\cite{Herzog:2020bqw} because we have a complex scalar, but we also need to divide by two to normalize the charge strength using the method of images. We have also included a standard factor of $(d-2) \text{vol}(\mathds{S}^{d-1})$ in the normalization of the scalar two-point function.}
\begin{equation}
h  = - \frac{1}{2^{d} (d-1) \text{vol}(\mathds{S}^{d-1})},
\end{equation}
which for $p=4$ and $q=2$ agrees with eq.~\eqref{scalard1}, and was also noted in ref.~\cite{Herzog:2020bqw} to violate the ANEC. While conical spaces with negative deficit angle appear to be consistent with the ANEC, these orbifold examples with positive deficit angle violate the ANEC, and exhibit a curious violation of reflection positivity for the displacement operator. 
 
Next we consider the fermions, which are more involved because fermions pick up a minus sign under $2\pi$ rotation. The same strategy that worked for the scalars will work here, but we need a more general matrix to impose the boundary conditions (see e.g. ref.~\cite{Herzog:2016ohd})
\begin{equation}
T_F =  \left(
\begin{array}{cccccc}
0 & \omega  & 0 &  \cdots & & 0\\ 
 & 0 & \omega & & \cdots & 0 \\
\vdots  &&&\ddots &  & \vdots  \\
0  & & & \cdots & 0 & \omega \\
\omega & 0 &&  \cdots  & 0  & 0 
\end{array}
\right) \ .
\end{equation}
 We choose $\omega = \exp( \frac{\pi i (n-1)}{n} )$ so that $(T_F)^n = (-1)^{n-1}$. One can think of $(T_F)^n$ as transporting the fermions around the branch point $n$ times, giving rise to a factor of $(-1)^n$.  However, mapping the $n$-fold cover back to the plane introduces an extra factor of $-1$. In this case, we choose a branch cut for the eigenvalues of $T_F$ such that $\alpha \in ( - \frac{1}{2}, \frac{1}{2} )$,
\begin{eqnarray}
\alpha \in \left( - \frac{n-1}{2n}, - \frac{n-3}{2n}, \ldots, \frac{n-1}{2n} \right) \ .
\end{eqnarray}
We also set $\xi = 0$, again because for $\alpha < \frac{1}{2}$, $\xi = 0$ corresponds to the IR fixed point theory. Summing the monodromy results, we find for the $n$-fold cover that
  \begin{eqnarray}
  \bcc  & = & \frac{(n^2-1) (1221 n^4 + 276 n^2 + 31)}{241920 n^5}\,, \\
  & & \nonumber \\
  d_1 & = & -\frac{(n^2-1)(367 n^4 + 178 n^2 + 31)}{12096  n^5}\,.  \label{hfermionorbifold}
  \end{eqnarray}
A ${\mathbb Z}_m$ orbifold can again be obtained under the replacement $n \to 1/m$. Note $d_2$ again has the wrong sign to obey the ANEC for the orbifolded cases, $n<1$, and $\dcc{1}$ will again violate reflection positivity for the displacement operator.
 
For the orbifold, the value of $\dcc{1}$, or equivalently $h$, is straightforward to check using the method of images, at least for odd $m$.  The two point function for the fermion in this case can be written as~\cite{Herzog:2015cxa}
  \begin{equation}
  \langle \psi(x) \psi^\dagger(x') \rangle = - \frac{1}{m} \frac{1}{(d-2) \text{vol}(\mathds{S}^{d-1})} \gamma^\mu \partial_\mu \sum_{k=0}^{m-1} \frac{(-1)^k e^{- \gamma^1 \gamma^2 \frac{\pi k}{m}} }
  {(|z - e^{2 \pi i k/m} z'|^2 + (y-y')^2)^{(d-2)/2}} \, .
  \end{equation}
  Note we have divided the two-point function by a factor of $m$ to normalize the strength of the inserted charges to one. Inserting this expression into a point-split version of the stress tensor, regulating by subtracting the stress tensor for the unorbifolded case, and carefully taking the limit $x' \to x$ yields $h$, and hence via eqs.~\eqref{eq:h-d2-q=2} and~\eqref{d1d2rel} also $\dcc{1}$ in eq.~\eqref{hfermionorbifold}.

The coefficients $\bcc$, $\dcc{1}$, and $\dcc{2}$ for the $n$-fold covers compute universal contributions to the R\'enyi entropies associated with a spatial region bounded by a curved surface, $\Sigma_4$. In the $n \to 1$ limit, via the replica trick, they also compute universal contributions to the EE. Furthermore, in the $n\to 1$ limit \cite{Safdi:2012sn}, these coefficients are related to the central charges of the ambient $d=6$ CFT, which were computed for free fields in ref.~\cite{Bastianelli:2000hi}: in terms of the $d=6$ CFT Weyl anomaly in eq.~\eqref{eq:6d-Weyl-anomaly}, for a free real scalar in $d=6$,
  \begin{equation}
  \label{sixdscalartrace}
 (a^{\text{\tiny{(6d)}}},c_1^{\text{\tiny{(6d)}}},c_2^{\text{\tiny{(6d)}}},c_3^{\text{\tiny{(6d)}}})=\frac{1}{ 7!}\left(\frac{5}{9},- \frac{28}{3},\frac{5}{3},2\right),
  \end{equation}
  while for a free Dirac fermion in $d=6$,
  \begin{equation}
  \label{sixdfermiontrace}
(a^{\text{\tiny{(6d)}}},c_1^{\text{\tiny{(6d)}}},c_2^{\text{\tiny{(6d)}}},c_3^{\text{\tiny{(6d)}}})=\frac{1}{ 7!} \left( \frac{191}{9},- \frac{896}{3},-32,40\right).
  \end{equation}
The Euler density is normalized such that on a unit sphere, $E_6 = 720$. From ref.~\cite{Safdi:2012sn}, we learn that for the $n$-fold cover in the $n \to 1$ limit, $a_\Sigma$, $\dcc{1}$, $\dcc{19}$, $\dcc{20}$, $\dcc{21}$, and $\dcc{22}$ can be expressed as combinations of $a^{\text{\tiny{(6d)}}}$, $c_1^{\text{\tiny{(6d)}}}$, $c_2^{\text{\tiny{(6d)}}}$, and $c_3^{\text{\tiny{(6d)}}}$. Since we have not independently computed $\dcc{19}$, $\dcc{20}$, $\dcc{21}$, and $\dcc{22}$, we focus on the first two, $a_\Sigma$ and $\dcc{1}$, for which we have the predicted relations
  \begin{subequations}
  \label{eq:eerelations}
  \begin{eqnarray}
  \frac{\partial}{\partial n} \bcc |_{n = 1} &=& 3 a^{\text{\tiny{(6d)}}} \ , \\
   \frac{\partial}{\partial n}\dcc{1} |_{n=1} &=& -12c_3^{\text{\tiny{(6d)}}} \ .
  \end{eqnarray}
   \end{subequations}
  These agree with the $E_6$ and $I_3$ coefficients in eqs.~\eqref{sixdscalartrace} and~\eqref{sixdfermiontrace}, remembering to double eq.~\eqref{sixdscalartrace} because the monodromy defects involve complex scalars.

\subsection{Weyl anomalies for probe AdS$_5$ branes in AdS$_{d\geq7}$}
\label{sec:probe-brane}

In this subsection, we will compute the defect central charges for a $p=4$ defect described holographically by a probe brane wrapping an AdS$_5$ submanifold in an ambient AdS$_{d+1}$ background with $d \geq 6$. In particular, using the results obtained by Graham and Reichert in ref.~\cite{Graham:2017bew},\footnote{A similar analysis, extending ref.~\cite{Graham:1999pm} to a co-dimension $q=d-4$ submanifold, was done in ref.~\cite{zhang2017}. See also ref.~\cite{Blitz:2021qbp} and references therein.} we will be able to map a higher-dimensional generalization of the Willmore energy~\cite{Willmore:1965tj}, computed for co-dimension $q=d-4$ branes, to the integrated Weyl anomaly of the holographically dual defect at the boundary of the AdS$_5$ submanifold. We will briefly review the construction in ref.~\cite{Graham:2017bew}, and draw on the analogous computation of holographic central charges in $p=2$ DCFTs following from the Graham-Witten anomaly~\cite{Graham:1999pm}, to read off the defect central charges from~\eq{defect-Weyl-anomaly}.

Following ref.~\cite{Graham:2017bew}, let $(\CX_{d+1}, g_+)$ be a $(d+1)$-dimensional Poincar\'e-Einstein space with metric $g_+$, and with conformal boundary $(\partial \CX_{d+1}, g_+|_\partial )= (\CM_d, g)$, where $\CM_d$ is a Riemannian manifold with metric $g$. In an asymptotic region near $\partial \CX_{d+1}$, we can express $g_+$ in normal form as
\begin{align}
g_+ = \frac{1}{r^2}(dr^2 + g).
\end{align}
Of course, $(\CM_d,g)$ exists as an element of the equivalence class of boundary conformal geometries. In particular, using the defining function $r^2$, and denoting $\hat g = r^2g_+ = dr^2 + g$, the boundary metric at $r=0$ is precisely $g$.  

Let $(\CY_{p+1}, \bar{g}_+)$ be a $(p+1)$-dimensional embedded submanifold of $(\CX_{d+1},g_+)$, transverse to $\CM_d$, with induced metric $\bar{g}_+$ obtained from the pullback of $g_+$ to $\CY_{p+1}$. The boundary submanifold is  $(\partial \CY_{p+1}, \bar{g}_{+}|_{\partial})  = (\Sigma_p, \bar{g})$ where $\CY_{p+1} \cap \CM_{d} = \Sigma_p$ . The boundary metric $\bar{g}$ is induced by the pullback of $\hat g$ (or equivalently the pullback of $\hat{\bar{g}} = r^2\bar{g}_+$), and the local coordinates on $\Sigma_p$ are denoted $y^a$, as in \sn{review}.

Assuming that $\Sigma_p$ is a compact submanifold embedded in $\CM_d$, in ref.~\cite{Graham:2017bew} Graham and Reichert show that the renormalized area of $\Sigma_p$, denoted $A_{\Sigma}$, computed in the collar neighborhood of $\partial \CY_{p+1}$ with cutoff parameter $\e$, is
\begin{align}\label{eq:area-functional}
A_{\Sigma} = \sum_{n=0}^{\frac{p}{2}-1}a_{2n}\,\e^{-p+2n} - \mc{E}_p\log \e + \ldots
\end{align}
as $\e\to0$.  The coefficient of the log divergent term, $\mc{E}_p$, is the Graham-Reichert (or generalized Willmore) energy.  $\CE_p$ encodes the conformally invariant part of the anomaly for even dimensional conformal defects $\Sigma_p$ embedded in $\CM_d$.  

Broadly, $\mc{E}_p$ is computed by first writing the area density, $dS_\CY$, on $\CY_{p+1}$ in the collar neighborhood of the boundary as $dS_\CY = r^{p-1} \phi(y,r) dr dS_\Sigma$, where $dS_\Sigma$ is the area density on $\Sigma_p$ induced by the pullback of the metric $g$ on $\CM_d$.  The invariant function $\phi(y,r)$ is determined by the extension of the induced metric $\bar{g}$ on $\Sigma_p$ into the neighborhood $\Sigma_p\times [0,\e)$, where $\phi$ has an expansion in even powers of $r$.   Computing the expansion of $\phi$ and extracting the $r^{p} \phi^{(p)}$ part, the Graham-Reichert energy is
\begin{align*}
\mc{E} _p= \int_{\Sigma_p} dS_{\Sigma}~\phi^{(p)}(y) .
\end{align*}
By choosing a convenient coordinate system at a point $y^\prime \in \Sigma_p$, namely geodesic normal coordinates at $y^\prime$, and making the assumption that the normal bundle frame is covariantly constant along the $y^a$-directions at $y^\prime$, the authors of ref.~\cite{Graham:2017bew} were able to compute $\phi^{(4)}$ in the case $p=4$. Ultimately, their result for the Graham-Reichert energy is, for compact $\Sigma_4$,
\begin{align}
\begin{split}\label{eq:GR-Energy-1}
\mc{E}_4 = \frac{1}{128}\int_{\Sigma_4} dS_\Sigma \Big(& \overline{D}^a \II^i \overline{D}_a \II_i - \II_i\II^j \II^i_{ab}\II^{ab}_j + \frac{7}{16}(\II_i\II^i)^2-\II^i\II^j W^a{}_{iaj} \\&- 8P^a{}_i \overline{D}_a \II^i - 8 C^a{}_{ai}\II^i - 8P^{ab}\II^i{}_{ab}\II_i + 5P^a{}_a (\II^i\II_i)\\
& -16 P^{ab}P_{ab} +16 P^{ai}P_{ai}+16(P^a{}_a)^2 - \frac{16}{d-4}B^a{}_a \Big).
\end{split}
\end{align}

We can rewrite $\CE_4$ using the basis in~\eq{defect-Weyl-anomaly}.  After a bit of algebra and replacing $d\to q+4$ for convenience, we find that the Graham-Reichert energy takes the form 
\begin{align}\begin{split}\label{eq:GR-Energy-2}
\hspace{-0.5cm}{\cal{E}}_4 = \frac{1}{128}\int_{\Sigma_4} d^4y \Big(&2\bar{E}_4 + 8\CJ_1 + \frac{8}{q}\CJ_2 - 2 W_{abcd}W^{abcd}-2 (W_{ab}{}^{ab})^2 -\frac{16}{3q}W_{aibj}W^{aibj}  \\
&+ \frac{8(3q+14)}{9q}W^b{}_{iab}W_c{}^{iac}+\frac{16}{3q}W_{abjk}W^{abjk}+\frac{4(q+2)}{3q}W_{iabc}W^{iabc} \\
&+\frac{8(3q+2)}{3q}W^c{}_{acb}W_d{}^{adb}+ \frac{16}{3q}W^a{}_{iaj}W_b{}^{ibj} +4W_{ab}{}^{ab}\oII^i_{cd}\oII_i^{cd} \\
& +\frac{8(q-2)}{3q} W^{a}{}_{bij} \oII^{i}_{ac}\oII^{jbc}-\frac{16(q+4)}{3q}W^a{}_{ibj}\oII^i{}_{ac}\oII^{jbc}  \\
&-\frac{16(2q+1)}{3q} W^{abcd}\oII^i_{ac}\oII_{ibd}-\frac{8(7q+6)}{3q}W_a{}^{bac}\oII^i_{bd}\oII_{ic}{}^d\\
&+ \frac{8(q+6)}{3q}W^c{}_{icj}\oII^i_{ab}\oII^{jab}+8\Tr~ \oII^i\oII_i\oII^j\oII_j +4\Tr~\oII^i\oII^j\oII_i\oII_j\\
& - 2 (\Tr~\oII^i\oII_i)^2-4 \Tr~\oII^i \oII^j \Tr~\oII_i \oII_j \Big) .
\end{split}\end{align}

\begin{table}
\hspace{-0.75cm}
\begin{tabular}{|c|c|c|c|c|c|c|c|c|c|c|c|}
\hline
$\bcc$ & $\dcc{1}$ &$\dcc{2}$ &$\dcc{3}$ &$\dcc{4}$ &$\dcc{5}$ &$\dcc{6}$ &$\dcc{7}$ &$\dcc{8}$ &$\dcc{9}$ &$\dcc{10}$ &$\dcc{11}$ \\\hline
$ \frac{1}{4}$ & $-1$ & $ -\frac{1}{q}$ & $ \frac{1}{4}$ & $ \frac{1}{4}$ & $\frac{2}{3q}$& $-\frac{(3q+14)}{9q}$& 0& 0&$-\frac{2}{3q}$ &$ -\frac{(q+2)}{6q}$&$-\frac{(3q+2)}{3q}$ \\\hline\hline
$\dcc{12}$ & $\dcc{13}$ &$\dcc{14}$ &$\dcc{15}$ &$\dcc{16}$ &$\dcc{17}$ &$\dcc{18}$ &$\dcc{19}$ &$\dcc{20}$ &$\dcc{21}$ &$\dcc{22}$ & --\\\hline
$-\frac{2}{3q}$ & $-\frac{1}{2}$ &$\frac{2-q}{3q}$& $ \frac{2(q+4)}{3q}$ & $\frac{2(2q+1)}{3q}$ & $\frac{(7q+6)}{3q}$ & $-\frac{(q+6)}{3q} $  &$-1 $ & $-\frac{1}{2}$ &$ \frac{1}{4}$ & $\frac{1}{2}$&-- \\\hline
\end{tabular}
\caption{Defect central charges of a $p=4$ defect holographically dual to an AdS$_5$ probe brane in AdS$_{d+1}$ (i.e. co-dimension $q = d-4$), in units of $\pi^2L_{\text{\tiny AdS}}^5 T_{\text\tiny br}$.}\label{tab:probe-brane-central-charges}
\end{table}

We want to use this expression to read off the defect Weyl anomaly arising from an AdS$_5$ probe brane in AdS$_{d+1}$, with $d\geq 6$. We define a holographically dual action with an Einstein-Hilbert term, with negative cosmological constant, and a probe brane term,
\begin{align}\label{eq:probe-brane-action}
I = -\frac{1}{16\pi G_N}\int dr d^{d}x \sqrt{g_+} \left(\CR + \frac{d(d-1)}{L_{\text{\tiny AdS}}^2}\right) + T_{\text{\tiny{br}}} \int dr d^{p}y \sqrt{\bar{g}_+}\,,
\end{align}
where $L_{\text{\tiny AdS}}$ is the curvature scale of the bulk AdS$_{d+1}$ geometry, $\mathcal{R}$ is the scalar curvature of the holographic space $\CX_{d+1}$, $T_{\text{\tiny{br}}}$ is the brane tension, and $g_+$ and $\bar{g}_+$ denote the determinants of the bulk AdS$_{d+1}$ metric and induced probe brane metric, respectively.  The holographically renormalized on-shell action of the probe brane takes the asymptotic form in~\eq{area-functional}, with the identification of the log divergent piece as
\begin{equation}\label{eq:GR-energy-trace}
{\cal{E}}_4 = - \int \sqrt{\bar g}\langle\left. T^{\mu}{}_{\mu} \right|_{\Sigma_4}\rangle\,.
\end{equation}
Thus, comparing~\eq{GR-Energy-2} to~\eq{defect-Weyl-anomaly}, we find 20 non-vanishing defect central charges, listed in~\tbl{probe-brane-central-charges}, in units of ($\pi^2 L_{\text{\tiny AdS}}^5T_{\text\tiny{br}}$).

We can check our results for $\bcc$ and $\dcc{2}$ in \tbl{probe-brane-central-charges} by considering the defect contribution to the EE. When the entangling region is a sphere of radius $L$ centered on the defect, the defect contribution to EE has been obtained for generic $p$ and $q$ in ref.~\cite{Kobayashi:2018lil}, using the method described in ref.~\cite{Jensen:2013lxa}, valid for probe branes dual to conformal defects. The same result can be obtained using the method of ref.~\cite{Karch:2014ufa}, based on adapting the generalized gravitational entropy of Lewkowycz and Maldacena~\cite{Lewkowycz:2013nqa} to the case when probe branes are embedded in the bulk spacetime (see ref.~\cite{Chalabi:2020tlw} for a generalization to the case of non-conformal branes). When $p=4$, the result is
\begin{equation}
\label{eq:EE_probe_brane}
 S_{A,\Sigma_4} = -\frac{4 \pi^2}{d-1} L_{\text{\tiny AdS}}^5T_{\text\tiny{br}} \log \left(\frac{L}{\epsilon}\right).
\end{equation}
By plugging the values of $\bcc$ and $\dcc{2}$ listed in \tbl{probe-brane-central-charges} into the general expression \eq{defect-EE} we find perfect agreement. In addition, when $q=1$, or equivalently $d=5$, the universal part of EE reduces to $-4 \bcc$, as expected since $h =0$ in that case.

When $d=6$ we can interpret our probe brane as the minimal-area surface used in Ryu and Takayanagi's holographic prescription for computing EE~\cite{Ryu:2006bv,Ryu:2006ef}. From that point of view, to obtain the universal part of the EE from eq.~\eqref{eq:GR-Energy-2}, we must set $T_{\text{\tiny{br}}}=1$ and divide by $4G_N$. With those changes, our results in~\tbl{probe-brane-central-charges} can be interpreted as holographic results for the twist defect in a $d=6$ CFT holographically dual to Einstein-Hilbert gravity in $AdS_7$, in the limit where the replica index $n\to1$. Our result of course agrees with those in refs.~\cite{Ryu:2006bv,Ryu:2006ef}, for example for the EE of a spherical region. More generally, as mentioned in secs.~\ref{sec:intro},~\ref{sec:review}, and~\ref{sec:orbifolds}, the defect central charges appearing in the universal part of EE should be linear combinations of the $d=6$ CFT central charges, computed holographically from the Einstein-Hilbert action in refs.~\cite{Henningson:1998gx,Henningson:1998ey}. Our result eq.~\eqref{eq:GR-Energy-2} by itself is insufficient to determine those linear combinations, but should provide useful data point for doing so.

\section{Boundary central charges in $d=5$}
\label{sec:bcc}
In this section, we restrict our attention to examples where the $p=4$ defect has co-dimension $q=1$, including cases with a $d=5$ ambient manifold with boundary. The Weyl anomaly for $p=4$ and co-dimension $q=1$ in eq.~\eqref{eq:boundary-Weyl-anomaly} is clearly much simpler than that of a $p=4$ defect of general co-dimension in eq.~\eqref{eq:defect-Weyl-anomaly}, and in particular we will only need to compute, at most, 9 independent boundary central charges. To do so, we will consider two separate examples, each requiring different modes of analysis. First, in subsection~\ref{sec:probe-brane-5d}, we will study a defect holographically dual to an AdS$_5$ probe brane in AdS$_6$. Second, in subsection~\ref{sec:AdS-BCFT}, we will study Takayanagi's AdS/BCFT~\cite{Takayanagi:2011zk,Fujita:2011fp} with $d=5$ where, by solving for the back-reaction, we will be able to compute several of the boundary central charges. Being bottom-up holographic models, both examples describe a $p=4$ defect or boundary in a strongly-coupled, large-$N$ CFT in $d=5$, respectively.

\subsection{Weyl anomalies for AdS$_5$ branes in AdS$_6$}

In this subsection, we reconsider the computation of the Graham-Reichert energy for AdS$_5$ probe branes in~\sn{probe-brane} in the limiting case where $q=1$. Compared to the $q>1$ case, for AdS$_5$ probe branes in AdS$_6$, the number of structures that can appear is greatly reduced.

For an AdS$_5$ probe brane embedded in AdS$_6$,
\begin{align}\begin{split}\label{eq:q=1-GR-energy}
\CE_4 = \frac{1}{128} \int_{\partial\CM_5} \Big(&(\ovl{D}_a K)^2 +(5P^{a}{}_a- K^2) \Tr K^2 + \frac{7}{16} (\Tr K^2)^2 +8P^a{}_n\ovl{D}_a K +8C^a{}_{an}K \\
&+8P^{ab}K_{ab}K+16 (P^{an}P_{an}+(P^a{}_a)^2 -P^{ab}P_{ab}-B^{a}{}_a)  \Big)\,.
\end{split}\hspace{-0.1cm}\end{align}
By using the Gauss-Codazzi-Ricci equations in~\App{relations}, and after partial integration on the defect, we can expand~\eq{q=1-GR-energy} in the same basis as \eq{boundary-Weyl-anomaly}, which gives
\begin{align}\begin{split}
\CE_4 = \frac{1}{128} \int_{\partial\CM_5} \Big( &2\ovl{E}_4 +8\CI_1 -6(\Tr \oK^2)^2 +12 \Tr\oK^4-2 W_{abcd}W^{abcd}-16\oK^{ac}\oK^{bd}W_{abcd}\\&+8W_{anbn}W^{anbn}+8\oK^{ac}\oK^b{}_cW_{anbn}+4 W_{nabc}W^{nabc}\Big).
\end{split}\hspace{-0.1cm}\end{align}
From this expression for $\CE_4$, the relation to the Weyl anomaly~\eq{GR-energy-trace}, and the conventions in \eq{probe-brane-action}, we can extract the $q=1$ defect central charges for the AdS$_5$ probe brane, which are listed in \tbl{boundary-central-charges-probe} in units of $\pi^2 L_{\text{\tiny AdS}}^5 T_{\text{\tiny{br}}}$.  Note that using the relations in~\eq{defect-boundary-central-charge-map}, taking the $q\to 1$ limit of the central charges in~\tbl{probe-brane-central-charges} agree with those found in~\tbl{boundary-central-charges-probe}.

\label{sec:probe-brane-5d}
\begin{table}
\centering
\begin{tabular}{|c|c|c|c|c|c|c|c|c|}
\hline
 $\bcc$ & $\dbcc{1}$ &$\dbcc{2}$ &$\dbcc{3}$ &$\dbcc{4}$ &$\dbcc{5}$ &$\dbcc{6}$ &$\dbcc{7}$ &$\dbcc{8}$ \\\hline
$\frac{1}{4} $ &$-1 $ & $\frac{3}{4}$  & $-\frac{3}{2}$ & $\frac{1}{4}$ & $-1$ & $2$ & $-1$ & $-\frac{1}{2}$\\\hline
\end{tabular}
\caption{ Defect central charges for the AdS$_5$ probe brane in AdS$_6$ in units of $\pi^2 L_{\text{\tiny AdS}}^5T_{\text\tiny{br}}$. }\label{tab:boundary-central-charges-probe}
\end{table}
\subsection{Weyl Anomalies in AdS/BCFT}
\label{sec:AdS-BCFT} 

In this subsection we study the boundary central charges in the AdS/BCFT setup proposed by Takayanagi~\cite{Takayanagi:2011zk,Fujita:2011fp}. This is a bottom-up model where,  in the Euclidean-signature formulation, the holographic space is determined by the action
\begin{equation}
\label{eq:AdSBCFT_action}
S_{\text{grav}} = -\frac{1}{16 \pi G_N} \int_\mathcal{N} \sqrt{\mathfrak{g}} \left(\mathcal{R}-2 \Lambda \right) - \frac{1}{8 \pi G_N} \int_\mathcal{Q} \sqrt{\bar{\mathfrak{g}}} \left( \mathcal{K} - T \right)-  \frac{1}{8 \pi G_N} \int_{\mathcal{M}_d} \sqrt{\bar{\mathfrak{g}}} \, \mathcal{K}\,.
\end{equation}
In the above equation $\mathcal{N}$ denotes the holographic space equipped with the metric $\mathfrak{g}$. Its boundary is $\partial \mathcal{N} = \mathcal{Q} \cup \mathcal{M}_d$, with induced metric denoted by $\bar{\mathfrak{g}}$. $\mathcal{M}_d$ coincides with asymptotic boundary where the BCFT is defined, while $\mathcal{Q}$ is an end-of-the-world brane. By construction we require $\mathcal{Q} \cap \mathcal{M}_d  = \Sigma_{d-1} \ne \emptyset$, which is the boundary of the BCFT. $\CR$ is the curvature on $\CN$, and ${\Lambda = - d(d-1)/(2 L_{\text{\tiny AdS}}^2)}$ is the cosmological constant. The extrinsic curvature of $\partial\CN$ is denoted by $\CK$, and $T$, which we assume to be a constant, arises from matter fields on $\mathcal{Q}$. In this framework, $T$ is related to the boundary condition of the BCFT.

\begin{figure}
	\begin{center}
		\includegraphics[width=\textwidth]{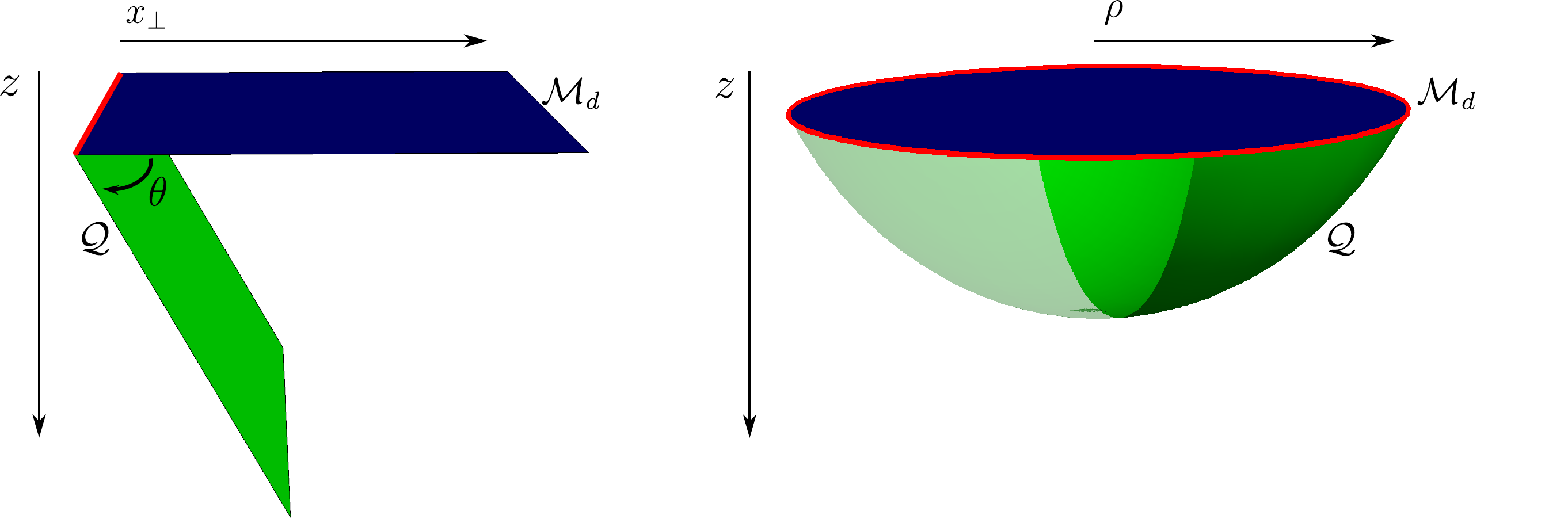}
		\caption{The holographic space dual to the BCFT. The BCFT is defined on the  manifold ${\cal M}_d$, depicted in blue, and its boundary is represented by the red curve. The green surfaces correspond to the end-of-the-world brane $\mathcal{Q}$. Left: BCFT on half space in the vacuum state, with BCFT boundary at $x_\perp =0$ and angle $\theta$ between $\mathcal{Q}$ and ${\cal M}_d$. Right: BCFT on a ball in the vacuum state, with BCFT radial coordinate $\rho$. \label{fig:adsBCFT_fig}}
	\end{center}
\end{figure} 

In the action principle we impose Dirichlet and Neumann boundary conditions on $\mathcal{M}_d$ and $\mathcal{Q}$, respectively. Requiring a stationary action yields the bulk equations of motion
\begin{eqnarray}
 \CR_{MN} - \frac{1}{2}\mathcal{R} \, \mathfrak{g}_{MN}+ \Lambda \mathfrak{g}_{MN} = 0\,, 
\end{eqnarray}
and the boundary condition
\begin{eqnarray}
 \mathcal{K}_{MN}  = \left(\mathcal{K}-T \right)\bar{\mathfrak{g}}_{MN} \qquad \text{only on} \qquad \mathcal{Q} \label{eq:K_bdy_eq}\,,
\end{eqnarray}
where $M, N = 1, \ldots d+1$ are indices on $\CN$.
The simplest solution to these equations is 
\begin{eqnarray}
\label{AdSmetrPoinc}
&& \mathcal{N}: \qquad ds^2 = \frac{L^2_{\text{\tiny AdS}}}{z^2} \left( dz^2 + d \boldsymbol{y}^2 + d x^2_\perp \right), \\
&& \mathcal{Q}: \qquad x_\perp = (\cot \theta) \, z\,, \qquad \theta \in (0,\pi)\,,
\end{eqnarray}
where $z$ is the holographic coordinate, and we set $T =- (d-1)/L_{\text{\tiny AdS}} \cos \theta$. The angle $\theta$ describes how the end-of-the-world brane extends into $\CN$, see the left side of \fig{adsBCFT_fig}. This solution has an $SO(d,1)$ isometry, and corresponds to a BCFT in the vacuum state when the boundary is the $(d-1)$-dimensional hyperplane at $x_\perp=0$, with coordinates ${\boldsymbol{y}} = (y_1,\ldots,y_{d-1})$. We use this solution to compute the A-type boundary central charge in subsection~\ref{sec:ads-bcft-atype}, and some B-type boundary central charges in subsection~\ref{sec:ads-bcft-btype}.

\subsubsection{Boundary A-type Weyl anomaly from AdS/BCFT}
\label{sec:ads-bcft-atype}

The A-type boundary central charge, $\bcc$, can easily be determined by computing the partition function of the theory on the $5$-ball, $\mathds{B}^5$, and singling out the logarithmically divergent contribution. According to the usual AdS/CFT dictionary, this corresponds to the value of the on-shell action of the holographic space. As discussed in refs.~\cite{Takayanagi:2011zk,Fujita:2011fp}, the relevant metric and embedding of $\mathcal{Q}$ can be found by applying a coordinate transformation. This results in the same background metric, but with a new embedding for $\mathcal{Q}$, given by the BCFT radial coordinate $\rho$ as a function of $z$ as (see the right side of~\fig{adsBCFT_fig}),
\begin{equation}
\label{eq:sphericalQ}
\mathcal{Q}: \qquad \rho_{\mathcal{Q}}(z) \equiv \sqrt{\frac{L^2}{\sin^2 \theta}-\left( z + L \cot\theta  \right)^2},  \qquad \theta \in (0,\pi).
\end{equation}
Note that we re-expressed the metric on $\CM_d$ in spherical coordinates, such that its boundary is mapped from $x_\perp = 0$ to $\rho=L$, and $\Sigma_4= \partial \mathds{B}^5$, which we take to be an $\mathds{S}^4$ with round metric. Using~\eq{sphericalQ} and introducing a UV cutoff hypersurface at $z = \e$, the on-shell action~\eq{AdSBCFT_action} can be expressed as
\begin{equation}
\label{eq:S_grav_onshell}
\begin{split}
S_{\text{grav}} & =\frac{8 \pi^2}{3}\frac{L_{\text{\tiny AdS}}^4}{16 \pi G_N}\int_\epsilon^{R \tan \frac{\theta}{2}}  dz \, \left[ 10\int_0^{\rho_{\mathcal{Q}}(z)} d\rho \,  \frac{\rho^4}{z^6}  + 2 \cos\theta \frac{\rho^4_\mathcal{Q}(z)}{z^5}  \sqrt{1+\rho'^2_\mathcal{Q}(z)} \right] \\
& =  \frac{\pi L_{\text{\tiny AdS}}^4}{6  G_N} \left[ - \frac{1+2 \sin^2 \theta}{\sin^2 \theta} \cot \theta   \log \left( \frac{L}{\epsilon} \right)  + \dots \right],
\end{split}
\end{equation}
where in the last step we isolated the logarithmic contribution. It is the only important term for our analysis, as it encodes the boundary Weyl anomaly. From~\eq{S_grav_onshell} we find
\begin{equation}
\int_{{\cal M}_5} \langle T^\mu_{~\mu} \rangle =\frac{\pi L_{\text{\tiny AdS}}^4}{6  G_N}  \frac{1+2 \sin^2 \theta}{\sin^2 \theta} \cot \theta.  
\end{equation}
For $\Sigma_4 = \mathds{S}^4$, $\int_{{\cal M}_5} \langle T^\mu_{~\mu} \rangle = - 4\bcc$, which implies
\begin{equation}
\label{eq:a4AdSBCFT}
\bcc = -\frac{\pi L_{\text{\tiny AdS}}^4}{24  G_N}  \frac{1+2 \sin^2 \theta}{\sin^2 \theta} \cot \theta .
\end{equation}
This result is consistent with the one found in ref.~\cite{Kobayashi:2018lil} using dimensional regularization. Furthermore, as a function of $\theta$, $\bcc$ is monotonically increasing with increasing $\theta$. Ref.~\cite{Fujita:2011fp} found that there are boundary RG flows between different values of $\theta$ which must have $\theta_{\text{\tiny{UV}}}\geq \theta_{\text{\tiny{IR}}}$, assuming that the matter fields localized on $\mathcal{Q}$ obey the null energy condition. Thus, the holographic central charge $\bcc$ in~\eq{a4AdSBCFT} decreases under a boundary RG flow, which is consistent with the boundary $a$-theorem~\cite{Wang:2021mdq}. Finally, we observe that $\bcc$ is not positive definite, and vanishes at $\theta=\pi/2$, similarly to the AdS$_4$/BCFT$_3$ case~\cite{Fujita:2011fp}.

\subsubsection{Boundary B-type Weyl anomalies from AdS/BCFT}
\label{sec:ads-bcft-btype}

We next study the B-type Weyl anomalies in AdS/BCFT. To do so, we need to find the solution to the Einstein equation when the BCFT has a non-trivial background metric. As in sec.~\ref{sec:boundary-stress-tensor}, we need the expansion of the metric of the BCFT found in \eq{bdy_metric}, and we will again assume a flat boundary metric $\bar g_{ab}$ on $\Sigma_4$. In addition, in order to obtain tractable equations of motion, we assume that $K_{ab}, R_{anbn}$, and $\partial_n R_{anbn}$ are constant. 

Our strategy is to solve the equations of motion perturbatively in the dimensionless quantity $x_\perp K_{ab}$ (and $x_\perp^2 R_{anbn}$, $x_\perp^3 \partial_n R_{anbn}$, etc.), following refs.~\cite{Miao:2017aba,Seminara:2017hhh,Miao:2018dvm}.  We use the following ansatz for the bulk metric~\cite{Miao:2017aba}
\begin{equation}
\label{eq:bulkmetric_ansatz}
\begin{split}
 ds^2  = & \frac{L^2_{\text{\tiny AdS}}}{z^2} \left[ dz^2 + \left(1 + \eta^2 x_\perp^2 \mathcal{G}^{(2)}(u)+\eta^3 x_\perp^3 \mathcal{G}^{(3)}(u)  \right)dx_\perp^2 + \right. \\
 & \left. + \left( \delta_{ab} + \eta \, x_\perp \mathcal{F}^{(1)}_{ab}(u) + \eta^2 \, x_\perp^2 \mathcal{F}^{(2)}_{ab}(u) + \eta^3 x_\perp^3 \mathcal{F}^{(3)}_{ab}(u) \right)  d y^a d y^b \right] + \mathcal{O}(\eta^4),
 \end{split}
\end{equation}
where $u\equiv z/x_\perp$, $\eta$ is a dimensionless parameter to keep track of the perturbative order, and the functions $\mathcal{G}^{(i)}(u)$ and $\mathcal{F}^{(i)}(u)$ are to be determined. Furthermore, we take the embedding function for $\mathcal{Q}$ to be of the form 
\begin{equation}
\label{eq:emb_pert}
u_\mathcal{Q}(x) = \tan \theta \,   + \eta B^{(1)} x_\perp + \eta^2 B^{(2)} x_\perp^2 + \eta^3 B^{(3)} x_\perp^3 + \mathcal{O}(\eta^4)\,,
\end{equation}
where the $B^{(i)}$ are constants to be determined. The first order result was found in refs.~\cite{Seminara:2017hhh,Miao:2017aba} for any dimension $d$, while the following orders are novel. Since the solution at higher order is quite cumbersome, we report only the calculation at first order to illustrate the method, while the higher orders appear in the supplementary Mathematica file. 

Requiring that at $z = 0$ the boundary metric reduces to~\eq{bdy_metric} for $u>0$ at first order in $\eta$, one finds the following solution to the Einstein equations 
\begin{equation}
\label{eq:ads-bcft-sol-1}
\mathcal{F}_{ab}^{(1)} (u) = -2 \left( \oK_{ab} f(u) + \frac{1}{4} K \delta_{ab} \right), \quad f(u) \equiv 1 + \frac{C_1}{2} \left(\frac{u}{ 1 + u^2}+ 2 u- 3 \arctan u  \right), \quad u>0,
\end{equation}
with free coefficient $C_1$. This solution, however, does not give a bulk metric that is smooth across $x_\perp = 0$. This is problematic when $\theta \geq \frac{\pi}{2}$. By solving the Einstein equations again in the region $u<0$, and asking for continuity of the metric and its first derivative with respect to $x_\perp$, we find a unique solution, which can be recast as
\begin{equation}
\label{eq:ads-bcft-sol-2}
f(u) = 1 + \frac{C_1}{2} \left(\frac{u}{ 1 + u^2}+ 2 u- 3 \left[\frac{\pi}{2}-2\arctan \left( \frac{1/u}{1+\sqrt{1+1/u^2)}} \right) \right] \right).
\end{equation}   
For $u>0$, the two functions in eqs.~\eqref{eq:ads-bcft-sol-1} and~\eqref{eq:ads-bcft-sol-2} are identical. The smooth solution was found first in ref.~\cite{Seminara:2017hhh} in any $d$, by employing a different coordinate system. However, even though the method of ref.~\cite{Seminara:2017hhh} is more convenient when studying the first order, for the discussion to higher orders we find it easier to work in Poincar\'e coordinates. 

The last step is to employ the boundary condition in~\eq{K_bdy_eq} to fix the remaining free coefficients, $C_1$ and $B^{(1)}$. Plugging the embedding in~\eq{emb_pert} into~\eq{K_bdy_eq}, we find at first order
\begin{equation}
C_1 = \frac{2}{3 \left( \theta - \sin \theta \cos \theta \right)}, \qquad B^{(1)} = \frac{K}{8 \cos^2 \theta}\tan \theta .
\end{equation}
As anticipated above, the explicit results for the functions in~\eq{bulkmetric_ansatz} are quite cumbersome already at first order. For the higher-order expressions we refer to our supplementary Mathematica notebook.

We will now show how to extract the boundary central charges in this setup. Our strategy will be to obtain the one-point function of the stress-tensor holographically, from which we can extract the boundary central charges using the method of sec.~\ref{sec:boundary-stress-tensor}. To this end, we employ the standard AdS/CFT dictionary, i.e. if we have the metric expressed in Fefferman-Graham coordinates,
\begin{equation}
ds^2 = \frac{L_{\text{\tiny AdS}}^2}{z^2} \left( dz^2 +  g_{\mu\nu} (z, x_\perp, \bm{y}) d x^\mu d x^\nu \right),
\end{equation}
where $g_{\mu\nu}$ has the Fefferman-Graham expansion
\begin{equation}
 g =  g_{(0)} +z^2  g_{(2)} + \dots + z^d  g_{(d)} + \dots,
\end{equation}
then the stress tensor one-point function can be extracted as~\cite{deHaro:2000vlm}
\begin{equation}
\label{eq:stress_tensor_generic}
\langle T_{\mu\nu} \rangle =  \frac{d L_{\text{\tiny AdS}}^{d-1}}{16 \pi G_N} \, g_{(d)\mu\nu}  + \ldots,
\end{equation}
where $\ldots$ contains the Weyl anomaly for the ambient CFT on $\CM_d$. When $d=5$, the ambient CFT has no Weyl anomaly, so we only need the first term in~\eq{stress_tensor_generic}. To begin, we focus on the result up to second order in $\eta$. By applying the same procedure discussed for the first order, we find that the form of $\langle T_{\mu\nu}\rangle$ is in perfect agreement with the general one in eqs.~\eqref{eq:T_exp},~\eqref{eq:T_coe1}, and~\eqref{eq:T_coe2}, with coefficients
\begin{equation}
\begin{split}
& A_T = -\frac{1}{3} \frac{4}{(\theta - \sin \theta \cos \theta )} \frac{\pi L_{\text{\tiny AdS}}^4}{G_N}, \\
&\beta_1 = \beta_3 = \frac{2}{3}\frac{1}{\theta - \tan\theta}  \frac{\pi L_{\text{\tiny AdS}}^4}{G_N}, \\
&\beta_2 =-\frac{1}{3} \left( \frac{3}{\theta - \cos \theta \sin \theta} + \frac{1}{\theta-\tan \theta} \right) \frac{\pi L_{\text{\tiny AdS}}^4}{G_N}.
\end{split}
\end{equation}
Up to second order, we directly apply the general result of \eq{T_anom_rel}, obtaining the following linear combinations of boundary central charges,
\begin{subequations}
	\begin{align}\label{eq:b1-constraint}
	& \dbcc{1}  =-\frac{1}{3} \frac{1}{ \theta -\sin \theta  \cos \theta  }\frac{\pi L_{\text{\tiny AdS}}^4}{G_N}, \\\label{eq:b4-b5-constraint}
	&2 \dbcc{4}+\dbcc{5} = \frac{1}{4} \frac{ 1 }{ \tan \theta-\theta } \frac{\pi L_{\text{\tiny AdS}}^4}{G_N}, \\\label{eq:b6-b7-constraint}
	& \dbcc{6} + \dbcc{7} =\frac{1}{24} \frac{13 \sin \theta- 3 \sin (3 \theta ) - 4 \theta \cos \theta}{ (\theta - \sin\theta\cos\theta) \left(\sin\theta-\theta \cos \theta \right)}\frac{\pi L_{\text{\tiny AdS}}^4}{G_N}.
	\end{align}
\end{subequations}

By applying again the method used in sec.~\ref{sec:boundary-stress-tensor} at third order, we find two additional linear combinations,
\begin{equation}\label{eq:b3-b4-b6-constraint}
\frac{\dbcc{3}}{2} - 2 \dbcc{4} + \dbcc{6} = \frac{1}{ 48}  \frac{ 13 - 3 \cos (2 \theta ) - 10 \theta \cot \theta }{ \left( 1 - \theta \cot \theta \right) (\theta - \sin \theta \cos \theta  )} \frac{\pi L_{\text{\tiny AdS}}^4}{G_N},
\end{equation}
\begin{align}\label{eq:b2-b3-constraint}
\begin{split}
\dbcc{2} +\frac{7}{12}\dbcc{3} = & \frac{1}{93312}\frac{ 1}{ (\sin (2 \theta )-2 \theta )^4 (\theta
	\cos \theta -\sin \theta )}\times\\
\times\Big[& 12 \left(7963 + 16996 \theta^2 - 6144 \theta^4\right) \cos \theta - 24 (5023 + 36 \theta^2) \cos (3 \theta )\\
&+8 (3719 + 4644 \theta^2) \cos (5 \theta )-4699  \cos (7 \theta )-57 \cos (9 \theta
)   \\
& + 24 \theta (-13933 + 1488 \theta^2) \sin\theta + 144 \theta (600 \theta^2 -421 ) \sin(3 \theta) \\
& + 21968 \theta \sin(5 \theta) - 11131 \theta \sin(7 \theta) + 
429 \theta \sin(9 \theta))\Big]\frac{\pi L_{\text{\tiny AdS}}^4}{G_N}.
\end{split}
\end{align}
We stress that, even though the solution of the bulk metric has been found perturbatively in the curvature tensors, the expressions of the boundary central charges we obtained are not in a probe limit.

To fix all the boundary central charges uniquely we would need $\mathcal{O}(\eta^4)$. However, due to the increasing complexity of the differential equations involved in the computation, extracting the remaining constraints at fourth order is not as straightforward as in the cases analyzed above, and so we will postpone the study of higher orders to future research.

\begin{figure}
	\begin{center}
		\includegraphics[width=\textwidth]{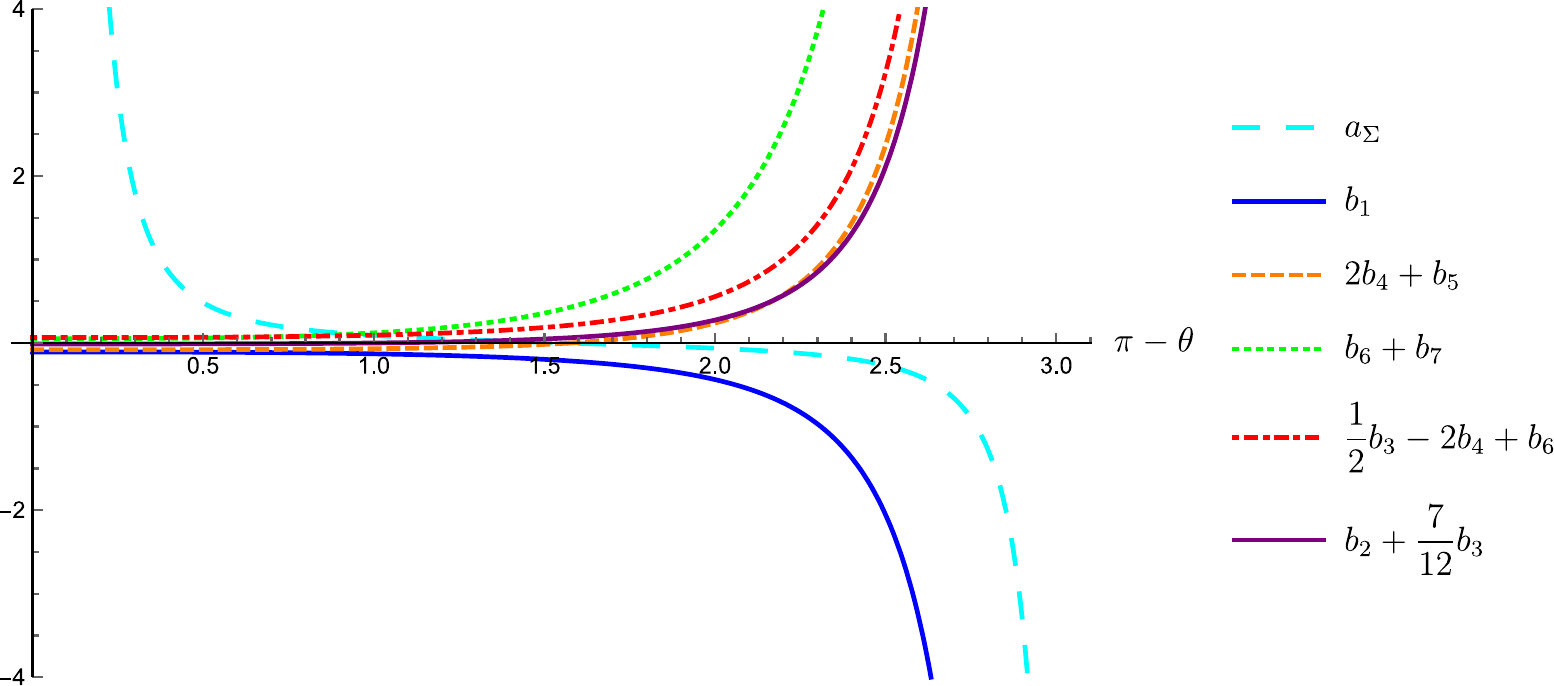}
		\caption{Behavior of the (linear combinations of) boundary central charges computed via AdS/BCFT,  in units of $\pi L^4_{\text{\tiny{AdS}}}/G_N$, as functions of the angle, $\pi - \theta$, at which the end-of-the-world brane intersects the AdS$_6$ boundary.  The UV values of $\theta$ are on the left hand side of the figure, while the IR values are on the right.  Note, though, that at all values of $0\leq\theta\leq\pi$ are conformal. \label{fig:constraints}}
	\end{center}
\end{figure} 

Curiously, as mentioned at the end of sec.~\ref{sec:correlators}, all of the linear combinations of boundary central charges in eqs.~\eqref{eq:b2-b3-constraint} to \eqref{eq:b2-b3-constraint} are invariant under the change to a $Q$-curvature basis for the boundary Weyl anomaly. That is, after using the mapping from defect to boundary central charges in eq.~\eqref{eq:defect-boundary-central-charge-map}, they are invariant under the mapping in eq.~\eqref{eq:Q-curvature-shifts}. We discuss the possible significance of this in sec.~\ref{sec:summary}.

It is revealing to study the behavior of the boundary central charges in eqs.~\eqref{eq:a4AdSBCFT} and~\eqref{eq:b1-constraint}, and the linear combinations in eqs.~\eqref{eq:b4-b5-constraint} to \eqref{eq:b2-b3-constraint}, as we vary the angle at which the end-of-the-world brane intersects the conformal boundary. Indeed, varying $\theta$ from $0$ to $\pi$ changes the conformal boundary conditions, and boundary RG flows obey $\theta_{\text{\tiny{UV}}} \geq \theta_{\text{\tiny{IR}}}$. Plots of our results for the (linear combinations of) boundary central charges as functions of $\pi - \theta$ appear in~\fig{constraints}. 

Several notable features appear along this boundary RG flow. First, we clearly see that $a_{\Sigma,{\text{\tiny{UV}}}} > a_{\Sigma,{\text{\tiny{IR}}}}$, and that $\bcc$ decreases monotonically as a function of $\pi -\theta$, which nicely demonstrates the $c$-theorem for a $p=4$ boundary~\cite{Wang:2021mdq}.  Further, $\bcc$ has a zero at $\theta =\pi/2$. The central charge $b_1$ appears to be always negative, in agreement with the relation \eqref{cDDd1rel} and the bound $c_{\CD \CD} \ge 0$, and a monotonically decreasing function of $\pi-\theta$.   In addition, we can clearly see that the combination $2 \dbcc{4} + \dbcc{5}$ is a monotonically increasing function of $\pi -\theta$ on the interval $[0,\pi]$, with a zero at $\theta = \pi/2$.  Finally, the linear combinations in eqs.~\eqref{eq:b6-b7-constraint},~\eqref{eq:b3-b4-b6-constraint}, and~\eqref{eq:b2-b3-constraint} are also monotonically increasing with increasing $\pi -\theta$, but do not have a zero on $\theta\in [0,\pi]$.

\section{Summary and Outlook}\label{sec:summary}

In this paper, we determined the most general form of the Weyl anomaly for a conformal defect of dimension $p=4$ in an arbitrary CFT of dimension $d\geq6$. To do so, we used a standard algorithm, finding a complete basis of conformal invariants localized to the submanifold supporting the conformal defect, and then eliminating terms by imposing WZ consistency and introducing local counterterms in the effective action. Our main result for the Weyl anomaly of a $p=4$ conformal defect with co-dimension $q \geq 1$ is eq.~\eqref{eq:defect-Weyl-anomaly}, with 23 parity-even terms and 6 parity-odd terms, plus the additional parity-odd terms in eqs.~\eqref{eq:parity-odd-q=2} and~\eqref{eq:parity-odd-q=4} for $q=2$ and $q=4$, respectively. Among the 23 parity-even terms, one is A-type while all the others are B-type. The parity-odd terms are all B-type. Each of these terms comes with a coefficient that defines a defect central charge.

For $p=4$ conformal defects with $q=1$, our result reduces to eq.~\eqref{eq:boundary-Weyl-anomaly}, which reproduces the 9 parity-even terms first obtained by Astaneh and Solodukhin in ref.~\cite{FarajiAstaneh:2021foi}. This served as a non-trivial check of our results. Moreover, beyond the parity-even anomalies, we found 3 parity-odd terms that were previously unknown.

We subsequently showed in sec.~\ref{sec:correlators} how some of the defect central charges appear in physical observables (besides the Weyl anomaly itself), namely the two-point function of the displacement operator, the one-point function of the stress tensor, and the universal contribution to EE of a spherical region centered on a flat defect. In sec.~\ref{sec:dcc} we computed many of the defect central charges in examples, including monodromy and conical defects in $d=6$ free field CFTs, and defects holographically dual to probe branes extended along an AdS$_5$ submanifold inside AdS$_{d\geq 7}$. This last example provided an important check on our main result, eq.~\eqref{eq:defect-Weyl-anomaly}, namely agreement with the Graham-Reichert energy of a probe brane~\cite{Graham:2017bew}. Correspondingly, we computed all $23$ parity-even defect central charges only for the probe brane. In sec.~\ref{sec:bcc} we computed the parity-even boundary central charges in two holographic systems, namely defects holographically dual to AdS$_5$ probe branes inside AdS$_6$ where we were able to find all the $9$ coefficients, and Takayanagi's AdS/BCFT with $d=5$ where we fixed $6$ of them.

Our results raise a host of questions, and suggest many directions for future research. We will discuss only some of these here.

Many questions remain about how defect/boundary central charges appear in physical observables. For example, in the calculation of EE in a $d=6$ CFT for a region with arbitrary shape, the defect contribution to the universal part will have the form of the Weyl anomaly in eq.~\eqref{eq:defect-Weyl-anomaly}, but with defect central charges fixed by the ambient CFT's central charges in eq.~\eqref{eq:6d-Weyl-anomaly}. We partially determined two of these EE defect central charges in eq.~\eqref{eq:eerelations}, however what about the rest? We saw that the central charges can help determine correlation functions of the stress tensor and displacement operator in flat space through their effect on contact terms.  Counter-terms in the action can also affect these contact terms, and an important open question is how to separate the effect of the counter terms from the effect of the central charges, especially in a curved space context where even defining the one point function of the stress tensor can become problematic. Moreover, how do defect/boundary central charges appear in higher-point functions of the displacement operator, mixed correlators like $\langle T \CD\CD\rangle$, not to mention thermal entropy, heat capacity, conductivities, and so on? Answering these questions could be especially enlightening for the many parity-odd defect/boundary central charges we found, which remain particularly mysterious. More generally, answering these questions could allow us to compute all the remaining defect/boundary central charges in our examples, and in many other important examples, including free-field CFTs, like monodromy and conical defects of a $d=6$ free, complex, self-dual $3$-form, as well as interacting CFTs, such as $p=4$ defects in $d=6$ SCFTs.

If we can determine how defect/boundary central charges enter physical observables, then we can ask whether any general principles provide bounds on them. In eq.~\eqref{eq:CD-d1-2} we found that the normalization of the displacement operator's two-point function was $\propto -d_1$ for a defect and $\propto -b_1$ for a boundary, hence reflection positivity requires $d_1 \leq 0$ or $b_1\leq0$, respectively. In eq.~\eqref{eq:h_d_2_rel} we found that, for $q\geq 2$, the normalization of the stress-tensor's one-point function was $\propto -d_2$, so that if the ANEC is valid in the presence of a defect, then $d_2 \leq 0$. Since the A-type central charge $\bcc$ obeys a $c$-theorem for boundary/defect RG flows~\cite{Wang:2021mdq}, and hence counts defect/boundary degrees of freedom, we can ask whether it is bounded from below. This bound cannot be zero, since $\bcc<0$ in explicit (reflection-positive) examples, including a free scalar BCFT in $d=5$ with Dirichlet boundary conditions (see~\cite{FarajiAstaneh:2021foi}), and AdS/BCFT (see~\fig{constraints}). Perhaps defect/boundary central charges are bounded by other defect/boundary central charges, in a similar fashion to the Hofman-Maldacena bounds on $a_\CM^{\text{\tiny(4d)}}/c^{\text{\tiny(4d)}}$~\cite{Hofman:2008ar,Hofman:2016awc}.

Do any of the defect and boundary central charges obey $c$-theorems, for either defect/boundary or ambient RG flows? To date, for $p=4$ a $c$-theorem has been proven only for the A-type central charge, $\bcc$, for a defect/boundary RG flow~\cite{Wang:2021mdq}. Several of our examples provided non-trivial tests of this $c$-theorem, including the $d=6$ free field monodromy defects in~\sn{monodromy} and AdS/BCFT with $d=5$ in~\sn{AdS-BCFT}. However, what about the B-type defect/boundary central charges? Nothing {\emph{a priori}} forbids them from also obeying $c$-theorems. The known examples provide useful data points. For example, for a free scalar BCFT in any $d$, a boundary mass term triggers a boundary RG flow from a Robin boundary condition in the UV to Dirichlet in the IR. The $d=5$ results in ref.~\cite{FarajiAstaneh:2021foi} show that under this RG flow $b_1$ is invariant, while $b_2$, $b_4$, $b_5$, $b_6$, and $b_8$ decrease, and $b_3$ and $b_7$ increase. The AdS/BCFT results in~\fig{constraints} show that $b_1$ is a monotonically decreasing function of $\pi-\theta$, and hence will monotonically decrease under a boundary RG flow from $\theta_{\text{\tiny{UV}}}$ to $\theta_{\text{\tiny{IR}}}\leq \theta_{\text{\tiny{UV}}}$. As a result, $b_1$ could possibly obey a $c$-theorem $b_{1\text{\tiny{UV}}}\geq b_{1\text{\tiny{IR}}}$, while $b_3$ could obey $b_{3\text{\tiny{UV}}}\leq b_{3\text{\tiny{IR}}}$, and similarly for $b_7$. Clearly, exploring more examples, and developing methods to prove $c$-theorems for defect/boundary central charges, remain important open questions.

Our holographic examples in~\sn{probe-brane},~\ref{sec:probe-brane-5d}, and~\ref{sec:AdS-BCFT} raise their own special questions. For example, in AdS/BCFT we computed the BCFT's stress-tensor one-point function to third order in perturbations about a flat ambient CFT geometry, which provided us with $\bcc$ and $\dbcc{1}$ exactly as functions of the angle at which the end-of-the-world brane intersects the conformal boundary of AdS$_6$. However, this provided only five equations for the remaining six parity-even boundary central charges. Moreover, all of these linear combinations are invariant under the change to a $Q$-curvature basis for the boundary Weyl anomaly, that is, upon using the mapping from defect to boundary central charges in eq.~\eqref{eq:defect-boundary-central-charge-map}, they are invariant under the mapping to the $Q$-curvature basis in eq.~\eqref{eq:Q-curvature-shifts}. In AdS$_5$/CFT$_4$, holography also naturally seems to provide the Weyl anomaly in the $Q$-curvature basis~\cite{Henningson:1998gx}. Do our results reveal some deeper principle or pattern at work in holography? It would be interesting to see if we pursue our strategy beyond third order in perturbations whether again the linear combinations of boundary central charges that we obtain are invariant under the shift to the Q-curvature basis. We could also explore other AdS/BCFT models, such as those of refs.~\cite{FarajiAstaneh:2017ghr,Miao:2017gyt,Chu:2017aab,Miao:2018qkc}, or the supergravity constructions of $p=4$ defects in refs.~\cite{Lin:2004nb,DHoker:2008wvd}, which, being top-down, could provide a path beyond large $N$ and strong coupling.

Adding SUSY also raises a host of special questions. For example, in refs.~\cite{Wang:2020xkc,Wang:2021mdq}, Wang showed for superconformal defects with $p=2$ and $p=4$ that $\bcc$ is fixed by certain 't Hooft anomaly coefficients, and obeys a form of $c$- or $a$-extremization~\cite{Intriligator:2003jj,Benini:2012cz,Benini:2013cda}, respectively. Are other defect/boundary central charges fixed by 't Hooft anomaly coefficients? Are defect B-type central charges also extremized or maxmized along RG flows to IR SCFTs? In ref. \cite{Chalabi:2020iie} the contribution of a superconformal defect to the SUSY Casimir energy of an SCFT on $S^1 \times \mathbb{R}^{d-1}$ was conjectured to be a (linear combination of) defect central charges. This conjecture was supported by several examples of p = 2 and p = 4 superconformal defects. However, whether this conjecture is true, and if so how to prove it, remain open questions.

As mentioned in sec.~\ref{sec:intro}, one of our main motivations was to study the $p=4$ superconformal defect in the $d=6$ $\CN=(2,0)$ SCFT, arising from the $1/2$-BPS intersection of M5-branes. Our main result for a $p=4$ defect Weyl anomaly in eq.~\eqref{eq:defect-Weyl-anomaly} provides a starting point for constructing the full defect super-Weyl anomaly, which should be crucial for characterizing these important defects. More generally, fully characterizing both co-dimension $2$ and $4$ defects in $d=6$ SCFTs through their defect central charges will be crucial not only to exploring M- and F-theory, but also exploring how, through partially twisted dimensional reduction, these data determine lower dimensional superconformal defects in class-$\mc{S}$ theories in $d=4$ \cite{Alday:2009aq, Alday:2009fs} and class-$\mc{R}$ theories in $d=3$ \cite{Dimofte:2011ju}.

Thinking more broadly, aside from the defect/boundary Weyl anomalies we reviewed in sec.~\ref{sec:review}, and our novel results for $p=4$ defects/boundaries in sec.~\ref{sec:anomalies}, what other defect/boundary Weyl anomalies are possible? Two obvious cases have yet to be studied. The first is $p=3$ in $d>4$ (example appear for instance in refs.~\cite{Gaiotto:2014ina,Gutperle:2020rty}). In this case, our preliminary analysis suggests that a $p=3$ defect in $d>4$ has parity-odd Weyl anomalies, whose form depends on the co-dimension, similar to what we found for $p=4$ in $d\geq 5$. The second is $p=5$ in $d\geq 6$. Crucially, when $d=6$ these must be non-SUSY, since the $d=5$ superconformal algebra does not embed in any $d=6$ superconformal algebra.

In short, our results open up many new possibilities for characterizing and classifying defects, boundaries, and CFTs with various $p$ and $d$. We intend to explore many of these possibilities in future research.

\section*{Acknowledgements}
The authors would like to thank J.~S.~Dowker, N.~Drukker, J.~Gauntlett, K.~Jensen, M.~Lemos, M.~Meineri, A.~Waldron and T.~Wiseman for useful discussions
and correspondence.  A.C. is funded by the Royal Society award RGF/EA/180098.  C.P.H. is funded by a Wolfson Fellowship from the Royal Society
  and by the U.K.\ Science \& Technology Facilities Council Grant ST/P000258/1. A.O'B. is a Royal Society University Research Fellow. B.R. is funded by in part by the KU Leuven C1 grant ZKD1118 C16/16/005.  J.S. is funded in part by the Royal Society award RGF/EA/181020, and in part by Knut and Alice Wallenberg Foundation under grant KAW 2016.0129 and by VR grant 2018-04438.
\hspace{0.51cm}

\appendix
 \section{4d defect Weyl anomaly basis}\label{app:basis_terms}

In this appendix, we present some details of the algorithm of sec.~\ref{sec:algorithm} applied to the case of a $p=4$ conformal defect. For the complete workings we refer to our supplemental Mathematica notebook.

\subsection{Arbitrary co-dimension} \label{app:basis-any-q}

We begin with the part of the anomaly that is admissible in any co-dimension $q\geq 2$. The $q=1$ case is a simple adaptation of the $q\geq 2$ case.

\paragraph{Step 1:} First, we determine the basis of terms. There are 36 scalars that transform non-trivially under Weyl transformations, and that do not have any derivatives acting on $\delta\omega$. They are
\begin{align}
&\CB_1=R^2\,,
&& \CB_2=R_{ab}R^{ab}\,,
&&\CB_3=(N^{ij}R_{ij})^2\,,\nonumber\\
& \CB_4=W_{ab}{}^{ab} N^{ij}R_{ij}\,,
&&\CB_5=R W_{ab}{}^{ab}\,,
&&\CB_6=R N^{ij}R_{ij}\,,\nonumber\\
&\CB_7=R_{ab}W_c{}^{acb}\,,
&&\CB_8=R_{ij}W^{iaj}{}_a\,,
&&\CB_9=R_{ij}R^{ij}\,,\nonumber\\
&\CB_{10}=R\II^i\II_i\,,
&&\CB_{11}=W_{ab}{}^{ab}\II^i\II_i\,,
&&\CB_{12}=N^{jk}R_{jk}\II^i\II_i\,,\nonumber\\
&\CB_{13}=R\oII^i_{ab}\oII^{ab}_i\,,\nonumber
&&\CB_{14}=N^{jk}R_{jk}\oII^i_{ab}\oII^{ab}_i\,,
&&\CB_{15}=R_{ij} \II^i\II^j\,,\nonumber\\
&\CB_{16}=R_{ij}\oII^i_{ab}\oII^{abj}\,,
&&\CB_{17}=R^{ab}\oII^i_{ab}\II_i\,,
&&\CB_{18}=R_a{}^b\oII^i_b{}^c\oII_{ic}{}^a\,,\nonumber\\
&\CB_{19}=W^a{}_i{}^b{}_j \oII^{(i}_{ab}\II^{j)}\,,
&&\CB_{20}=W^{acb}{}_c\oII^i_{ab}\II_i\,,
&&\CB_{21}=W_{iaj}{}^a \II^i\II^j\,,\\
&\CB_{22}=W_{ic}{}^{ac} \oD^b\oII^i_{ab}\,,
&&\CB_{23}=W_{ic}{}^{ac}\oD_a \II^i\,,
&&\CB_{24}=\II^i D_i R\,,\nonumber\\
&\CB_{25}=\oII^{abi}D_i W_{ a c b}{}^c\,,
&&\CB_{26}=\II^i D_i W_{ab}{}^{ab}\,,
&& \CB_{27}=D^i D_i R\,,\nonumber\\
&\CB_{28}=D^i D_i W_{ab}{}^{ab}\,,
&& \CB_{29}=\II^i \Tr \oII_i\oII^j\oII_j\,,
&& \CB_{30}=\II^i \II_i \Tr \oII^j\oII_j\,,\nonumber\\
& \CB_{31}=\II^i \II^j \Tr \oII_i\oII_j\,,
&& \CB_{32}=(\II^i\II_i)^2\,,
&& \CB_{33}=\oD_a \oII^{ab}_i\oD_b \II^i \,,\nonumber\\
&  \CB_{34}=\oD_a \II_i \oD^a \II^i\,,
&& \CB_{35}=\oD_a \oII^i_{bc} \oD^a \oII^{bc}_i\,,
&& \CB_{36}=\oD^b \oII^i_{ba} \oD^c \oII^i_c{}^a\,,\nonumber
\end{align}
where $\oD_a \oII^{ab}_i\oD_b \II^i= N_{\mu\nu}\oD_a \oII^{ab\mu}\oD_b \II^\nu$, and similarly for the other terms of the form $(\oD\II)^2$. There are 24 trivial Weyl invariants without derivatives on $\delta\omega$:
\begin{align}
&\tilde{\CB}_1=W_{abcd} W^{abcd}\,,
&&\tilde{\CB}_2=(W_{ab}{}^{ab})^2\,,
&&\tilde{\CB}_3=W_{abcd} W^{ab}{}_{ef} \e^{cdef}\,,\nonumber\\
&\tilde{\CB}_4=W_{ijcd} W^{ij}{}_{ef} \e^{cdef}\,,
&& \tilde{\CB}_5=W_{iajb}W^{iajb}\,,
&&\tilde{\CB}_6=W_{abi}{}^bW^{aci}{}_c\,,\nonumber\\
&\tilde{\CB}_7=W_{ijk\ell}W^{ijk\ell}\,,
&& \tilde{\CB}_{8}=W_{aijk}W^{aijk}\,,
&& \tilde{\CB}_{9}=W_{abjk}W^{abjk}\,,\nonumber\\
&\tilde{\CB}_{10}=W_{acb}{}^cW^{adb}{}_d\,,
&&\tilde{\CB}_{11}=W_{iaj}{}^aW^{ibj}{}_b\,,
&&\tilde{\CB}_{12}=W_{ab}{}^{ab}\oII^i_{cd}\oII^{cd}_i\,,\nonumber\\
&\tilde{\CB}_{13}=W_a{}^b{}_{ij} \oII^i_b{}^c\oII^j_c{}^a\,,
&&\tilde{\CB}_{14}=W_{aibj}\oII^{ibc}\oII^{j}_c{}^a\,,
&&\tilde{\CB}_{15}=W^{abcd}\oII^i_{ac}\oII_{bdi}\,,\\
&\tilde{\CB}_{16}=W_{ab}{}^a{}_d \oII^b{}_f{}^i \oII^{df}_i\,,
&&\tilde{\CB}_{17}=W_{icj}{}^c\oII^i_{ab}\oII^{abj}\,,
&&\tilde{\CB}_{18}=W^{abcd}\oII_a{}^{ei}\oII_b{}^f{}_i \epsilon_{cdef}\,,\nonumber\\
& \tilde{\CB}_{19}=\Tr \oII^i\oII_i\oII^j\oII_j \,,
&& \tilde{\CB}_{20}=\Tr \oII^i\oII^j\oII_i\oII_j \,,
&& \tilde{\CB}_{21}=\Tr \oII^i\oII_i \,  \Tr \oII^j\oII_j\,,\nonumber\\
& \tilde{\CB}_{22}=\Tr \oII^i\oII^j \,  \Tr \oII_i\oII_j\,,
&& \tilde{\CB}_{23}=\oII^i_a{}^e\oII^j_{be}\oII_c{}^f{}_i\oII_{dfj}\epsilon^{abcd}\,,
&&\tilde{\CB}_{24}=W_{ij cd}\oII^b{}_{e}^{i}\oII_{bf}^j \epsilon^{cdef}\,.\nonumber
\end{align}
Finally, there are 41 terms with derivatives acting on $\delta\omega$
\begin{align}\label{eq:D4d}
&\CD_{1}=W_{ic}{}^{ac}\oII^i_{ab}\oD^b\delta\omega\,,
&&\CD_{2}=W_{ic}{}^{ac} \II^i\oD_a\delta\omega\,,
&&\CD_{3}=\oII^i_{ab}R^{ab}D_i\delta\omega\,,\nonumber\\
&\CD_4=\II^i R D_i\delta\omega\,,
&&\CD_5=\II^i N^{jk} R_{jk}D_i\delta\omega\,,
&&\CD_6=\oII^{abi} W_{ a c b}{}^c D_i\delta\omega\,,\nonumber\\
&\CD_7=\oII^{abi} W_{i a j b}D^j\delta\omega\,,
&&\CD_8=\II^i W_{ab}{}^{ab} D_i \delta\omega\,,
&&\CD_{9}=\II^i R_i{}^j D_j\delta\omega\,,\nonumber\\
&\CD_{10}=\II^{i} W_{i a j }{}^aD^j\delta\omega\,,
&&\CD_{11}=R\oB \delta\omega\,,
&& \CD_{12}=N^{ij}R_{ij}\oB \delta\omega\,,\nonumber\\
& \CD_{13}=W_{a c b}{}^c\oD^a \oD^b\delta\omega\,,
&&\CD_{14}=D^i R D_i\delta\omega\,
&&\CD_{15}=R D^i D_i\delta\omega\,,\nonumber\\
&\CD_{16}=N^{ij}R_{ij} D^k D_k\delta\omega\,,
&&\CD_{17}=W_{ab}{}^{ab} \oB\delta\omega\,,
&&\CD_{18}=W_{ab}{}^{ab} D^i D_i\delta\omega\,,\nonumber\\
&\CD_{19}=D^i  W_{ab}{}^{ab}D_i\delta\omega\,,
&&\CD_{20}=W_{ aci}{}^c  \oD^a D^i\delta\omega\,,
&&\CD_{21}=W_{icj}{}^c D^i D^j \delta\omega\,,\nonumber\\
&\CD_{22}=R^{ij} D_i D_j\delta\omega\,,
&& \CD_{23}=\Tr \oII^i\oII_i\oII^j D_j\delta\omega\,,
&& \CD_{24}=\II^j \Tr \oII^i\oII_i D_j\delta\omega\,,\\
& \CD_{25}=\II^i \Tr \oII_i\oII^j D_j\delta\omega\,,
&& \CD_{26}=\II^i\II_i\II^j D_j\delta\omega\,.
&& \CD_{27}=\oII^{ab}_i \oD_a \II^i \oD_b \delta\omega\,,\nonumber\\
& \CD_{28}=\oII^{ab}_i \II^i \oD_a \oD_b\delta\omega\,, 
&& \CD_{29}=\II_i \oD_a \II^i \oD^a \delta\omega\,,
&& \CD_{30}=\oII^i_{ab} \oII^{ab}_i\oB \delta\omega\,,\nonumber\\
&\CD_{31}=\oII^{ab}_i\oD^c \oII^i_{cb}\oD_a\delta\omega\,,
&&\CD_{32}=\oII_i^{ac}\oII_{c}{}^{bi}\oD_a\oD_b\delta\omega\,,
&&\CD_{33}=\Tr \oII^i\oII^j D_i D_j\delta\omega\,,\nonumber\\
&\CD_{34}=\Tr \oII^i\oII_i \, D^j D_j\delta\omega\,,
&&\CD_{35}=\II^i\II^j  D_i D_j\delta\omega\,,
&&\CD_{36}=\II^i\II_i D^j D_j\delta\omega\,,\nonumber\\
&\CD_{37}=\II^i D_i D^j D_j\delta\omega\,,
&& \CD_{38}=\II^i \oB D_i  \delta\omega\,,
&& \CD_{39}=\oII^i_{ab} \oD^a\oD^b D_i \delta\omega \,,\nonumber\\
&\CD_{40}=(D_i D^i)^2\delta\omega\,,
&&\CD_{41}=\oII^i_{bf}\oD_c \oII_{d}{}^f{}_i\epsilon^{abcd} \oD_a\delta\omega\,.\nonumber
\end{align}
In total, our basis is 101-dimensional. For $q=1$ the basis is over-complete because some terms that are distinct in $q\geq 2$ may reduce to the same term in $q=1$. Moreover some terms vanish identically when $q=1$.

\paragraph{Step 2:} Next, we find solutions to WZ consistency. Computing a second Weyl variation produces terms of the form $\mathcal{D}_i^{WZ}= \delta\omega_2 \mathcal{D}_i^\partial \delta\omega_1 - (1 \leftrightarrow 2)$ for $i=1, \ldots 41$, where $\mathcal{D}_i^\partial$ corresponds to the operator version of $\CD_i$ in \eq{D4d} with the variation parameter $\delta\omega$ omitted. In addition, there are the following terms
\begin{align}
&\CD^{WZ}_{42}=W_{a c i}{}^c\oD^a\dw D^i \dww-(1\leftrightarrow 2)\,, 
&&\CD^{WZ}_{43}=\II^i D_i\dw \oB \dww-(1\leftrightarrow 2)\,,\nonumber\\
&\CD^{WZ}_{44}=\II^i \oD_a D_i\dw \oD^a\dww-(1\leftrightarrow 2)\,,
&&\CD^{WZ}_{45}=\oII^i_{ab} D_i\dw \oD^a\oD^b \dww-(1\leftrightarrow 2)\,,\nonumber\\
&\CD^{WZ}_{46}=\oII^i_{ab} \oD^a D_i\dw \oD^b \dww-(1\leftrightarrow 2)\,,
&&\CD^{WZ}_{47}=\II^i D_i\dw D^j D_j \dww-(1\leftrightarrow 2)\,,\\
&\CD^{WZ}_{48}=\II^i D_j  D_i\dw D^j \dww-(1\leftrightarrow 2)\,,
&&\CD^{WZ}_{49}=\oB \dw D^i D_i \dww-(1\leftrightarrow 2)\,,\nonumber\\
&\CD^{WZ}_{50}=D^i D_i  \dw D^j D_j  \dww-(1\leftrightarrow 2)\,,
&&\CD^{WZ}_{51}=D^i\dw D_i D^j D_j \dww-(1\leftrightarrow 2)\,.\nonumber
\end{align}

The 24 $\tilde\CB$'s trivially solve WZ consistency. In addition, one also finds four non-trivial linear combinations of $\CB$'s. Three of them correspond to $\ovl E_4$, $\CJ_1$, $\CJ_2$. The final one can be written as a linear combination of other invariants and $W_{iabc}W^{iabc}$. In \eq{defect-Weyl-anomaly}, we redefine our basis to include $W_{iabc}W^{iabc}$ instead of this extra conformal invariant. 

\paragraph{Step 3: } We also find 33 linear combinations of $\CD$'s that naively solve WZ consistency. However, 32 of them can be rendered WZ inconsistent by a choice of scheme. One can introduce counterterms that remove at least one of the constituent terms of the linear combination such that the remainder is inconsistent. The anomaly, however, must be WZ consistent in any scheme. Thus, we must insist that the overall coefficients of these 32 linear combinations of $\CD$'s must be zero in any scheme. Therefore, they are absent in the anomaly. The one remaining linear combination of $\CD$'s, which just consists of the single term $\CD_{41}$ in \eq{D4d}, is unaffected by local counterterms. It is related to the term in \eq{defect-Weyl-anomaly} whose coefficient is $\tdcc{3}$. As we comment in the main body, it is a genuine anomaly coefficient. 

This leaves us with 23 parity-even contributions to the trace anomaly. One of them is A-type and the remaining 22 are B-type in the classification of ref.~\cite{Deser:1993yx}. There are 6 terms that are parity-odd along the defect that are admissible in any co-dimension $q$, although 3 of them vanish identically when $q=1$.

\subsection{Parity-odd terms in the normal bundle when $q=2$}\label{app:basis-q=2}

In addition to the terms that are parity-odd along the defect, there are terms that break parity in the normal bundle. First, consider the co-dimension $q=2$ case.

\paragraph{Step 1:} The extra parity-odd terms in the normal bundle are
\begin{align}
&\CB^{(6)}_1 = R_i{}^j n_{jk} \II^i \II^k\,,
&&\CB^{(6)}_2 = R_i{}^j n_{jk} \oII_{ab}^i \oII^{abk}\,,
&&\CB^{(6)}_3 = R^{ab}\II^i\oII^j_{ab} n_{ij}\,,\nonumber\\
&\CB^{(6)}_4 = W_{aibj}\oII^{abi}\II_k n^{jk}\,,
&&\CB^{(6)}_5 = W_{aibj}\oII^{ab}_k\II^i n^{jk}\,,
&&\CB^{(6)}_{6}=W_{ic}{}^{ac} n^{ij}\oD^b\oII_{abj}\,,\nonumber\\
&\CB^{(6)}_{7}=W_{ic}{}^{ac}n^{ij}\oD_a \II_j\,,
&&\CB^{(6)}_{8}=\II_i n^{ij}D_j R\,,
&&\CB^{(6)}_{9}=\oII^{abi}n_{ij}D^j W_{ a c b}{}^c\,,\\
&\CB^{(6)}_{10}=\II^i n^{kj}D_j R_{ki}\,,
&&\CB^{(6)}_{11}=\II^i n_{ij} \Tr\oII^k \oII_k \oII^j\,,
&&\CB^{(6)}_{12}=\II^i\II^j n_{jk} \Tr \oII^k \oII_i\,,\nonumber\\
&\CB^{(6)}_{13}=n_{ij} \oD^a \II^i \oD^b \oII_{ab}^j\,,\nonumber
\end{align}
and
\begin{align}
&\tilde{\CB}^{(6)}_1 = W_{ajbk}\oII^{ac}_i\oII_c{}^{bj}n^{ik}\,,
\end{align}
as well as
\begin{align}
&\CD^{(6)}_{1}=W_{ic}{}^{ac}\oII_{abj}n^{ij}\oD^b\,,
&&\CD^{(6)}_{2}=W_{ic}{}^{ac} \II_jn^{ij}\oD_a\,,
&&\CD^{(6)}_{3}=\oII_{abj}R^{ab}n^{ij}D_i\,,\nonumber\\
&\CD^{(6)}_4=\II_i n^{ij} R D_j\,,
&&\CD^{(6)}_5=\oII^{ab}_k n^{ik} W_{i a j b}D^j\,,
&&\CD^{(6)}_6=\oII^{abi} W_{i a j b}  n^{jk}D_k\,,\nonumber\\
&\CD^{(6)}_{7}=\II_k n^{ki} W_{i a j }{}^aD^j\,,
&&\CD^{(6)}_{8}=\II^i R_{ik}n^{jk} D_j\,,
&&\CD^{(6)}_{9}=n^{ki}\II_k R_i{}^j D_j\,,\nonumber\\
&\CD^{(6)}_{10}=n^{ij} D_i R D_j\,,
&&\CD^{(6)}_{11}=W_{ aci}{}^c n^{ij} \oD^a D_j\,,
&&\CD^{(6)}_{12}=R^{ij} n_{jk} D^k D_i\,,\nonumber\\
&\CD^{(6)}_{13} = n^{ij} D_j R_{ia} \oD^a\,,
&&\CD^{(6)}_{14}=n^{jk}D_k R_{ij}D^i\,,
&&\CD^{(6)}_{15}=\Tr \oII^i\oII_i \oII^j n_{jk}D^k\,,\\
& \CD^{(6)}_{16}=n_{jk}\II^j\Tr \oII^i\oII_i D^k\,,
&&\CD^{(6)}_{17}=n_{jk}\II^i\Tr \oII_i\oII^j D^k\,,
&&\CD^{(6)}_{18}=n_{jk}\II^i\II_i\II^j D^k\,,\nonumber\\
&\CD^{(6)}_{19}=n_{ij} \oD^a \II^i \oII_{ab}^j \oD^b \,,
&&\CD^{(6)}_{20}=n_{ij}  \II^i \oII_{ab}^j \oD^a\oD^b \,,
&&\CD^{(6)}_{21}=n_{ij} \II^i \oD_a \II^j \oD^a\,,\nonumber\\
&\CD^{(6)}_{22}=n_{ij} \oII^i_{ab}\oD_c \oII^{cbj} \oD^a\,,
&&\CD^{(6)}_{23}=\Tr \oII^i \oII^j n_{jk} D^k D_i\,,
&&\CD^{(6)}_{24}=\II^i\II^j n_{jk} D^k D_i\,,\nonumber\\
&\CD^{(6)}_{25}=n_{ik}\II^i D^k D^j D_j\,,
&& \CD^{(6)}_{26}=n_{ik}\II^i \oB D^k  \,,
&& \CD^{(6)}_{27}=n_{ik}\oII^i_{ab} \oD^a\oD^b D^k  \,.\nonumber
\end{align}
Note that the identity $n_{ij}n^{k\ell}=N_i^k N_j^\ell-N_j^k N_i^\ell$ is very restrictive in combination with various symmetry properties of tensors, and it sets to zero many candidate terms.

\paragraph{Step 2:} In addition to $\mathcal{D}_i^{(6),WZ}= \delta\omega_2 \mathcal{D}_i^{(6),\partial} \delta\omega_1 - (1 \leftrightarrow 2)$ for $i =1,\ldots, 27$, the following terms are generated when computing a second Weyl variation:
\begin{align}
&\CD^{(6),WZ}_{28}=W_{aibj}n^{ij}\oD^a \dw \oD^b \dww-(1\leftrightarrow 2)\,,
&&\CD^{(6),WZ}_{29}=W_{ai}{}^{aj}n_{kj} D^k\dw D^i\dww-(1\leftrightarrow 2)\,,\nonumber\\
&\CD^{(6),WZ}_{30}=W_{ac}{}^{aj}n_{ij} D^i\dw \oD^c\dww-(1\leftrightarrow 2)\,,
&&\CD^{(6),WZ}_{31}=R n_{ij} D^i\dw D^j\dww-(1\leftrightarrow 2)\,,\nonumber\\
&\CD^{(6),WZ}_{32}=R_{ij}n^{kj}D^i\dw D_k\dww-(1\leftrightarrow 2)\,,
&&\CD^{(6),WZ}_{33}=\oII_{cd}^j\oII^{cdk}n_{ij}D_k\dw D^i \dww-(1\leftrightarrow 2)\,,\nonumber\\
&\CD^{(6),WZ}_{34}=\II^i\II^j n_{jk} D_i\dw D^k \dww-(1\leftrightarrow 2)\,,
&& \CD^{(6),WZ}_{35}=\oII_a{}^{cj} \oII_{cb}^i n_{ij} \oD^a\dww \oD^b \dw-(1\leftrightarrow 2)\,,\nonumber\\
&\CD^{(6),WZ}_{36}=\oII_{ab}^j n_{ij} \oD^b D^i\dw\oD^a \dww-(1\leftrightarrow 2)\,,
&&\CD^{(6),WZ}_{37}=\oII^{cd i}n_{ki} D^k\dww \oD_d\oD_c\dw-(1\leftrightarrow 2)\,,\nonumber\\
& \CD^{(6),WZ}_{38}=\II^i n_{ij}D^j \dww D^kD_k \dw-(1\leftrightarrow 2)\,,
&& \CD^{(6),WZ}_{39}=\II^i n_{ij}D^j \dww \oB \dw-(1\leftrightarrow 2)\,,\nonumber\\
&\CD^{(6),WZ}_{40}=\II^i n^{jk}D_k D_i \dw D_j \dww-(1\leftrightarrow 2)\,,
&& \CD^{(6),WZ}_{41}=\II^i n_{ij} \oD^a D^j \dww \oD_a\dw-(1\leftrightarrow 2)\,,\nonumber\\
&\CD^{(6),WZ}_{42}=n^{ij} D_j D^k D_k \dw D_i \dww-(1\leftrightarrow 2)\,,
&& \CD^{(6),WZ}_{43}=n_{ij} D^i\dww \oB D^j \dw-(1\leftrightarrow 2)\,,\nonumber\\
&\CD^{(6),WZ}_{44}=n^{ij}D_i D_k \dw D_j D^k \dww-(1\leftrightarrow 2)\,.
\end{align}
There are no solutions of WZ consistency involving $\CB$'s. Only $\tilde{\CB}^{(6)}_1$ and linear combinations of $\CD$'s (naively) solve WZ consistency. 

\paragraph{Step 3:} All of the linear combinations of $\CD$'s are scheme-dependent. There is only one genuine scheme-independent solution to WZ consistency that is parity-odd in the normal bundle: the trivial Weyl invariant $\tilde{\CB}^{(6)}_1$ which appears in \eq{parity-odd-q=2}.

\subsection{Parity-odd terms in the normal bundle when $q=4$}\label{app:basis-q=4}

The co-dimension $q=4$ case is more restrictive. The only new terms that one can write down are trivial Weyl invariants:
\begin{align}
&\tilde{\CB}^{(8)}_1=\e^{abcd}n^{ijk\ell} W_{abij}W_{cdk\ell}\,,
&&\tilde{\CB}^{(8)}_2=n^{ijk\ell}W_{aibj}W^a{}_k{}^b{}_\ell\,,\nonumber\\
&\tilde{\CB}^{(8)}_3=n^{ijk\ell}W_{piqj}W^p{}_k{}^q{}_\ell\,,
&&\tilde{\CB}^{(8)}_4=n^{ijk\ell}W_{aipj}W^a{}_k{}^p{}_\ell\,,\\
&\tilde{\CB}^{(8)}_5=n_{ijk\ell}\e_{abcd} W^{abij} \oII_f{}^{ck} \oII^{fd\ell}\,,
&&\tilde{\CB}^{(8)}_{6}=n_{ijk\ell}W^{abij} \oII_{ac}^{k} \oII_b{}^{c\ell}\,.\nonumber
\end{align}
These trivially solve WZ consistency and cannot be removed by counterterms. They appear in \eq{parity-odd-q=4}.

\section{Isolating the anomaly: $p=2$ conformal defect in a $d=4$ CFT}\label{app:WZ-example}

In this appendix we illustrate the algorithm presented in sec.~\ref{sec:algorithm} with a simple example: the case of a $p=2$ defect in a $d=4$ ambient CFT. We will reproduce the known result of refs.~\cite{Berenstein:1998ij,Graham:1999pm,Henningson:1999xi,Gustavsson:2003hn,Asnin:2008ak,Schwimmer:2008yh,Cvitan:2015ana,Jensen:2018rxu}, stated in eq.~\eqref{eq:2d-defect-Weyl-anomaly}. The structure of this section follows our supplemental Mathematica notebook, and we use notation that mirrors the notation in the notebook. Readers interested in reproducing our results for $p=4$ defects may wish to consult this appendix first, as a warm-up exercise, before diving into our notebook.

\paragraph{Step 1:} We begin by finding a basis of terms for the defect Weyl anomaly. The terms need to be scalars built out of the metric $g_{\mu\nu}$, the pullback $e_a^\mu$, the Weyl variation parameter $\delta\omega$, and two derivatives. There are 10 of them. 

It is convenient to separate the terms into three categories. The first one involves scalars with a non-trivial Weyl transformation. There are three such terms:
\begin{align}\label{eq:B2d-4d}
\mathcal{B}_1 &= \ovl R\, , & \mathcal{B}_2 &= R\, , & \mathcal{B}_3 &=  \II^i \II_i \,.
\end{align}
The second category involves scalars that are trivially Weyl invariant. There are four linearly-independent such terms:
\begin{align}\label{eq:tB2d-4d}
\tilde{\mathcal{B}}_1 &= \mathring{\II}^i_{ab}\mathring{\II}_i^{ab}\, ,&\tilde{\mathcal{B}}_2 &=  \ovl g^{ac} \ovl g^{bd} W_{abcd} \, , & \tilde{\mathcal{B}}_3 &= n_{ij} \epsilon^{ab}\oII^i_{ac} \oII^{jc}_{b} \, , &\tilde{\mathcal{B}}_4 &=n^{ij} \epsilon^{ab}  W_{iajb} \,.
\end{align}
Finally, we list terms with derivatives acting on the variation parameter. There are three of them:
\begin{align}\label{eq:D2d-4d}
\mathcal{D}_1 &= \II^i D_i \delta\omega\, , & \mathcal{D}_2 &=  N^{ij} D_i D_j \delta\omega\, , & \mathcal{D}_3 = n_{ij} \II^i D^j \delta\omega\, . 
\end{align}
In general, these will all appear in the anomaly as
\begin{equation}
\delta\CW^{(1)} = \int_{\Sigma_2} \sqrt{\ovl g}\left( \sum^3_{i=1}\mathfrak{b}_i \CB_i \delta\omega+ \sum^4_{i=1}\tilde{\mathfrak{b}}_i \tilde{\CB}_i \delta\omega+ \sum^3_{i=1}\mathfrak{d}_i \CD_i\right).
\end{equation}

A few comments are in order. In constructing this basis, and making sure that it is not over-complete, we have used a number of geometric relations. Since the Weyl tensor $W_{\mu\nu\rho\sigma}$ is a linear combination of Riemann tensor, Ricci tensor and Ricci scalar, we are free to disregard any terms built out of $R_{\mu\nu\rho\sigma}$. Similarly, we have traded in a $\II^\mu_{ab}$ for its traceless part $\oII^\mu_{ab}$ and its trace $\II^\mu$. Moreover, any term containing two copies of either $\epsilon^{ab}$ or $n^{ij}$ can be re-written without any epsilon symbols using $\epsilon_{ab}\epsilon^{cd}=\delta_a^c\delta_b^d-\delta_b^c\delta_a^d$, and similarly $n_{ij}n^{k\ell}=N_i^k N_j^\ell-N_j^k N_i^\ell$. Further, we only consider quantities that do not need to be extended into the bulk. An example is $N^{ij}D_i \II_j$, which crucially depends on how $\II_j$ is extended into the ambient geometry. Since there is no canonical way to do so, we do not consider such terms here. In general, we don't expect such terms be part of the physical Weyl anomaly. In the present 2d case, WZ consistency eliminates them. One may have also expected the following terms:
\begin{align}
&\mathcal{C}_1=\ovl g^{ab}R_{ab}\,,&&\mathcal{C}_2=N^{ij}R_{ij}\,, &&\mathcal{C}_3=N^{ik}N^{j\ell}W_{ijk\ell}\,, &&\mathcal{C}_4=\ovl g^{ab}N^{ij}W_{aibj}\,,\nonumber\\
& \mathcal{C}_5=R^\perp_{ijab}\epsilon^{ij}\epsilon^{ab}\,, && \mathcal{C}_6 = E^a_\mu\ovl D_a \II_{}^\mu\,, && \mathcal{C}_7 = E^b_\mu\ovl D^a \oII_{ab}^\mu\,,
\end{align}
where $R^\perp$ is the normal bundle curvature, i.e. the curvature associated with the connection induced from $D$ that maps normal vectors to normal vectors by parallelly transporting them along the submanifold. However, these are not linearly independent. Firstly, $\mathcal{C}_1 = \mathcal{B}_2 - \mathcal{C}_2$. We can use this relation to eliminate $\mathcal{C}_1$. Now, $\mathcal{C}_2$ appears in the twice contracted Gauss \eq{gauss-2}. Trading the Riemann tensor for the Weyl tensor, it reads
\begin{equation}
3 \mathcal{C}_2 =-\mathcal{B}_1+ \mathcal{B}_2 - \tilde{\mathcal{B}}_1 +\tilde{\mathcal{B}}_2+ \frac{3}{4} \tilde{\mathcal{B}}_3  \,.
\end{equation}
Thus we can use it to eliminate $\mathcal{C}_2$. Furthermore, the fact that the trace of any two indices of $W_{\mu\nu\rho\sigma}$ with the ambient metric $g_{\mu\nu}$ vanishes implies that $\mathcal{C}_3=-\mathcal{C}_4 =\tilde{\mathcal{B}}_2$. $\mathcal{C}_5$ can be removed by the Ricci identity \eq{ricci}. Finally, $\mathcal{C}_6 = - \mathcal{B}_3$ using $E^a_\mu \II^\mu =0$ and the definition of $\II$, and similarly $\mathcal{C}_7=-\tilde{\CB}_1$. This leaves us with the terms in eqs.~\eqref{eq:B2d-4d},~\eqref{eq:tB2d-4d} and~\eqref{eq:D2d-4d}.

\paragraph{Step 2:} WZ consistency reduces the number of terms to 7. To compute $\left[\delta_1, \delta_2\right]\CW^{(1)}$ we take a second Weyl variation of the above basis. We anti-symmetrise in the variation parameters, $\delta\omega_1$ and $\delta\omega_2$, which generates linear combinations of the following terms
\begin{align}
\mathcal{D}^{WZ}_1 &= \delta\omega_2\II^i D_i \delta\omega_1 - (1 \leftrightarrow 2)\, , & \mathcal{D}^{WZ}_2 &= \delta\omega_2N^{ij} D_i D_j \delta\omega_1 - (1 \leftrightarrow 2)\,,\nonumber\\
 \mathcal{D}^{WZ}_3 &= \delta\omega_2n_{ij} \II^i D^j\delta\omega_1- (1 \leftrightarrow 2)\,,&\mathcal{D}^{WZ}_4 &=  n^{ij} D_j\delta\omega_2  D_i\delta\omega_1- (1 \leftrightarrow 2)\, . 
\end{align}
The second Weyl variations are
\begin{align}
(\sqrt{\ovl g})^{-1}\delta_1 (\sqrt{\ovl g}\,\CB_1\delta\omega_2) - (1 \leftrightarrow 2) &= 0,\nonumber\\
(\sqrt{\ovl g})^{-1}\delta_1 (\sqrt{\ovl g}\,\CB_2\delta\omega_2) - (1 \leftrightarrow 2) &=4\,\left(  \mathcal{D}^{WZ}_1- \mathcal{D}^{WZ}_2\right)\,,\nonumber\\
(\sqrt{\ovl g})^{-1}\delta_1 (\sqrt{\ovl g}\,\CB_3\delta\omega_2) - (1 \leftrightarrow 2) &=-4 \mathcal{D}^{WZ}_1\,,\nonumber\\
(\sqrt{\ovl g})^{-1}\delta_1 (\sqrt{\ovl g}\,\tilde{\CB}_1\delta\omega_2) - (1 \leftrightarrow 2)&=0\,,\nonumber\\
(\sqrt{\ovl g})^{-1}\delta_1 (\sqrt{\ovl g}\,\tilde{\CB}_2\delta\omega_2) - (1 \leftrightarrow 2)&=0\,,\nonumber\\
(\sqrt{\ovl g})^{-1}\delta_1 (\sqrt{\ovl g}\,\tilde{\CB}_3\delta\omega_2) - (1 \leftrightarrow 2)&=0\,,\\
(\sqrt{\ovl g})^{-1}\delta_1 (\sqrt{\ovl g}\,\tilde{\CB}_4\delta\omega_2) - (1 \leftrightarrow 2)&=0\,,\nonumber\\
(\sqrt{\gamma})^{-1}\delta_1 (\sqrt{\ovl g}\,\CD^{\partial}_{1}\delta\omega_2) - (1 \leftrightarrow 2)&=0\,,\nonumber\\
(\sqrt{\ovl g})^{-1}\delta_1 (\sqrt{\ovl g}\,\CD^{\partial}_{2}\delta\omega_2) - (1 \leftrightarrow 2)&=0\,,\nonumber\\
(\sqrt{\ovl g})^{-1}\delta_1 (\sqrt{\ovl g}\,\CD^{\partial}_{3}\delta\omega_2) - (1 \leftrightarrow 2)&=-4\mathcal{D}^{WZ}_4\,,\nonumber
\end{align}
where $(1\leftrightarrow 2)$ only exchanges the subscripts on the Weyl variation parameters, and $\CD_i^\partial$ denotes the operator version of $\CD_i$ in eq.~\eqref{eq:D2d-4d} with the variation parameter $\delta\omega$ omitted. In the above we have dropped total derivatives along the submanifold $\ovl D (\ldots)$ since the above terms appear in $\left[\delta_1, \delta_2\right]\CW$ under an integral.

Solving WZ consistency reduces to a simple problem in linear algebra. Let $(\underline{\mathcal{B}}^{WZ} )^T= (\CB_1, \ldots, \CB_3, \tilde{\CB}_1, \ldots, \tilde{\CB}_4, \CD^\partial_1, \ldots, \CD^\partial_3)$ and $(\underline{\mathcal{D}}^{WZ} )^T= (\mathcal{D}^{WZ}_1, \ldots, \mathcal{D}^{WZ}_4)$. Let $M^{WZ}$ be the $10\times 4$ matrix that implements the transformation,
\begin{equation}
\left(M^{WZ}\right)^T= 
\begin{pmatrix*}[r]
 0 & 4 & -4 & ~~0 & ~~0 & ~~0 & ~~0 & ~~0 & ~~0 & 0 \\
0 & -4 & 0 & 0 & 0 & 0 & 0 & 0 & 0 & 0 \\
0 & 0 & 0 & 0 & 0 & 0 & 0 & 0 & 0 & 0 \\
0 & 0 & 0 & 0 & 0 & 0 & 0 & 0 & 0 & -4
\end{pmatrix*}\,,
\end{equation}
i.e. $\mathscr{S}^2 \underline{\mathcal{B}}^{WZ} =M^{WZ} \underline{\mathcal{D}}^{WZ} $, where the operator $\mathscr{S}^2$ acts like $\mathscr{S}^2 A = (\sqrt{\ovl g})^{-1}\delta_1 (\sqrt{\ovl g}\,A\delta\omega_2) - (1 \leftrightarrow 2)$ for some $A$. The (right) null space of $(M^{WZ})^T$ is the general solution to the condition $\left[\delta_1, \delta_2\right]\CW^{(1)}=0$. In this example, the solutions are particularly simple: they just correspond to $\CB_1\delta\omega$, $\tilde{\CB}_1\delta\omega$, $\tilde{\CB}_2\delta\omega$, $\tilde{\CB}_3\delta\omega$, $\tilde{\CB}_4\delta\omega$, $\CD_1$ and $\CD_2$, each added with an arbitrary coefficient in the anomaly polynomial, i.e.
\begin{equation}\label{eq:2d-4dWZ}
\delta\CW^{(2)} = \int_{\Sigma_2} \sqrt{\ovl g}\left( \mathfrak{b}_1 \CB_1 \delta\omega+ \sum^4_{i=1}\tilde{\mathfrak{b}}_i \tilde{\CB}_i \delta\omega+ \sum^2_{i=1}\mathfrak{d}_i \CD_i\right).
\end{equation}
The other coefficients must vanish, i.e. $\mathfrak{b}_2=\mathfrak{b}_3=\mathfrak{d}_3=0$. 

More generally, the solutions to WZ consistency may involve a linear combination of terms with fixed relative coefficients, i.e. WZ consistency forces the coefficients of some terms to be determined by one another up to a single overall number. Each such linear combination would appear in $\delta\CW$ with this unfixed coefficient. For example, for a $p=4$ defect, WZ consistency fixes the relative coefficients appearing in eqs.~\eqref{eq:J1} and~\eqref{eq:J2} up to a single overall coefficient for each of them, $\dcc{1}$ and $\dcc{2}$, respectively. 

One sometimes also finds linear combinations of $\CD$'s at this stage. Indeed, this is the case for a $p=4$ defect. As we explained in app.~\ref{app:basis_terms}, however, they are typically not genuine solutions of WZ consistency, but can be made inconsistent by addition of local counterterms.

\paragraph{Step 3:} Finally, we introduce local counterterms in $\CW$. By adjusting their coefficients we can set to zero some of the remaining coefficients in \eq{2d-4dWZ}. We find 5 scheme-independent terms in the anomaly which cannot be removed by such counterterms.

Let $\CW_{CT}$ be the counterterm action. We cannot add any of the $\CD$'s as counterterms because $\CW_{CT}$ does not involve the variation parameter $\delta\omega$. In principle, we could introduce the $\tilde{\CB}$'s. However, they are Weyl invariant and, therefore, cannot remove any terms from the anomaly. We are thus left with the $\CB$'s. The counterterm action reads
\begin{equation}
\CW_{CT} = \int_{\Sigma_2} \sqrt{\ovl g} \left(\sum^3_{i=1} \mathfrak{c}_i \CB_i \right).
\end{equation}
The first Weyl variation of these terms is
\begin{align}
(\sqrt{\ovl g})^{-1}\delta\left(\sqrt{\ovl g}\,\mathcal{B}_1\right) &= 0\, , \nonumber\\
(\sqrt{\ovl g})^{-1}\delta\left(\sqrt{\ovl g}\,\mathcal{B}_2\right) &=4\,(  \mathcal{D}_1- \mathcal{D}_2)\delta\omega\, , \\
 (\sqrt{\ovl g})^{-1}\delta\left(\sqrt{\ovl g}\,\mathcal{B}_3\right) &=-4 \CD_1 \delta\omega\, ,\nonumber
\end{align}
where we have again dropped total derivatives $\ovl D (\ldots)$.

Determining the terms in the anomaly that cannot be removed by local counterterms reduces again to a linear algebra problem. Let $\underline{\CB} = (\CB_1, \ldots, \CB_3)^T$ and $\underline{\CD}=(\CD_1,\ldots,\CD_3)^T$. We introduce the $3\times 3$ matrix $M$,
\begin{equation}
M = 
\begin{pmatrix*}[r]
0 & 4 & -4  \\
0 & -4 & 0  \\
0 & 0 & 0  \\
\end{pmatrix*}\,,
\end{equation}
which implements the first Weyl variation with appropriate factors of $\sqrt{\ovl g}$, i.e. $\mathscr{S}\underline{\CB} = M \underline{\CD} $, where $\mathscr{S}$ acts like $\mathscr{S} A=(\sqrt{\ovl g})^{-1}\delta\left(\sqrt{\ovl g}\,A\right)$. The terms that cannot be removed by local counterterms are given by the (right) null space of $M$. Generally, a null vector is a linear combination of $\CD$'s, and one must choose a scheme in which one of the terms in that linear combination cannot be set to zero.

In the present case, however, the null space is just the span of $\CD_3$. Therefore, all but $\CD_3$ can be removed unambiguously by adjusting the values of the coefficients $\mathfrak{c}_i$. In particular, 
\begin{equation}
\delta\CW=\delta\CW^{(2)}+\delta\CW_{CT}
\end{equation}
with arbitrary $\mathfrak{c}_1$, $\mathfrak{c}_2 = \frac{1}{4}\mathfrak{d}_2$, and $\mathfrak{c}_3=\frac{1}{4}(\mathfrak{d}_1+\mathfrak{d}_2)$ sets the coefficients of $\CD_1$ and $\CD_2$ to zero. Since WZ consistency requires that the coefficient of $\CD_3$ vanishes, the scheme-independent part of the anomaly is therefore
\begin{equation}
\delta\CW=\int_{\Sigma_2} \sqrt{\ovl g}\left(\mathfrak{b}_1 \CB_1 + \tilde{\mathfrak{b}}_1\tilde{\CB}_1+ \tilde{\mathfrak{b}}_2\tilde{\CB}_2+ \tilde{\mathfrak{b}}_3\tilde{\CB}_3+ \tilde{\mathfrak{b}}_4\tilde{\CB}_4 \right)\delta\omega\,.
\end{equation}
After appropriately relabelling the coefficients, we find~\eq{2d-defect-Weyl-anomaly} with $q=2$.

\section{Gauss, Codazzi, Ricci, and differential relations}\label{app:relations}

In this appendix, we collect various useful geometric relations for embedded submanifolds and their ambient geometries. 

On the submanifold $\Sigma_p$, the metric decomposes as follows
\begin{equation}
g_{\mu\nu} = h_{\mu\nu} + N_{\mu\nu}\,,
\end{equation}
which, implies for example
\begin{align}
\ovl g^{ab}R_{ab} =h^{\mu\nu} R_{\mu\nu} =  R - N^{\mu\nu}R_{\mu\nu}\,,
\end{align}
and
\begin{equation}\label{eq:box}
\Box f = h^{\mu\nu} D_\mu D_\nu f+N^{\mu\nu}D_\mu D_\nu f= \ovl\Box f-\II^\mu D_\mu f+N^{\mu\nu}D_\mu D_\nu f\,,
\end{equation}
for some scalar function $f$. In the second equality of \eq{box} we used $h^{\mu\nu}=e_a^\mu e_b^\nu \,\ovl g^{ab}$, $e_a^\mu D_\mu = \ovl D_a$, the product rule, and the definition $\II^\mu =  \ovl g^{ab} \oD_a e_b^\mu$. 

The Gauss equation relates the intrinsic Riemann tensor $\ovl R_{abcd}$ on the submanifold and the pullback of the ambient Riemann tensor $R_{abcd}$. Together with the equation's contractions with the induced metric $\ovl g^{ab}$, they read
\begin{subequations}
\begin{align}
R^a{}_{bcd} &= \ovl{R}^a{}_{bcd} - 2 \II_{\mu [c}{}^a \II^\mu_{d]b}\,, \label{eq:gauss}\\
R_{ab} &=\ovl{R}_{ab}-\II_{\mu ab} \II^\mu + \II_{\mu ac} \II^{\mu}{}_b{}^c +  N^{\rho\sigma} R_{a\rho b\sigma} \, ,\\
R &=\ovl{R}-\II_\mu \II^\mu + \II_{\mu ab} \II^{\mu ba} +2N^{\mu\nu} R_{\mu\nu}- N^{\mu\rho} N^{\nu\sigma} R_{\mu\nu\rho\sigma} \, . \label{eq:gauss-2}
\end{align}
\end{subequations}
The Codazzi relation and its contraction with $\ovl g^{ab}$ read
\begin{subequations}
\begin{align}
N^\mu_\nu R^\nu{}_{abc}&=N^\mu_\nu \left(\oD_b \II^\nu_{ca}-\oD_c \II^\nu_{ba}\right)\,,\label{eq:codazzi}\\
N^\mu_\nu R^\nu{}_b &= N^\mu_\nu \left(\oD_b \II^\nu - \oD_c \II^\nu_{b}{}^c\right) + N^{\mu\nu} N^{\rho\sigma} R_{b\sigma \nu\rho} \,.
\end{align}
\end{subequations}
Finally, we also have the Ricci equation
\begin{equation}\label{eq:ricci}
N^\mu_\rho N^\sigma_\nu R^\rho{}_{\sigma ab} = (R^\perp)^\mu{}_{\nu ab} - \II^\mu_{ac} \II_{\nu b}{}^c + \II^\mu_{bc}\II_{\nu a}{}^c \,.
\end{equation}

Together with the first and second Bianchi identities, the Gauss-Codazzi-Ricci relations above can be used to derive differential equations involving intrinsic and extrinsic curvature tensors. The ones we find for co-dimension $q\geq 2$ are similar to the ones listed in appendix~A of~\cite{FarajiAstaneh:2021foi} for $q=1$. For $q\geq 2$, however, the number of such relations is larger, and they involve many more terms, so we do not list them here. Instead we refer the interested reader to our supplemental Mathematica notebook.

\section{$\dbcc{1}$ in $\langle T \CD\rangle$}\label{app:TD}

In this appendix, we relate the scale anomaly in $\langle T_{\mu\nu} (x) \CD (0) \rangle$ to the curvature invariant $\dbcc{1} {\mathcal I}$ in the trace anomaly for a $d=5$ CFT with a boundary in eq.~\eqref{eq:boundary-Weyl-anomaly}.  (The higher co-dimension case is more involved, because the correlator involves an additional tensor structure.) Of course, we have already shown that $\dbcc{1}$ is determined by the displacement operator's two-point function, and specifically its coefficient $c_{\CD \CD}$: see eq.~\eqref{cDDd1rel}. Here we take a different route to the same result. Instead of varying the scale anomaly in the effective action twice with respect to the defect's embedding function, $X^i(\mathbf{y})$, we will vary once with respect to $X^i(\mathbf{y})$, thus introducing a displacement operator $\CD$, and once with respect to the metric, thus introducing $T_{\mu\nu}$.

The two-point function $\langle T_{\mu\nu} (x) \CD (0) \rangle$ is completely fixed by conformal symmetry, up to a single number. That number is $c_{\CD \CD}$ due to the identification $\lim_{x_\perp \to 0} T_{nn} (x) = \CD({\bf y})$. In preparation for a Fourier transform along the defect, we write $\langle T_{\mu\nu} (x) \CD (0) \rangle$ in components, and for general $d$:
\begin{eqnarray}
\langle T_{ab}(x) \CD(0) \rangle &=& \frac{d}{d-1}\frac{c_{\CD \CD}}{({\bf y}^2 + x_\perp^2)^{d+2}} \left(4 x_\perp^2 y_a y_b - \frac{1}{d} \delta_{ab} ({\bf y}^2 + x_\perp^2)^2 \right) \nonumber \ , \\
\langle T_{na} (x) \CD(0)  \rangle &=& \frac{d}{d-1}\frac{2 c_{\CD \CD} x_\perp y_a}{({\bf y}^2 + x_\perp^2)^{d+2}}(x_\perp^2 - {\bf y}^2) \ , \\
\langle T_{nn} (x) \CD(0) \rangle &=& \frac{c_{\CD \CD}}{({\bf y}^2 + x_\perp^2)^{d+2}} \left(- \frac{4 d}{d-1} x_\perp^2 {\bf y}^2 +  ({\bf y}^2 + x_\perp^2)^2 \right) \nonumber \ .
\end{eqnarray}

A Fourier transform is perhaps the quickest, if not the most elegant, way of identitying the anomalous contribution to the two point function.
Specializing to $d=5$, we find\footnote{
To perform the Fourier transform, we need
\[
\int d^4 y \frac{e^{i k \cdot {\bf y} }}{({\bf y}^2 + x_\perp^2)^\alpha} = 2 \pi^2 \left( \frac{k}{2 x_\perp} \right)^{\alpha-2} \frac{1}{\Gamma(\alpha)} K_{2-\alpha} (kx_\perp)  \ .
\]
}
\begin{eqnarray}
\frac{1}{c_{\CD \CD}} \int d^4 y e^{i k \cdot {\bf y}} \langle T_{ab}(x) \CD(0) \rangle &=&  \frac{\pi^2(\delta_{ab} k^2 - 4 k_a k_b) }{1152 x_\perp^4}
- \frac{\pi^2 (\delta_{ab} k^2 - 2 k_a k_b) k^2 }{4608 x_\perp^2} + \CO(k^6 \log x_\perp)  \nonumber \ , \\
\frac{1}{c_{\CD \CD}} \int d^4 y e^{i k \cdot {\bf y}} \langle T_{na}(x) \CD(0) \rangle &=& \frac{i \pi^2 k_a k^2}{1152 x_\perp^3} - \frac{i \pi^2 k_a k^4}{4608 x_\perp} + \CO(k^7 x_\perp \log x_\perp) \ ,  \\
\frac{1}{c_{\CD \CD}} \int d^4 y e^{i k \cdot {\bf y}} \langle T_{nn}(x) \CD(0) \rangle &=& \frac{\pi^2 k^4}{2304 x_\perp^2} + \CO(k^6 \log x_\perp )  \nonumber \ .
\end{eqnarray}
By writing $1/x_\perp^m$ as a derivative operator acting on a logarithm of $x_\perp$,
\[
\frac{1}{x_\perp^m} = \frac{(-1)^{m-1}}{(m-1)!} \frac{\partial^m}{\partial x_\perp^m} \log (x_\perp \mu)  \Theta(x_\perp) \ ,
\]
 we can identify the scale anomaly from these Fourier transforms. The Heaviside theta function $\Theta(x_\perp)$ implements the boundary condition that nothing is beyond the boundary at $x_\perp = 0$. By first performing the variation $\mu\, \partial/\partial\mu$ and then taking a $\partial_{n}$ derivative, we will generate a Dirac delta function $\delta(x_\perp)$.  We can then read off from the Fourier transforms the anomalous contributions to the two-point function:
\begin{eqnarray}
\mu \frac{\partial}{\partial\mu}  \langle T_{ab}(x) \CD(0) \rangle
&=&
\frac{\pi^2 c_{\CD \CD}}{1152} \biggl[ -\frac{1}{6}\partial_n^3 ( -\delta_{ab} \ovl{\Box}+ 4 \partial_a \partial_b)   
\nonumber \\
&&
- \frac{1}{4}\partial_n (-\delta_{ab} \ovl{\Box}+ 2 \partial_a \partial_b) \ovl{\Box}\biggr]  \delta ({\bf y}) \delta(x_\perp) \ ,  \nonumber \\
\mu \frac{\partial}{\partial\mu}  \langle T_{na}(x) \CD(0) \rangle 
& = &
\frac{\pi^2 c_{\CD \CD}}{2304} \left[\partial_n^2 \partial_ a \ovl{\Box}
+ \frac{1}{2} \partial_a\ovl{\Box}^2 \right] \delta ({\bf y}) \delta(x_\perp)  \ , \\
\mu \frac{\partial}{\partial\mu}   \langle T_{nn}(x) \CD(0) \rangle 
&= & - \frac{\pi^2 c_{\CD \CD}}{2304}\partial_n\ovl{\Box}^2  \delta ({\bf y}) \delta(x_\perp) \nonumber \ .
\end{eqnarray}

We want to compare this anomaly in $\langle T \CD \rangle$ with the corresponding pieces of the anomaly in the effective action:
\[
\left. \delta_\omega {\mathcal W} \right|_{\dbcc{1}} = \frac{\dbcc{1}}{(4 \pi)^2}  \int_{\partial M} \left( \oK^{ab} D_n W_{nanb}  + \frac{2}{9} \bar D^b \oK_{ba} \bar D^c {\oK_c}{}^{a}  + \ldots\right) \  .
\]
The variation of the $K^2$ term is straightforward:
\begin{eqnarray}
\lefteqn{ \delta_g \delta_{x_\perp} ((\bar D^b \oK_{ab} )(\bar D^c \oK_{cd}) \ovl{g}^{ad}) =
2 (\delta_{x_\perp} \bar D^a \oK_{ab} )(\delta_g \bar D^c \oK_{cd}) \ovl{g}^{bd} } \\
&=& \frac{3}{2} \left[ - \frac{3}{4} \ovl{\Box}^2 \partial^a \delta g_{a n} 
+ \frac{1}{2} \ovl{\Box} \partial_n \partial^a \partial^b \delta g_{ab} - \frac{1}{8} \ovl{\Box}^2 \partial_n \delta g^c{}_c
 \right]  \ .
\nonumber
\end{eqnarray}
The variation of the $KW$ term is a bit more involved. Combining the two, we find
\begin{eqnarray}
\delta_\omega  \langle T_{nn} (x) \CD(0) \rangle &=& 2 \frac{\dbcc{1}}{ (4 \pi)^2}\frac{1}{4} 
\partial_{n}\ovl{\Box}^2 \delta^{(4)}({\bf y}) \delta(x_\perp) \ , \nonumber  \\
\delta_\omega \langle T_{na} (x) \CD(0) \rangle &=&  - 2 \frac{\dbcc{1}}{(4 \pi)^2}  \left(\frac{1}{4}\partial_n^2 \partial_a\ovl{\Box}+
 \frac{1}{8}\ovl{\Box}^2 \partial_a \right)
 \delta^{(4)}({\bf y}) \delta(x_\perp) \ , \\
\delta_\omega \langle T_{ab} (x) \CD(0) \rangle &=& 2 \frac{\dbcc{1}}{ (4 \pi)^2} 
\biggl( 
\frac{1}{3}\partial_n^3 \partial_a \partial_b - \frac{1}{12} \delta_{ab}\partial_n^3 \ovl{\Box} \nonumber 
+ \left( \frac{1}{12}  + \frac{1}{6}\right)\partial_n \ovl{\Box} \partial_a \partial_b
\\
&& 
- \left( \frac{1}{12} +\frac{1}{24} \right)\delta_{ab}\partial_n \ovl{\Box}^2 
  \biggr)  \delta^{(4)}({\bf y}) \delta(x_\perp) \nonumber \ .
\end{eqnarray}
This recovers the previous relationship between the $\langle \CD \CD \rangle$ two-point function and the anomaly coefficient of the ${\mathcal I}$ invariant in eq.~\eqref{cDDd1rel}, $c_{\CD \CD}= - 72/\pi^4 \dbcc{1}$.

\bibliographystyle{JHEP}
\bibliography{4dAnomalies}
\end{document}